\documentclass[article,12pt]{JHEP}
\usepackage{amssymb,amsfonts,bm}
\usepackage[vcentermath,enableskew]{youngtab}
\usepackage{epsfig}
\usepackage{longtable}
\relax
\renewcommand{\theequation}{\arabic{section}.\arabic{equation}}
\def\be{\begin{equation}}
\def\ee{\end{equation}}
\def\bs{\begin{subequations}}
\def\es{\end{subequations}}
\newcommand{\een}{\end{subequations}}
\newcommand{\ben}{\begin{subequations}}
\def\beq{\begin{equation}}
\def\eeq{\end{equation}}

\def\g{\gamma}

\def\m{\mu}
\def\n{\nu}

\def\a{\alpha}
\def\b{\beta}

\def\d{\partial}

\def\sp{\;\;\;,\;\;\;}

\def\e{\epsilon}
\def\brn{\bf}
\def\diag{{\rm diag}}
\def\eqp{\;\;.}

\newcommand\fverb{\setbox\pippobox=\hbox\bgroup\verb}
\newcommand\fverbdo{\egroup\medskip\noindent%
                        \fbox{\unhbox\pippobox}\ }
\newcommand\fverbit{\egroup\item[\fbox{\unhbox\pippobox}]}
\newbox\pippobox

\def\beq{\begin{equation}}
\def\eeq{\end{equation}}
\newcommand{\bea}{\begin{eqnarray}}
\newcommand{\eea}{\end{eqnarray}}
\def\nn{\nonumber}
\def\d{\delta}

\def\4R{{{}^{(4)}R}}

\def\K5{{\kappa}}
\def\K52{{\kappa^2}}

\def\hre#1#2{\href{http://arxiv.org/abs/#1/#2}{[ArXiv:#1/#2]}}

\title{Orientifolds, hypercharge embeddings and the Standard Model}
\author{P. Anastasopoulos$^1$, T. Dijkstra$^2$,
E. Kiritsis$^{3,4}$, B. Schellekens$^{2,5}$\\
~\\
$^1$Department of Physics,
University of Athens,\\
Panepistimiopolis, Zografou
157 84 Athens, Greece\\
~\\
$2$NIKHEF,
Kruislaan 409, 1009DB Amsterdam,
The Netherlands\\
~\\
$^3$CPHT, Ecole Polytechnique,
 91128, Palaiseau, France\\
 ( UMR du CNRS 7644).\\
~\\
$^4$Department of Physics, University of Crete\\
71003 Heraklion, Greece\\
~\\
$^5$IMAPP, Radboud Universiteit Nijmegen, The Netherlands\\
~\\
{\tt http://www.cpht.polytechnique.fr/cpth/kiritsis/}}

\preprint{\hepth{0605226} \\ CPHT-RR015.0306\\
NIKHEF/2006-003}

\abstract{The embedding of the SM hypercharge into an orientifold
gauge group is studied. Possible embeddings are classified, and a
systematic construction of bottom-up configurations and top-down
orientifold vacua is achieved, solving the tadpole conditions in
the context of Gepner orientifolds. Some hypercharge embeddings
are strongly preferred compared to others. Configurations with
chiral antisymmetric tensors are suppressed. We find among others,
genuine examples of supersymmetric SU(5), flipped SU(5),
Pati-Salam and trinification vacua with no chiral exotics.}

\begin{document}

\section{Introduction}

During the past twenty years it has become clear that the
information String Theory gives us about the Standard Model (SM)
is extremely complex. It does not seem that String Theory, or some
selection principle on top of it, will gives us a unique
four-dimensional gauge theory that is identical to the Standard
Model. On the other hand, there may be non-trivial restrictions on
the kind of gauge theories that can emerge from String Theory, and
it is not {\it a priori} obvious that the Standard Model satisfies
those restrictions.

In this situation, there are several approaches one can follow. It
would be a tremendous success if one could find a string vacuum
that precisely matches all current experimental constraints. For
many years it seemed plausible that those constraints would be
restrictive enough to reduce the number of vacua to (at most) a
single one.  The hope was that, having found that vacuum, we would
see all the remaining pieces of the puzzle fall into place, and we
could start making falsifiable predictions for future experiments.

However, it now seems wishful thinking to believe that this will
actually be true. Although reliable estimates cannot be made,
naive guesses suggest that the number of vacua meeting all current
experimental constraints may well be much larger than 1. Even in
that situation, finding just one of those would be a huge success,
at least  as an existence proof. But making predictions based on
such a vacuum is a rather delicate affair if one does not know the
complete ensemble of  vacua satisfying all current constraints.
This does not mean that no further predictions can be made.  There
is no reason to expect the moment of discovery of the landscape of
string theory, \cite{Susskind:2003kw}, \cite{Schellekens:2006xz}
to coincide with the end of such successful theoretical
predictions. But all successes of the past (such as the relation
between $\alpha$ and $g-2$) can be understood in terms of the
quantum field theory description of the Standard Model. There may
be further successes of this kind, but what one would really hope
to find is a genuine string prediction.

At present too little is known about the set of String Theory
vacua to be able to say how far this programme can be pushed. One
extreme might be that the problem is too (NP)hard for us to solve
\cite{Denef:2006ad},  and that we will have to be satisfied with
having a certain degree of confidence that the Standard Model does
indeed exist somewhere in the landscape, just as we are confident
that the DNA molecule is a solution of QED, without being able to
write it down explicitly. To accept such an outcome would require,
at the very least, some kind of confirmation that String Theory is
the correct theory of Quantum Gravity. The other extreme is that
the potentially huge set of unfixed degrees of freedom do not
actually exist for the Standard Model, or are confined to an
irrelevant sub-sector, such as a barely observable (``hidden")
sector.

In either case it is clearly essential to expand our knowledge of
the landscape of string vacua by all means at our disposal and to
understand the possible realizations of vacua that have the SM as
a low energy limit. There are two approaches to that end. The
first is the top-down approach that constructs string vacua using
CFT techniques and then checks whether their low energy limit
compares favorably to the SM. This approach has been used
extensively in heterotic model building, and more recently in
orientifold model building.

The other approach is the bottom up approach that has been
especially suited to the orientifold context, \cite{dsm1,aiqu}.
This is because the back-reaction of a brane-configuration
comes-in at the next order in the coupling constant expansion. It
has slowly become clear that in searching for the SM-like vacua, a
combination of the two approaches may be the most efficient one.

In this paper, we want to make some modest steps towards
understanding the complexity of the landscape, and in particular
the different possibilities for realizing SM-like vacua. In
particular, it is known that there are several possible embeddings
of the SM hypercharge into the orientifold gauge group
\cite{dsm1,Ibanez:2001nd,AD}. Such embeddings affect crucial
phenomenological properties of the vacua. It is therefore
important to analyze such embeddings. For this, instead of
focusing on a particular model we will try to broaden the scope as
much as possible. In \cite{Dijkstra:2004ym},
\cite{Dijkstra:2004cc} two of the authors presented a detailed
investigation of a piece of the landscape that until then was
barely accessible: orientifolds of Gepner models. The approach of
these papers can be described as a mixture of a top-down and a
bottom-up method. On the one hand, exact string solutions were
looked for and found. But on the other hand, the kind of solutions
that were searched were limited {\it a priori} by a choice of a
``bottom up" realization of the Standard Model, constructed out of
intersection sets of branes\rlap.\footnote{This terminology is
used here only to guide the intuition. In reality the models are
described algebraically in terms of annulus coefficients in
boundary CFT.}

The scope of the RCFT method, even when restricted to Gepner
models, seems to be considerably larger than that of the much more
extensively studied orbifold models. Indeed, the first example of
a supersymmetric spectrum that matches the standard model exactly
(in the chiral sense) was found using an RCFT construction in
\cite{Dijkstra:2004ym}.  This was an amazing eight years after the
first steps towards realistic model building with orientifolds
were taken \cite{Angelantonj:1996uy}, using orbifold methods.
Since that pioneering paper, the orbifold/orientifold method has
been explored extensively by many authors (see
\cite{Blumenhagen:2005mu} and references therein) who succeeded in
getting ever closer to the supersymmetric standard model spectrum,
until that goal was finally reached in \cite{Honecker:2004kb} for
the $Z_6$ orbifold.

During the same period there has been relatively little work on
Gepner orientifolds
\cite{Angelantonj:1996mw}-\cite{Aldazabal:2004by}, and with
relatively little success, the first paper finding a chiral
spectrum being \cite{Brunner:2004zd} in 2004. However, it is now
clear that the lack of success was due to the fact that until
recently only a limited number of partition functions and boundary
states was accessible. A recent investigation
\cite{Gmeiner:2005vz} of $Z_2 \times Z_2$ orbifolds has shown that
the three family standard model spectrum is just beyond the limit
of statistics in that case. By contrast, with Gepner models more
than 200.000  standard model realizations were found in
\cite{Dijkstra:2004cc}, despite the fact that the average success
rate is actually lower (empirically, ``one in a billion" for $Z_2
\times Z_2$ orbifolds, and about $4 \times 10^{-14}$ for Gepner
models).

On the other hand, RCFT methods also have clear disadvantages in
comparison to orbifold methods. In particular, they do not allow
continuous variations of moduli, and are not suitable for
discussions of flux compactifications and moduli stabilization, at
least not without radically new ideas. But their larger scope
makes them ideally suited for scanning a substantially larger part
of the landscape than was possible up to now, provided one focuses
only on issues related to spectroscopy. This is precisely our goal
in this paper. Our main phenomenological input will be the chiral
spectrum of the standard model. Our intention is to loosen
considerably the bottom up assumptions made in
\cite{Dijkstra:2004cc}, and investigate a large number of other
ways of realizing the standard model with D-branes (or boundary
states).

The kind of bottom up models considered in \cite{Dijkstra:2004cc}
were variations on the model first proposed in
\cite{Ibanez:2001nd}. They are characterized by four stacks of
branes with a Chan-Paton group $U(3)_{\brn a} \times U(2)_{\brn b}
\times U(1)_{\brn c} \times U(1)_{\brn d}$, with the standard
model generator $Y$ embedded as $Y=\frac16 Q_{\brn a} -\frac12
Q_{\brn b} -\frac12 Q_{\brn c}$. The variations include  the
possibility of choosing the second and third Chan-Paton factor
real, and allowing the $B-L$ abelian vector boson to be either
massive or massless in the exact string theory. These models have
a perturbatively unbroken baryon and lepton number.

Many other brane realizations exist, and some of those have been
discussed in the literature. To obtain the results of
\cite{Dijkstra:2004cc} a huge effort was required in terms of
computer time. In principle, this project could be redone for
anyone's favorite bottom-up model. However, it seems preferable to
try  to remove the bias implied by a particular choice of model,
and try to repeat the computation assuming as little as possible
about the bottom-up realization.

In principle, the only feature we assume is the most robust part
of what we presently know about the Standard Model: that there are
three chiral families of quarks and leptons in the familiar
representations of $SU(3)\times SU(2) \times U(1)$. In practice,
we still have to make a few concessions. In particular, we will
have to limit the number of participating branes and forbid
non-chiral mirror pairs of arbitrary charge. This will be
discussed in more detail in the next section.

The new features that we do allow include
\begin{itemize}
\item{Anti-quarks realized as anti-symmetric tensors of $U(3)$}
\item{Charged leptons  and neutrinos realized as anti-symmetric tensors}
\item{Non-standard embeddings of the $Y$-charge}
\item{Embeddings of $Y$ in non-abelian groups}
\item{Strong-Weak unification ({\it e.g.} $SU(5)$)}
\item{Baryon-lepton unification ({\it e.g.} Pati-Salam models)}
\item{Trinification}
\item{Baryon and/or lepton number violation}
\item{Family symmetries}
\end{itemize}
We are not claiming that all of these features are desirable, but
our strategy is to allow as many possibilities in an early stage,
and leave the final selection to the last stage, so that it will
not be necessary to restart the entire search procedure if new
insights emerge.

Some of these options may address unsolved problems that occur
for the standard realization  \cite{Ibanez:2001nd} of the standard
model. For example, the perturbatively unbroken lepton number of
these models makes it hard to implement a see-saw like mechanism
to give small masses to neutrinos. Coupling constant unification,
if it is indeed a fundamental feature of nature and not a
semi-coincidence, is not automatic in the standard realization,
but it would be in $SU(5)$ models. This does not mean that the
standard realization cannot accommodate the current experimental
values of the couplings constants, but only that the fact that
they presently appear to converge (with gaugino contributions
taken into account) would be a mere coincidence. We have indeed
found some really simple and elegant realizations of $SU(5)$
models, but unfortunately we did not find a credible mechanism for
generating up-quark masses. We will comment on this and on the
viability of some of the other options in section \ref{Pheno}.

One of our goals is to analyze which model can be built from a
bottom-up point of view, and how many of them can be realized as
top-down models. By ``bottom-up" we mean here a brane realization
that produces the correct chiral standard model spectrum if the
gauge group is reduced to $SU(3)\times SU(2)\times U(1)$ (without
assuming a particular mechanism for that reduction). On the
``top-down" side two types of concepts should be distinguished:
standard model brane configurations and solutions to the tadpole
conditions. The focus in this paper is on the former, {\it i.e.}
choices of boundary labels\footnote{We label the complete set of
boundaries of a given modular invariant partition function of a
CFT as $a,b,c,d,\ldots$. The specific boundaries that participate
in a Standard Model configuration are denoted as {\brn a}, {\brn
b}, {\brn c} and {\brn d}. We allow a maximum of four (plus a
hidden sector), with the first two corresponding to $SU(3)_{\rm
color}$ and $SU(2)_{\rm weak}$.} {\brn a}, {\brn b}, {\brn c} and
{\brn d} such that with an appropriate choice of the Chan-Paton
gauge group and the appropriate embedding of $SU(3)\times SU(2)
\times U(1)$ one obtains the standard model. Here we also require
that the standard model $U(1)$ generator does not acquire  mass
due to bilinear axion couplings.

Given such a standard model configuration, there may still be
uncancelled tadpoles in RR closed string one-point functions on
the disk and the crosscap. Within this context, the only way to
cancel them is to add additional hidden matter, except in a few
cases where they already cancel among the standard model branes.
To see if this can happen is an extremely time-consuming, and
ultimately unsolvable problem. Furthermore for any given brane
configuration there may be many ways of cancelling the tadpoles.
In the continuum theory, background fluxes, not considered here,
contribute to the tadpoles. But perhaps more importantly, the set
of boundary states we consider here is limited by the choice of
rational CFT. We consider the complete set of boundaries allowed
by the RCFT, {\it i.e.} all boundaries that respect its chiral
algebra. But that chiral algebra is larger than the $N=2$
world-sheet algebra required to describe a geometric Calabi-Yau
compactification. Since we get the $c=9$ chiral algebra as a
tensor product of minimal $N=2$ algebras, the chiral algebra also
contains all  differences of the $N=2$ algebras of the factors. If
we would reduce the chiral algebra, additional boundary states are
allowed, and could contribute to tadpole cancellation. Of course
this also allows additional ways of constructing standard model
configurations, but we cannot make  regarding a complete
classification there anyway.

It is essentially impossible to conclude, with RCFT techniques
alone, that the tadpoles of a certain standard model configuration
cannot be cancelled. Positive results, on the other hand, imply
that one has a valid supersymmetric string vacuum. We see tadpole
cancellation therefore mainly as an existence proof of a given
string vacuum. Once that proof has been given, we do not continue
searching for additional tadpole solutions for the same chiral
configuration.  This gives  an enormous cut-off in computer time.
One should keep in mind that for the most frequent chiral model
considered in \cite{Dijkstra:2004cc},  we found a total of 16
million tadpole solutions (about 110000 of them distinct). We now
keep only one of those solutions. This also implies that we cannot
provide meaningful statistical results regarding tadpole
solutions, but only regarding brane configurations.

We summarize  briefly our results:
\begin{itemize}

\item We develop a detailed classification of allowed embeddings
of the SM hypercharge inside the orientifold gauge group. To do
this, we classify brane stacks according to how they contribute to
the hypercharge. The hypercharge embedding is then characterized
by a real variable $x$ which is quantized in half-integral units
in genuine non-orientable vacua.

\item We produce 19345 chirally distinct top-down SM spectra
(before tadpole cancellation) and 1900 chirally distinct models
solving the tadpole conditions and realizing the different
embeddings.

\item We find that the $x={1\over 2}$ hypercharge embedding
dominates by far all other choices. The Madrid embedding
\cite{Ibanez:2001nd} belongs to this class.

\item The presence of chiral symmetric and antisymmetric tensors
is highly suppressed. For some hypercharge embeddings, such
tensors  are crucial for anomaly cancellation and they may produce
anti-quarks and other weak singlets. This implies the associated
suppression of such embeddings.

\item We produce the first examples of supersymmetric SU(5) and
flipped SU(5) orientifold vacua, with the correct chiral spectrum
(no extra gauge groups and no exotic $G_{CP}$ chiral  states).
However, as we argue, all such orientifold models, as well as
models with quarks in the antisymmetric representation have a
serious phenomenological problem associated with masses.

\item We find some minimal supersymmetric Pati-Salam and
trinification vacua.

\item We have examples of spectra (but no tadpole solutions yet)
with extended ({\cal N}=4 or {\cal N}=8) supersymmetry in the bulk
and {\cal N}=1 supersymmetry on the branes.

\item We have found SM spectra solving the tadpole conditions on a
relative of the quintic CY.

\end{itemize}

This paper is organized in follows. In the next section, we define
precisely what our criteria for standard model realizations are.
In section \ref{ClassBU}, we work out the consequences of these
criteria from the bottom-up point of view. We classify these
models in terms of a parameter $x$ which determines the embedding
of $Y$ in the {\brn a} and {\brn b} stacks. We first identify a
class of models for which $x$ is undetermined by the quark and
lepton charges. In all other models $x$ is half-integer, with
$x=0$ corresponding to $SU(5)$-type models \cite{AD} as well as
models A,A'in \cite{dsm1,dsm3}, $x=\frac12$ to Madrid-type
configurations, \cite{Ibanez:2001nd}, and $x=1$ in models B,B' in
\cite{dsm1,dsm3}. We present some explicit realizations of each
type of model, and also identify possible lepton number symmetries
and Higgs boson realizations. In section \ref{TDspectra} we
present the results of  a search for orientifolds of Gepner
models. We limit ourselves to simple current modular invariant
partition functions (MIPFs) with at most 1750 boundaries. In
section \ref{TadSol} we present a small sample of the tadpole
solutions that we have found, providing concrete example of a
variety of types of models. In section \ref{Pheno} we discuss the
viability of various features of models we found, focusing mainly
on masses. Finally in section  \ref{CY} we examine correlations
between the topology of the CY manifold, and the features of the
SM realization.

In appendix \ref{Algo} we explain in detail  the search algorithm
for these configurations in RCFT orientifolds. In appendix
\ref{gg}, we examine the correlation between gauge coupling
constants, and the allowed values for the string scale for
different hypercharge embeddings.

\section{What we are looking for\label{WWALF}}

Our goal is to search for the most general embedding of the
standard model in the Chan-Paton gauge group of Gepner
Orientifolds.

We first introduce some notation. We denote the full Chan-Paton
group as $G_{\rm CP}$. This is the group obtained directly from
the multiplicities of the branes, without taking into account
masses generated by two-point axion-gauge boson couplings. We
require that the standard model gauge group, $G_{\rm SM} = SU(3)
\times  SU(2) \times   U(1)_Y$ is a subgroup of $G_{\rm CP}$.
Furthermore we require that the generator of $U(1)_Y$ does not get
a mass from axion-gauge boson couplings.

The main condition we impose on the spectrum is the presence of
three families of quarks and leptons, and the absence of chiral
exotics. Since chirality can be defined with respect to various
groups, and the term ``exotics" is used in different senses in the
literature, we will define this more precisely.
Group-theoretically, the standard-model-like spectra we allow are
described as follows. Denote the full set of massless
representations of $G_{\rm CP}$ as $R_{\rm CP}$. The subset of
these representations that is chiral with respect to $G_{\rm CP}$
is denoted $R^{\rm chir}_{CP}$. The  reduction of these
representations to the the group $G_{\rm SM}$ are denoted as
$R_{\rm SM}$ and $R^{\rm chir}_{\rm SM}$ respectively. By
``reduction" we mean here only that we decompose representations
in terms of representations of a subgroup. No assumptions are made
at this point regarding dynamical mechanisms (like the
Brout-Englert-Higgs mechanism) to achieve such a reduction.
Consider now the subset of either $R_{\rm SM}$ or $R^{\rm
chir}_{\rm SM}$ that is chiral with respect to $G_{\rm SM}$. The
result is required to be precisely the following set of
left-handed fermions (all fermions will be in left-handed form in
this paper) \beq \label{eq:SMrep} 3 \times  [(3, 2,\frac16) + (3^*
, 1,-\frac23) + (3^* , 1, \frac13)+ (1, 2,-\frac12) + (1, 1, 1)]
\eeq Any other particles must be non-chiral with respect to
$G_{\rm SM}$. This may include left-handed anti-neutrinos in the
representation $(1,0,0)$ and MSSM Higgs pairs, $(1,2,\frac12)+
(1,2,-\frac12)$. Anything else will be called exotic.

The foregoing describes the most general configuration one could
reasonably call an embedding of the standard model without chiral
exotics, but we will have to impose a few additional constraints
to make a search feasible. First of all we require that the
standard model groups $SU(3)$ and $SU(2)$ come each from a single
stack of branes, denoted {\brn a} and {\brn b} respectively. This
forbids diagonal embeddings of these groups in more than one CP
factor. In general by a stack we mean a single label for a real
(orthogonal or symplectic) boundary, or a pair of conjugate labels
for complex, unitary branes. The CP factor yielding $SU(3)$ must
be $U(3)$, whereas the weak gauge symmetry $SU(2)$ can come from
either  $U(2)$ or $Sp(2)$. The group $O(3)$ is not allowed,
because one cannot get spinor representations of orthogonal groups
in perturbative open string constructions.

The hypercharge generator $Y$ is a linear combination of the
unitary phase factors of $U(3)$,  $U(2)$ (if available) and any
other generator of one of the other factors in $G_{\rm CP}$. All
representations $(3, 2)$ must necessarily come from
bi-fundamentals of the {\brn a} and {\brn b} stacks, but not all
anti-quarks can come from those stacks. Although there can be
anti-quarks due to chiral anti-symmetric tensors of $SU(3)$, they
all have the same hypercharge. Hence there must be at least one
other stack of branes, labeled {\brn c}.

In principle there could be any number of additional stacks of
branes, but for purely practical reasons we allow at most one more
stack (labeled {\brn d}) to contribute to the standard model
representation (\ref{eq:SMrep})\footnote{In general we also expect
that the number of exotics to rise fast with the number of
additional stacks participating in the SM group.}. Additional
branes may be present, and may be required for tadpole
cancellation. They will be referred to as the ``hidden sector". If
stack {\brn d} does not contribute to  (\ref{eq:SMrep}) at all we
regard it as part of the hidden sector. The standard model branes
{\brn a}, {\brn b},{\brn c} (and {\brn d}, if present) will be
called the ``observable sector". Note that left-handed
anti-neutrinos\footnote{Since our convention is to represent all
matter in terms of left-handed fermions, right-handed neutrinos
are referred to as left-handed anti-neutrinos.} are not listed in
(\ref{eq:SMrep}). We do not impose an {\it a priori}
constraint\footnote{The minimum number is two in order to
accommodate the experimental data. We will comment further on
neutrino masses in section \ref{nm}.} on the number of left-handed
anti-neutrinos, although in some cases a certain number of such
states is required by anomaly cancellation in $G_{\rm CP}$. They
may in fact come from the hidden sector or the observable sector,
or even from strings stretching between the two sectors.

Our next condition concerns the precise definition of the standard
model generator $Y$. We allow it to be embedded in the most
general way possible in the Chan-Paton factors of brane {\brn c}
and {\brn d} (in addition to the unitary phases of {\brn a} and
{\brn b}). In principle it could also have components in the
hidden sector  without affecting any of the foregoing, as long as
all particles charged under those components of $Y$ are massive or
at least non-chiral. One could even try to use this as a mechanism
to cancel bilinear axion coupling of $Y$, which would give the
$Y$-boson a mass\footnote{Anomalous U(1) masses have been
calculated for general orientifolds in \cite{au1}.}. We will not
consider that possibility here. This is equivalent to a
restriction to standard model realizations with at most four
participating branes, except for one intriguing possibility: a
three brane realization with a fourth brane added purely to fix
the axion couplings of $Y$, without contributing to quarks or
leptons. This possibility was not included in our search. It
should be mentioned however, that a qualitatively similar
situation does indeed arise. There are orientifold vacua where
there is a U(1) arising from the SM stack of branes , under which
all SM particles are neutral. In this case there is a continuous
family of possible hypercharge embeddings. In some cases, the
masslessness condition breaks the degeneracy. This provides a
string realization of the field theory models in \cite{KN}. In
other cases, even the masslessness condition does not lift the
degeneracy.

The general form of $Y$ is
\beq \label{Yform}
Y=\sum_{\alpha}  t_{\alpha} Q_{\alpha} + W_{\brn c} + W_{\brn d} \ ,
\eeq
where $\alpha$ runs over the values {\brn a},{\brn b},{\brn
c},{\brn d}, $Q_{\alpha}$ is the brane charge of brane $\alpha$
($+1$ for a complex brane, $-1$ for its conjugate, and 0 for a
real brane), and $W_{\brn c}$ and $W_{\brn d}$ are generators from
the non-abelian part of the Chan-Paton group. Therefore $W_{\brn
c}$ and $W_{\brn d}$ are traceless. Such contributions to $Y$
occur for example in Pati-Salam and trinification models, and
therefore we want to allow this possibility.

There is one more condition we impose for practical reasons,
namely that $R^{\rm chir}_{CP}$ may only yield representations of
standard model particles or their mirrors. The main purpose of
this condition (as we will see in more detail below) is to prevent
an unlimited proliferation of $G_{\rm CP}$-chiral, but $G_{\rm
SM}$ non-chiral representations such as $(1, 1, q) + (1, 1,-q)$,
with $q$ arbitrary. In addition, this condition also forbids
triplets of $SU(2)_{\rm weak}$, which can be chiral with respect
to  $U(2)_{\brn b}$.

One may distinguish three types of matter in these models: OO, OH
and HH, where the two letters indicate if the endpoints of the
open string are in the observable or hidden sector. All conditions
on OO matter were already formulated above. The ``no chiral
exotics" constraint formulated above allows HH matter to be chiral
with respect to $G_{\rm CP}$. For OH matter we impose a somewhat
stronger constraint, namely that there cannot be any
bi-fundamentals between the standard model and the hidden sector
that are chiral with respect to $G_{\rm CP}$. This is a stronger
condition because the ``no chiral exotics" constraint allows
SM-Hidden sector bi-fundamentals as long as they are non-chiral
with respect to $G_{\rm SM}$. For example a mirror quark pair
$(3,V)+(3^*,V)$, where $V$ is a vector in a hidden sector $U(N)$
group, could be allowed under the more general rules. The
resulting $U(N)$ anomalies can be cancelled in various ways.

We will allow the brane stacks ${\brn a}$, ${\brn b}$, ${\brn c}$,
${\brn d}$ to have identical labels, with the exception of {\brn
c} and {\brn d} (if they are  identical, we might as  well regard
them as a single brane stack with a larger CP multiplicity). By
allowing identical labels we are able to obtain examples of
unified models, such as (flipped) SU(5) or Pati-Salam like models.
In the case of identical labels, we count them as follows: the QCD
and weak group count as one stack each, and the branes that remain
after removing the QCD and weak groups count as additional stacks,
such that  the total  does not exceed four. For example, we can
get $U(5)$ models with at most two additional CP-factors (plus any
number of hidden sector  branes).

We conclude this section with a summary of the kind of ``exotics"
(plus singlets and Higgs candidates) that may occur in generic
models, indicating which kind we do and do not allow. We split
$G_{\rm CP}$ into an observable and a hidden part as $G_{\rm O}
\times G_{\rm H}$. In all cases we combine representations into
non-chiral sets (usually, but not always pairs) if possible. We
can distinguish the following possibilities
\begin{enumerate}
\item{Matter of type OO
\begin{enumerate}
\item{Non-chiral with respect to $G_{\rm CP}$. This may include symmetric and anti-symmetric
tensors or adjoints of $SU(3)$ or of SU(2), mirror pairs of quarks and leptons, as well as
bi-fundamentals with unusual and in a few cases even irrational charges. All particles
in this class are allowed by our conditions. }
\item{Chiral with respect to $G_{\rm CP}$, non-chiral with respect
to $G_{SM}$. Examples are symmetric tensors of $U(2)_{\rm weak}$,
mirror pairs of quark and lepton doublets that are chiral with
respect to $U(2)_{\rm weak}$, mirror pairs where one member of the
pair is a rank-2 tensor and the other member a bi-fundamental. We
do allow such particles, except the symmetric $U(2)_{\rm weak}$
tensors, and non-chiral pairs of quarks and leptons with
non-standard charges. }
\item{Chiral with respect to $G_{\rm CP}$, chiral with respect to
$G_{\rm SM}$, non-chiral with respect to $\rm{QED} \times
\rm{QCD}$. An example of such exotics would be a fourth family.
Exotics of this type are not allowed by our conditions.}
\item{Chiral with respect to $G_{\rm CP}$, chiral with respect to
$G_{\rm SM}$, and chiral with respect to $\rm{QED} \times
\rm{QCD}$. Clearly this is not acceptable.}
\end{enumerate}
A mass term for exotics of type 1a is allowed by the full gauge
symmetry, and hence it is possible that such a term is generated
by shifting the moduli of the model. It is an interesting question
whether the appearance of such exotics is a special feature of
RCFT, or if they persist outside the rational points. It should be
possible to get some insight in this question by analyzing the
coupling of these particles to the moduli, but this is beyond the
scope of this paper. Exotics of type 1b may get a mass without
invoking the standard model Higgs mechanism, and hence may become
more massive than standard quarks and leptons. However, this will
always require some additional dynamical mechanism beyond
perturbative string theory. Exotics of type 1c require the
standard model Higgs mechanism to get a mass. This may not be
sufficient, since the Higgs couplings themselves may be forbidden
by string symmetries, in which case additional mechanisms must be
invoked. In any case it would be hard to argue that such particles
would be considerably more massive than known quarks and leptons.}
\item{Matter of type HH. These are standard model singlets. No
constraints are imposed on this kind of matter. One may
distinguish two kinds.
\begin{enumerate}
\item{Non-chiral with respect to $G_{\rm CP}$. These particles may
get a mass from continuous deformations of the model, as above.}
\item{Chiral with respect to $G_{\rm CP}$, non-chiral with respect
to $G_{\rm H}$. These particles may get a mass from hidden sector
dynamics.}
\end{enumerate}}
\item{Matter of type OH. In many cases particles in this class
have half-integer charge. This occurs if the electromagnetic
charge gets a contribution $\frac12$ from each observable brane,
which turns out to be the most frequently occurring kind of model.
There are many possibilities for the chiralities, which we list
here for completeness. We use a notation $(\chi_{G_{\rm
CP}},\chi_{G_{\rm H}},\chi_{G_{\rm O}},\chi_{G_{\rm SM}},
\chi_{\rm{QED} \times \rm{QCD}})$, where each $\chi$ indicates
chirality, and can be $Y$ (yes) or $N$ (no).
\begin{enumerate}
\item{(N,N,N,N,N).}
\item{(Y,N,N,N,N).}
\item{(Y,Y,N,N,N).}
\item{(Y,N,Y,N,N).}
\item{(Y,N,Y,Y,N).}
\item{(Y,N,Y,Y,Y).}
\item{(Y,Y,Y,N,N).}
\item{(Y,Y,Y,Y,N).}
\item{(Y,Y,Y,Y,Y).}
\end{enumerate}
An example of type 3b, chiral with respect to the full Chan-Paton
group, but not with respect to any of its subgroups, is
$(3,0,V)+(3^*,0,V)+3\times (1,1,V^*) + 3 \times (1,-1,V^*)$ in
$U(3)\times U(1) \times U(N)$, with the first two factors from
$G_{\rm O}$ and the last from $G_{\rm H}$. Of all these
possibilities, only 3a is allowed by our criteria. Types 3b, 3c
and 3g might be tolerated on more general grounds, and types 3f
and 3i are clearly unacceptable.}
\end{enumerate}

\section{Classification of bottom-up embeddings\label{ClassBU}}

Here we will discuss the possible values of the coefficients
$t_{\alpha}$ that occur in the brane decomposition of $Y$.
We will use the following expression for $Y$:
\beq\label{YLinComb}
Y=\sum_{\alpha}  x_{\alpha} Q_{\alpha} \ ,
\eeq
where $Q_{\alpha}$ is  the $U(1)$ charge of brane $\alpha$. In
contrast to (\ref{Yform}) the sum is here not {\it a priori}
restricted to a definite number of branes. In our search we will
allow also the possibility that diagonal Lie algebra generators
$W$ of $SO(N)$, $Sp(2N)$ or $SU(N)$ groups contribute to $Y$, but
this can always be taken into account by splitting those groups
into $U(m)$ factors according to the $W$ eigenvalues $e_i$. For
example, if there are two distinct eigenvalues\footnote{Two is the
maximum we allow. If there are more, this necessarily yields
unconventional quark or lepton charges. For more details, see
appendix A.} we get for symplectic groups $Sp(2N)$ a contribution
$W_{\alpha}=\diag(N\times (e),N\times(-e))$, which can be
accommodated by splitting $Sp(2N)$ into conjugate brane stacks
with a CP group $U(N)$ and a contribution $e Q_{\alpha}$.
Geometrically, this means that the $2N$ symplectic branes are
moved off the orientifold plane. The same reasoning applies to
$O(2N)$ branes. If there are $O(2N+1)$ stacks, the assumption of
at most two distinct eigenvalues only allows the traceless
generator $W=0$ in (\ref{Yform}), and hence such branes cannot
contribute to $Y$ at all. Finally, $U(N)$ branes can contribute
$t_{\alpha} Q_{\alpha} +\diag (n_1 \times e_1, n_2 \times e_2)$,
with $n_1+n_2=N$, $n_1 e_1+n_2 e_2=0$. This can be regarded as two
stacks $U(n_1) \times U(n_2)$  contributing $(t_{\alpha}+e_1)
Q_{\alpha_1}+(t_{\alpha}+e_2) Q_{\alpha_2}$, so that $x_{\alpha_1}
= t_{\alpha}+e_1$ and $x_{\alpha_2} = t_{\alpha}+e_2$ Therefore
formula \ref{YLinComb} covers all cases.

The brane configurations we consider here are subject to two
constraints: the spectrum must match that of the standard model in
the chiral sense, with chirality defined with respect to $SU(3)
\times SU(2) \times U(1)$. Furthermore all cubic anomalies in each
factor of the full Chan-Paton group must cancel. This must be true
because we want to be able to cancel tadpoles, and tadpole
cancellation imposes cubic anomaly cancellation (mixed anomalies
are cancelled by the generalized Green-Schwarz mechanism). The
tadpoles are usually cancelled by adding hidden sectors, which
adds new massless states to the spectrum. We do not allow these to
be chiral with respect to $SU(3) \times SU(2) \times U(1)$, and
hence they cannot alter the cubic anomalies. The cubic anomaly
cancellation conditions that are derived from tadpole cancellation
are the usual ones for the non-abelian subgroups of  $U(N)$,
$N>2$. Vectors contribute 1, symmetric tensors $N+4$ and
anti-symmetric tensors $N-4$, and conjugates contribute with
opposite signs. But the same condition emerges even if $N=1$ and
$N=2$. This means that for example a combination of three vectors
and an anti-symmetric tensor is allowed in a $U(1)$ factor. This
is counter-intuitive, because the anti-symmetric tensor does not
even contribute massless states, so that one is left with just
three chiral massless particles, all with charge 1. The origin of
the paradox is that it is incorrect to call this condition
``anomaly cancellation" if $N=1$ and $N=2$ and if chiral tensors
are present. It is simply a consequence of tadpole cancellation;
the anomaly introduced by the three charge 1 particles is
factorizable, and cancelled by the Green-Schwarz mechanism.

One might entertain the thought that this peculiar $U(1)$
cancellation might have something to do with the fact that we have
three families of standard model particles. For example, one could
assign the same  $U(1)$ charge to all quarks or leptons of a
certain type, and then cancel this anomaly with anti-symmetric
tensors. This would require this particle type to appear  with a
multiplicity divisible by three. Because the $U(1)$ is anomalous,
it would acquire a mass via the Green-Schwarz term. However,
although configurations of this kind can indeed be constructed,
they are complicated and unlikely to occur. We did indeed find
examples of $U(1)$ anomaly cancellations due to anti-symmetric
tensors, but usually with a more complicated family structure that
does not admit such an interpretation.

\subsection{Orientable configurations}

Let us now return to our goal of determining the possibilities for $Y$.
We begin by demonstrating that in principle all real values of
the leading coefficient $x_{\brn a}$ are allowed.
Using the quark doublet charges we may write $Y$ as follows
\beq
Y = (x-\frac13) Q_{\brn a} + (x-\frac12) Q_{\brn b} + {\rm rest }
\eeq
Here we assume (without loss of generality) that the quark
doublets all come from bi-fundamentals $(V,V^*)$ stretching
between the QCD and the weak brane. The second entry could also be
a $V$, but then we can conjugate $U(2)$ to obtain $(V,V^*)$. A
mixture of $V$ and $V^*$ is however not allowed if we want $x$ to
take all  real values; neither is a chiral anti-symmetric tensor
in either $U(3)$ or $U(2)$, or the option of using $Sp(2)$ instead
of $U(2)$. Here and in the following all representations are in
terms of left-handed spinors.

Now we need lepton doublets. They can only be bi-fundamentals
ending on the $U(2)$. The other end must be on a brane that
contributes to $Y$ in such a way that the total charge is either
$-\frac12$ or $\frac12$. The latter value is considered because in
addition to lepton doublets, we also allow mirrors, or  MSSM Higgs
pairs. Again we will write these bi-fundamentals exclusively as
$(V,V^*)$ (the first entry corresponds to $U(2)$). Mixtures of
$(V,V)$ and $(V,V^*)$ between the same branes would fix $x$, and
if there are no mixtures we can convert all bi-fundamentals to the
form $(V,V^*)$. The multiplicities of these bi-fundamentals may be
negative, in which case we interpret them as $(V^*,V)$.

Since we only allow $SU(2)$ doublets with charges $\pm\frac12$,
the possibilities for the charge coefficients of the new branes
are $x$ or $x-1$. We refer to branes with these charges as ``type
C" and ``type D" respectively (the QCD and weak branes are defined
to be of type A and B respectively. We use small letters {\brn a},
{\brn b}, {\brn c}, {\brn d}, {\brn e},$\ldots$ to label different
stacks, and capitals A,B,C,$\ldots$ to label their types, with
respect to the hypercharge embedding. Branes {\brn a} and {\brn b}
are always of type A and B, but there is no one-to-one
correspondence for the other branes). Note that these types C and
D become equivalent (up to conjugation) if and only if
$x=\frac12$. We are not requiring that the type C or D branes are
identical for all leptons or Higgs, or each other's conjugate,
even if their charges would allow that.

Let $n_1$ be the net number of chiral states between brane {\brn b} and all of the C-type
branes, and $n_2$ the same for type D. To be precise:
\beq
n_1 = \sum_i \left[(N(V,V^*)_{{\brn b}C_i}-N(V^*,V)_{{\brn b}C_i})\right]\ ,
\eeq
where $N$ is the absolute number of massless states with given
properties. We now impose anomaly cancellation in $U(2)$ (for
three families)
\beq -9 + n_1 + n_2 = 0 \ , \eeq
because no chiral tensors are allowed for generic $x$. We also
impose the requirement of having three chiral lepton doublets
\beq n_1 - n_2 = 3 \eeq
which can be solved to yield $n_1=6$ and $n_2=3$. Note that the
anomaly conditions for the Chan-Paton factors at the other end can
aways be satisfied for some of the solutions. This is because the
solution allows all multiplicities of $N(V,V^*)$ as well as
$N(V^*,V)$ to be multiples of three. If we make three open strings
end on the same $U(1)$ brane, the corresponding $U(1)$ anomalies
can always be cancelled by anti-symmetric tensors.

Next we need anti-quarks. Since for general $x$ anti-symmetric
$U(3)$ tensors are not allowed, they must be bi-fundamentals
between the $U(3)$ stack and other branes. If we introduce new
branes for the anti-quark strings to end on, we can always arrange
the configuration so that the anti-quarks are of the form
$(V^*,V)$. Then we need a brane of type C for down anti-quarks and
a brane of type D for up anti-quarks. One may also use one of the
already present branes of type C and D for this purpose, provided
that only combinations $(V,V^*)$ or $(V^*,V)$ are used. Anything
else implies a condition on $x$. Even if one uses distinct branes
for all particle types, there are  many ways to cancel the $U(1)$
anomalies using anti-symmetric tensors.

Finally we need charged lepton singlets and their mirrors.
They can occur in four different ways for generic $x$:
\begin{enumerate}
\item{With both ends on an existing brane of types C and D.}
\item{With one end on a previous C or D brane and one end on a new
one. This would require new branes with various possible charges.
In particular, it allows the following new charges: $x+1$, $x-2$
and their conjugates. We refer to these as types E and F. For
$x=\frac12$ these are each other's conjugates, and for
$x=\frac32,1,0$ and $-\frac12$ some of the types C,D,E and F are
equivalent.}
\item{With both ends on the same, new brane. This requires a new
brane with $t_{\alpha} = \pm\frac12$. We call this type G, unless
it coincides with a previous type.}
\item{With both ends on two distinct new branes. This would in
principle allow two new branes with contributions $y$ and $1-y$
to $Y$. Such branes (if they do not coincide with any previous
type) will be called type H.}
\end{enumerate}
There are even more possibilities if one allows arbitrary numbers
of additional branes for charged leptons. For example, one can
connect new branes to types E and F with charge contributions
$x-2$ or $x+3$, connect new branes to types G and H or add more
branes of type H. By allowing mirror leptons one can build
arbitrarily long chains of branes in this manner. However, this is
too baroque\footnote{It should be kept in mind that as the number
of branes participating in the SM configuration increases, the
number of chiral exotics, fractionally charged particles and other
unwanted states increases exponentially fast.  It is possible that
the lower success rate may be compensated by the potentially
larger number of such configurations. It is still true however,
that the effective field theory of such vacua, will be very
complicated or maybe intractable.} to consider seriously, and can
in any case not be realized with at most four branes, a
restriction we will ultimately impose. Already the fourth option
is then impossible.

Options three and four split the standard model into two chirally
disconnected sectors ({\it i.e.} there are no chiral strings
connecting the two). This implies that the $Y$ anomaly does not
cancel in each sector separately, and hence the two components of
the would-be $Y$-boson must have Green-Schwarz couplings to axions
that give it a mass. In principle these contributions could cancel
for $Y$, but that seems improbable, and hence reduces the
statistical likelihood of this sort of configuration in a search.
Furthermore lepton Yukawa  couplings are perturbatively forbidden
in such models.

The same four options exist for left-handed anti-neutrinos, but we
do not impose any requirements on our construction with regard to
their multiplicity. If they come from strings not attached to any
of the previous branes, we regard them as part of the hidden
sector\rlap.\footnote{In the actual search we have relaxed this
condition slightly, and allowed a brane {\brn d} that just yields
anti-neutrinos.} Furthermore, we do not allow $Y$ to have
contributions from branes that do not couple to charged quarks and
leptons. Otherwise one could extend $Y$ by arbitrarily large
linear combinations that only contribute non-chiral states. This
implies that we regard a brane configuration as complete (prior to
tadpole cancellation) if all charged quark and leptons exist
chirally, and if all cubic $U(N)$ anomalies cancel. This
configuration may already contain a few candidate right-handed
neutrinos, and additional ones may appear, after tadpole
cancellation, from hidden sector states, or strings between the
standard model and the hidden sector.

Clearly this still leaves a huge number of possibilities to
realize this kind of configuration, but there is an obvious
maximally economical choice, namely identifying all branes of
equal charge with each other, and the brane with opposite charge
with its conjugate. This then results in a $U(3)\times U(2) \times
U(1) \times U(1)$ model with the following chiral spectrum
\begin{eqnarray*}\noindent
3& \times &(V,V^*,0,0)  \\
3& \times &(V^*,0,V,0) \\
3& \times &(V^*,0,0,V) \\
6& \times &(0,V,V^*,0) \\
3& \times &(0,V,0,V^*) \\
3& \times &(0,0,V,V^*) \label{Oriented}
\end{eqnarray*}
Although we anticipated the possible need for anti-symmetric
tensors, it turns out that they are not needed at all in this
particular configuration. All anomalies are already cancelled.
This is a consequence of standard model anomaly cancellation. The
formula for $Y$ is
\beq \label{GenY}Y= (x-\frac13) Q_{\brn a} + (x-\frac12) Q_{\brn
b}  + x Q_{\brn c} + (x-1) Q_{\brn d} \eeq
This model has the feature that it can be realized entirely in
terms of oriented strings, which  of course implies that $x$ is
not fixed. The converse is not true because one can allow $U(1)$
anti-symmetric tensors; they do not yield massless particles and
hence give no restriction on $x$. By construction, this is the
minimal realization of the standard model in terms of oriented
strings. Oriented configurations (although more complicated than
the one shown above) were considered earlier in
\cite{leigh}\cite{aiqu} in the context of type-II theories.

One can generalize these orientable models further by allowing
stack {\brn c} and/or {\brn d} to consist of several type C and D
branes. The most general configuration can be denoted as
$U(3)\times U(2) \times U(p_1+q_1) \times U(p_2+q_2)$, where $p_1$
is the number of type C branes on stack c, etc. To achieve this
split we allow non-trivial generators $W_{\brn c}$ and $W_{\brn
d}$ in the definition of $Y$. This gives an infinite set of
solutions, all with at least three Higgs pairs (this follows from
$U(2)$ anomaly cancellation). All these models have in fact
precisely the same structure as the basic four-stack model above,
except for an additional possibility that occurs if type C or D
branes are in different positions ({\it i.e.} have different
boundary labels). If in total three open strings are needed ending
on brane C to get three anti-quarks, then if there are several
type C branes the total number of such strings must be three.
However, each multiplicity can be positive and negative, and hence
cancellations are possible, that show up in the spectrum as
additional mirrors on top of the basic configuration.

One of these cases corresponds to the ``trinification" model
\cite{trinification,d-trinif}. One starts with a gauge group
$SU(3)_{\rm color}\times SU(3)_L \times SU(3)_R$ and matter in
three copies of the representation
$(V,V^*,0)+(V^*,0,V)+(0,V,V^*)$. This configuration fits into our
construction by starting with four stacks ({\brn a}, {\brn b},
{\brn c}, {\brn d}) with a CP group $U(3)\times U(2) \times U(1)
\times U(3)$, and $Y=-\frac16 Q_{\brn b} +\frac13 Q_{\brn c} +
W_{\brn d}$, where $W_{\brn d}$ is the $SU(3)_{\brn d}$ generator
${\rm diag}(\frac13,\frac13,-\frac23)$. The spectrum is three
times
$(V,V^*,0,0)+(V,0,V^*,0)+(V^*,0,0,V)+(0,V,0,V^*)+(0,0,V,V^*)$. The
trinification model is obtained by putting stacks {\brn b} and
{\brn c} on top of each other. In terms of the foregoing
discussion, this model has $x=\frac13$, and three branes of type C
(one from stack {\brn c} and two from stack {\brn d}) plus one
brane of type D (from stack {\brn d}). The value $x=\frac13$ can
easily be understood as follows: in a standard trinification model
$Y$ is embedded entirely in $SU(3)$ factors, and cannot have
components in the brane charges. Therefore in particular it cannot
have any component in $U(3)_{\brn a}$.

The foregoing orientable standard model configurations can be
realized in principle in non-orientable string theories. In these
realizations the value of $x$ is often fixed by the requirement
that $Y$ does not get a mass due to bilinear couplings with
axions. Sometimes this yields rather bizarre looking solutions.
For example, in our set of solutions there is one with
$t_a=\frac1{33}$. There are also cases where $Y$ remains massless
for any value of $x$.

\subsection{Charge Quantization}

There are further constraints on $x$ if one considers unoriented
models. First of all, for generic values of $x$ the non-chiral
part of the string spectrum contains states of fractional or even
irrational charge, from $(V,V)$ bi-fundamentals or from rank-2
tensors. Since such states are always non-chiral, they may be
massive, or become massive under perturbations of the model. They
would however be stable and are not confined by additional gauge
interactions, because they live entirely within the standard model
sector. Therefore, although this possibility cannot be completely
ruled out, it certainly seems preferable to avoid it.

The foregoing discussion is quite general, and can be used to
analyse charge quantization for non-standard-model states in any
brane realization of the standard model. The dependence on
$Q_{\brn a}$ and $Q_{\brn b}$ in (\ref{GenY}) is the most general
one possible, up to an irrelevant sign choice. The complete string
spectrum contains states with charges of all sums and differences
of the components of $Y$, as well as all values multiplied by 2.
It is easy to see that just from branes {\brn a} and {\brn b}, we
get the charge quantization condition
\beq\label{QuantCond} x=0 \hbox{~mod~} \frac12 \  , \eeq
if we require that all massive open string states from
bi-fundamentals and rank two tensors between standard model branes
{\brn a} and {\brn b} to have integer charges (taking into account
QCD confinement). Clearly this condition also implies charge
integrality if branes of types C,D,E and F are present. Only if
charged leptons come from a chirally decoupled sector (the third
or fourth case listed earlier) further conditions may be needed.

A second type of fractional charges that may occur are those
coming from strings with a single end on a standard model brane,
and the other end on a hidden sector brane. Even if these states
are non-chiral, they certainly exist as massive excitations. In
principle, such charges could be confined by hidden sector gauge
groups, but to avoid them altogether, the following condition must
hold
\beq x=0 \hbox{~mod~} 1 \ . \eeq
Also this condition can be derived from just the {\brn a} and
{\brn b} branes. If it is satisfied, branes of types C, D, E and F
satisfy the hidden sector charge quantization condition, but types
G and H do not, in general.

Note that the first charge quantization condition (absence of
fractional charge within the standard model sector) is
automatically satisfied in oriented strings for any $x$, because
the strings that might violate it simply do not exist in oriented
string theories. However, quantization conditions do arise if one
wishes to include hidden branes. These should not contribute to
$Y$. This imposes the second charge quantization condition, $x=0
\hbox{~mod~1}$, for oriented strings.

\subsection{Non-orientable configurations}

The foregoing restrictions were necessary if one wishes to avoid
non-chiral fractionally charged matter. More severe restrictions
apply if some of the quarks and leptons themselves come from
states that break the orientability of the open string theory.

Note first of all that in most cases both type C and type D branes
are needed, in order to get up and down anti-quarks. The only way
out is to get either all down anti-quarks or all up anti-quarks
from anti-symmetric $U(3)$ tensors. The former possibility
requires $x=\frac12$, and then types C and D are the same. This
possibility is realized in flipped $SU(5)$ models, of which we
will give examples later. The second option leads to $x=0$. Then
no type D brane is needed for the quarks, and type C branes do not
contribute to $Y$. This possibility finds a natural realization in
$SU(5)$ GUT models. For all other values of $x$ at least one type
C and one type D brane is needed in addition to branes {\brn a}
and {\brn b}.

Consider now the possibility that a chiral state (a quark or
lepton, or a mirror) breaks the orientability of the
configuration. Obviously this sort of analysis applies to each
chirally decoupled subsector separately (connected components of
quiver diagrams), and we will only consider the component
connected to the {\brn a} and {\brn b} branes.

The possibilities for such a chiral state, and the resulting restrictions on $x$ are as follows
\begin{itemize}
\item{Chiral anti-symmetric tensors on brane {\brn a};  $x=0$  or $\frac12$}
\item{Chiral anti-symmetric tensors on brane {\brn b};
$x=0$, $\frac12$ or $1$}
\item{$(V,V)$ between on branes {\brn a} and {\brn b}; $x=\frac12$.}
\item{Chiral tensors on a brane of type C; $x=0$, $\frac12$ or $-\frac12$}
\item{Chiral tensors on a brane of type D; $x=\frac32$, $1$ or $\frac12$}
\item{$(V,V)$ between brane {\brn a} or {\brn b} and a type C brane; $x=0$ or $\frac12$.}
\item{$(V,V)$ between brane {\brn a} or {\brn b} and a type D brane; $x=\frac12$ or $1$}
\item{$(V,V)$ between type C and a type D brane; $x=0$, $\frac12$ or $1$}
\end{itemize}

Note that the occurrence of $(V,V)$ is automatic if one of the
endpoint branes is real, and that $(V,V)$ between two distinct
type C or type D branes is equivalent to  chiral tensors on a
single such brane. We can extend this list further by including
branes of types E and F, but this will just give similar numbers
modulo half-integers. Note that in all cases the quantization
condition (\ref{QuantCond}) is satisfied.

One important general observation can be made now. For values of
$x$ other than $0,\frac12$ and $1$ all quarks and lepton doublets
must be realized exactly as in the orientable four-stack model
discussed above, because anti-quark weak singlets can only come
from bi-fundamentals, and $U(2)$ anomaly cancellation cannot be
fixed with anti-symmetric tensors. This only leaves some freedom
for the leptonic weak singlets. On the other hand, for $x=0,
\frac12$ and $1$ the $U(2)$ anomaly condition  can always be
satisfied by adding anti-symmetric tensors. They contribute $\pm2$
to the anomaly, but since the total number of doublets is even, so
is the chiral number of doublets (the number of $V$'s minus the
number of $V^*$). (Note that is true for any $U(2)$ because of
cancellation of global anomalies).

If we limit ourselves to four stacks, the number of possibilities
is even smaller. For values of $x$ other than $0$ and $\frac12$
branes of both types C and D are needed. This means that there is
no room for E or F branes and the more exotic values for $x$ they
might allow. This is true even if branes C and D are ``unified"
into a single Chan-Paton group. In order to get a value of $x$
outside the range $-\frac12,\ldots,\frac32$ in a non-orientable
configuration, it must be the chiral strings between the unified
C/D brane and E or F type branes that break the orientability,
{\it i.e.}\ both $(V,V)$ and $(V,V^*)$ must occur. But it is easy
to see that in that case such states necessarily give rise to
leptons with charges $\pm2$, because they must couple to both the
type C and the type D brane.

This reduces the allowed range for $x$ to $-\frac12\ldots\frac32$,
and one can read off from the list which orientation breaking
chiral states are allowed in each case. In the following sections
we will show how to construct four-stack non-orientable
realizations of any of these, at least as ``bottom up" brane
configurations.

\subsection{The cases $x=-\frac12$ or $x=\frac32$}

To get the largest and smallest numbers in this range, the only
orientation breaking chiral states must be chiral tensors on a
type C or type D brane, respectively. This implies that the first
five representations (\ref{Oriented}) (those yielding quarks and
lepton doublets) must be identical to those of the four-stack
orientable model (up to mirror pairs due to distributing type C
and D branes over various positions, as discussed above for the
orientable configuration). In particular it means that we can only
vary the open string origin of the charged leptons. The values
$-\frac12$ and $\frac32$ are essentially ``dual" to each other
under interchange and conjugation of the type C and D branes.

To construct a non-orientable $x=-\frac12$ configuration we start
with four stacks ({\brn a}, {\brn b}, {\brn c}, {\brn d})
generating a CP group $U(3) \times U(2) \times U(1) \times U(1)$,
with the latter two  are type C and D branes respectively. The
only allowed deviation in comparison to the orientable
configuration are $S_{\brn c}$ symmetric tensors on brane {\brn
c}, $m$ bi-fundamentals $(V,V^*)$ between branes {\brn c} and
{\brn d}, $A_{\brn c}$ anti-symmetric tensors on brane {\brn c}
and $A_{\brn d}$ on brane {\brn d}.  Although the anti-symmetric
tensor can occur only  in non-orientable strings, they do not
break the orientability in the sense of fixing $x$, because they
do not yield massless particles imposing constraints on $x$. Their
only  r\^ole is to cancel chiral anomalies.

We get the following conditions from cubic anomaly cancellation
and the requirement that the net number of positively charged
leptons must be three:
\begin{eqnarray} \nonumber
5S_{\brn c}+m-3A_{\brn c} &=3 \\  \nonumber
-m-3A_{\brn d} &= -3 \\  \nonumber
m-S_{\brn c}&=3  \nonumber
\end{eqnarray}
The solution is $S_{\brn c}=-3A_{\brn d}, m=3-3A_{\brn d}, A_{\brn
c}=-6A_{\brn d}$. Hence $m$ and $S_{\brn c}$ must be multiples of
3, and since $S_{\brn c}=0$ brings us back to an orientable
configuration, the simplest non-trivial solution is $S_{\brn
c}=-3$, $m=0$, $A_{\brn c}=-6$ and $A_{\brn d}=1$. The analysis
for $x=\frac32$ is analogous, interchanging the r\^oles of branes
C and D.

Another set of possibilities  (for $x=-\frac12$) is obtained by
putting three type-C branes in stack c, with the CP multiplicity
providing the multiplicities of the anti-quarks and the lepton
doublets. Now anti-symmetric tensors on brane {\brn c} produce
chiral particles, and fix $x$. A simple sequence of solutions is
obtained for $S_{\brn c}=0$, $m=1-A_{\brn d}$, $A_{\brn
c}=-A_{\brn d}$. This is a $U(3)\times U(2) \times U(3) \times
U(1)$ solution with one anti-symmetric conjugate tensor on brane
{\brn c} (which provides the charged leptons) and an
anti-symmetric tensor on brane {\brn d}, just to cancel anomalies.

One can generalize this further by allowing $(p_1,q_1)$ type (C,D)
branes on stack {\brn c}, and $(p_2,q_2)$ type (C,D) branes. This
is accomplished  by having CP gauge groups $U(p_1+q_1)_{\brn c}$
and $U(p_2+q_2)_{\brn d}$, and splitting up their contribution to
$Y$ by means of non-trivial generator $W_{\brn c}$ and $W_{\brn
d}$ in (\ref{Yform}). Since there must be both type C and type D
branes, and they cannot come all from the same stack, we may
require $p_1 > 0$ and $q_2 > 0$. Solving the constraints then
yields solutions only in the following cases: $p_1=1$ or $3$,
$q_2=0$, $q_2=1$ and arbitrary $p_2$, each with a sequence  of
allowed values for the representation multiplicities. The spectra
with $p_2 \not=0$ are rather unappealing: they either have $G_{\rm
CP}$-chiral pairs of mirror anti-quarks, or large numbers of
rank-2 tensors. The  ones with $p_2=0$ were already discussed
above.

\subsection{The case $x=1$\label{XisOne}}

A simple way to obtain a configuration with $x=1$ is to replace
the fourth CP group in the orientable configuration by $O(1)$ in
order to break the orientability. In addition, there is a
possibility of allowing $k$ anti-symmetric tensors of $U(2)$,
yielding $k$ charged leptons. If brane {\brn c} has a Chan-Paton
group $U(1)$, the most general structure is, with CP-group $U(3)
\times U(2) \times U(1) \times O(1)$ is
\begin{eqnarray*}
3& \times &(V,V^*,0,0)\\
3& \times &(V^*,0,V,0)  \\
3&\times &(V^*,0,0,V)\\
m& \times &(0,V,V^*,0)\\
n& \times &(0,V,0,V)\\
l&\times &(0,0,V,V) \\
k&\times &(0,A,0,0)\\
t&\times &(0,0,A,0)
\end{eqnarray*}
with the conditions
\begin{eqnarray*}
m-n&=3 \\
-9+m+n-2k&=0\\
k+l&=3\\
9-2m+l-3t&=0
\end{eqnarray*}
These are respectively the requirements of three lepton doublets,
$U(2)$ anomaly cancellation, three charged leptons and brane {\brn
c} anomaly cancellation. This yields a one-parameter set of
solutions, $m=6+k, n=3+k, l=3-k, t=-k$. There are many more
possibilities if we allow larger CP-factors for {\brn c} and {\brn
d}. It is also possible to use a $U(1)$ CP-factor for {\brn d}.
This leads to an additional anomaly constraint, but there are many
ways to satisfy it by replacing some of the vectors by their
conjugates, and adding anti-symmetric and/or symmetric tensors.
The latter yield singlet neutrinos. The complete solution is too
complicated to present here.

\subsection{Realizations with three brane stacks for $x=0$}

The cases $x=0$ and $x=\frac12$ allow far more possibilities. We
will solve them here in general, in the special case that they are
realized with just three branes, yielding a group $U(3) \times
U(2) \times U(p,q)$, where $p$ and $q$ are the number of
eigenvalues $x$ and $x-1$.

Consider first $x=0$. We assume that  there are $t$ chiral rank-2
tensors on brane {\brn a}. Then the most general choice of
bi-fundamentals for anti-quarks and lepton doublets is as follows
\begin{eqnarray}\nonumber
n& \times &(V^*,0,V)  \\ \nonumber
m& \times &(V^*,0,V^*)\\ \nonumber
k& \times &(0,V,V^*)\\ \nonumber
l&\times &(0,V,V) \nonumber
\end{eqnarray}
Furthermore we allow $r$ chiral anti-symmetric $U(2)$ tensor, and
$a$ and $s$ chiral anti-symmetric and symmetric $U(p,q)$ tensors.
The latter are allowed only if $q=0$ (since otherwise one gets
charge 2 leptons), and if $q>1$ no $U(p,q)$ tensors are allowed at
all. Furthermore we must require $mq=lq=0$ to prevent particles
with unacceptable charges. To get three lepton doublets we need
$k(p-q)=3$, {\it i.e.} $p-q=\pm3$ or $\pm1$. The total number of
charged leptons is $-r-apq$.

Let us assume first that $q>1$. Then $a=s=0$, and $r=-3$, and
$m=l=0$. $U(2)$ anomaly cancellation then implies $(p+q)k-2r-9=0$,
and hence $(p+q)k=3$. But we have already seen that $k(p-q)=3$,
and hence this is not consistent with the assumption. Now assume
$q=1$. Also in this case $m$ and $l$  must vanish. Then the
condition for getting three anti-down-quarks is $np=3$. This
allows $p=1$ or $p=3$, but neither is consistent with $p-q=\pm3$
or $\pm1$.

Hence the only possibility is $q=0$. Then  $r=-3$. The third brane
does not contribute to $Y$, and the distinction between $V$ and
$V^*$ on that brane is irrelevant for all hypercharges. The
conditions for getting the right number of anti-down quarks is
$(n+m)p=3$, and for lepton doublets it is $(k+l)p=3$. Hence $p$ is
either 1 or 3. Anti-up quarks can only come from the $t$
anti-symmetric $U(3)$ tensors. Hence $t=3$. In the  $U(3) \times
U(2)$ subgroup we find the representation $3 \times (A,0) +
3\times (V,V^*) +  3\times (0,A^*)$, which of course fits
precisely in $3\times (10)$ of $U(5)$. The $U(1)$ generators $Y$
becomes an $SU(5)$ generator. Hence the only possibility for $x=0$
and at most three participating branes is broken $U(5)$. This can
be reduced to two participating branes by putting the {\brn a} and
{\brn b} branes on top of each other, to get unbroken $U(5)$.  The
CP group on the third brane can be $U(1)$ or $U(3)$, but since
this brane does not contribute to $Y$ one can also allow $O(1)$ or
$O(3)$. In that case there are no anomaly constraints to worry
about. If the c-brane group is unitary, the total anomaly is
$3(n-m)+2(l-k)$.  This leaves many possible values, and this
anomaly can be cancelled in many ways using symmetric or
anti-symmetric tensors. In the spectrum, these appear as standard
model singlets, {\it i.e.}  candidate anti-neutrinos.

\subsection{Realizations with three brane stacks for $x=\frac12$ }

Consider now $x=\frac12$. Then if $p=q$ the third brane could be
orthogonal or symplectic, in which case there is no anomaly
cancellation condition for it. Furthermore the weak group can then
be $Sp(2)$. This makes little difference, because $U(2)$ anomalies
can be cancelled by means of anti-symmetric tensors, which in this
case are standard model singlets (right-handed neutrinos) which we
do not constrain {\it a priori}.

We assume that  there are $t$ chiral rank-2 tensors on brane {\brn a}.
Then the most general structure is
as follows
\begin{eqnarray}\nonumber
n& \times &(V^*,0,V)  \\ \nonumber
m& \times &(V^*,0,V^*)\\ \nonumber
k& \times &(0,V,V^*)\\ \nonumber
l&\times &(0,V,V) \nonumber
\end{eqnarray}
We have to require
\begin{eqnarray}\nonumber
t + n p + m q &= 3 \\ \nonumber
n q + m p &=3\\ \nonumber
k p + l q - k q - lp &= 3
\end{eqnarray}
for getting the right anti-up, anti-down and lepton doublet count.
The first two equations imply $(n-m)(p-q)=-t$, and the last one
$(k-l)(p-q)=3$. Hence $p\not=q$, and brane {\brn c} cannot be
real. The only allowed values for $p-q$ are $-3,-1,1,3$, and $t$
must be a multiple of $p-q$. Given these four values, we can
compute $n-m$ and $k-l$. To cancel the anomalies on brane {\brn c}
and to provide charged leptons we introduce $a$ anti-symmetric and
$b$ symmetric tensors. The conditions for anomaly cancellation on
brane {\brn c}, and a net number of 3 charged leptons can be
combined to yield \beq 3(n-m)(p-q)-2(k-l)(p-q)-3(a-b)(p-q)=-6 \eeq
which together with the previous conditions implies $a-b=n-m$. The
remaining equations are
\begin{eqnarray}
(n+m)(p+q)&=&6-t\\
(a+b)(p+q)&=&(n-m)+2(k-l)={6-t\over p-q}
\end{eqnarray}
{}From their ratio we see that $(n+m)=(p-q)(a+b)$. Furthermore we
see that $p+q$ and $p-q$ must both be divisors of $6-t$. This
allows a limited number of values for $p+q$, and then $(a+b)$ and
$(n+m)$ are determined. Hence all solutions are specified in terms
of $t$ plus a limited number of values for $p+q$ and $p-q$. There
are three more parameters that are not yet specified: $k+l$, the
number of anti-symmetric tensors on brane {\brn b}, and the
difference between the number of $(V,V^*)$ and $(V,V)$ quark
doublets. One linear relation between them is imposed by $U(2)$
anomaly cancellation; in the $Sp(2)$ case there is no constraint.

\subsection{Solutions with type E and F branes}

Type E and F branes contribute to $Y$ with coefficients $x+1$ and
$x-2$ respectively. They cannot contribute to to quarks or lepton
doublets. We assume here that their contribution includes at least
one $(V,V^*)$ bi-fundamental; if they produce valid quarks or
lepton doublets (or mirrors) only as $(V,V)$ bi-fundamentals we
conjugate the E/F brane, and redefine its coefficients. Depending
on the actual value of $x$ an E or F brane then becomes a brane of
type C or D, and is already included in our foregoing discussion.

Furthermore an E/F brane must be connected, by definition, via
$(0,0,V,V^*)$ bi-fundamentals to the {\brn c}-brane. As discussed
above, in a four-stack configuration E or F branes can only be
allowed in principle for $x=0$ or $x=\frac12$. As in the rest of
the paper, we allow the {\brn c} and {\brn d} stacks to consist of
two brane types, with eigenvalues differing by one unit. The
options are then {\brn c}=(C,D), {\brn d}=(E,C) or {\brn c}=(C,D),
{\brn d}=(D,F), where each type can occur with an arbitrary
multiplicity, and E and F have to occur at least once. However, in
all cases one of the two branes on stack {\brn c} would give rise
to a charge-2 lepton. This reduces the possibilities to {\brn
c}=(C), {\brn d}=(E,C) for $x=\frac12$ (and its conjugate, {\brn
c}=(D), {\brn d}=(D,F)) or {\brn c}=(D), {\brn d}=(D,F) for $x=0$.
However, the latter possibility is ruled out, since at least one
C-type brane is needed to produce $d^c$ anti-quarks. The next
constraint is anomaly cancellation for stack {\brn d}. Since it
only shares bi-fundamentals $(0,0,V,V^*)$ with brane {\brn c} and
nothing with any other brane, the anomalies of the $V^*$'s must be
cancelled by rank-2 tensors. This forbids two distinct
Y-eigenvalues on stack {\brn d}, since the sums of these
eigenvalues would appears as invalid charges in the spectrum. It
also limits the multiplicity of the E or F branes to 1, and only
allows anti-symmetric tensors to cancel the anomaly. The
multiplicity of $(0,0,V,V^*)$ must then be a multiple of three.

Configurations of this type can indeed be constructed. The {\brn
c}-group can either be $U(1)$ or $U(3)$. In the former case, there
is a two-parameter series of solutions labelled by the number of
$SU(3)_{\brn a}$ anti-symmetric tensors, and the number of
$(0,0,V,V^*)$. The $U(1)_{\brn c}$ anomalies are cancelled by
anti-symmetric and/or symmetric tensors, and the latter also
contribute charged leptons. If {\brn c}-group is $U(3)$, there
must be three anti-symmetric {\it conjugate} tensors of
$SU(3)_{\brn a}$ (yielding three left-handed down quarks, which
must be combined with six left-handed down anti-quarks from
$(V^*,0,V,0)$), and there can be charged leptons from
$(0,0,V,V^*)$ as well as anti-symmetric $U(3)$ tensors.

Furthermore, one may use both $U(2)$ and $Sp(2)$ as the Chan-Paton
group of brane {\brn b}.

None of these models have appeared in our top-down search.

\subsection{Solutions with type G branes}

Type-G branes are defined as branes that contribute non-trivially
to $Y$ but that contribute to the chiral spectrum only through
rank-2 tensors. This implies that their $Y$-coefficient must be
$\pm \frac12$. If $x=\frac12$, this can be viewed as just a
standard type C or D brane. These cases are taken into account in
our bottom-up construction as standard $x=\frac12$ models. They do
indeed occur as brane configurations, although rarely. For
example, we have generated all brane configurations with four
unitary CP factors, at most three Higgs pairs, at most three
$G_{\rm CP}$ exotics and at most six $G_{\rm CP}$ chiral singlets.
Of the 10820995 unitary models with $x=\frac12$, only 338 have
type-G branes, {\it i.e.} a brane with only chiral tensors and no
bi-fundamentals.

A more interesting situation occurs when $x=0$ (the only other
value of $x$ where type-G branes might occur). In that case the
type-G brane has a non-canonical contribution $\pm \frac12$ to $Y$
(the canonical value is 0 or $\pm1$).

However, the foregoing analysis of three brane realizations with
$x=0$ shows that this possibility does not exist. The only
three-brane models are (broken) $SU(5)$ with a set of neutral
C-type branes. This result was obtained without requiring any
particular value for the number of charged leptons. The latter
came out uniquely as three. Since the {\brn c} stack is neutral,
it cannot provide charged leptons or mirrors either. Hence all
three-stack models already have precisely three charged leptons,
and all the G-brane could still do is add mirror pairs. This could
happen even with a chiral {\brn d} stack, for example with three
anti-symmetric tensors and a symmetric tensor of $U(2)$, with
$W_{\brn d}={\rm diag}(\frac12,-\frac12)$. However, this is not of
much interest, and furthermore these models are equivalent to
those where brane {\brn d} does not contribute to $Y$ at all, and
brane {\brn d} just yields $G_{\rm CP}$-chiral neutrinos.


\section{Statistics of bottom-up configurations\label{StatBU}}

In this section we will provide an enumeration of bottom-up
configurations, providing some numbers to the theoretical analysis
of the previous section. We will consider for simplicity the $\bf
c$ and $\bf d$ groups to be abelian. We will also impose
(generalized) anomaly cancellation.

The associated statistics is shown and compared in table
\ref{tbl:TableThree} of the next section, where detailed
definitions are also given.

\subsection{Three stacks: the $U(3)\times U(2)\times
U(1)$ models}

We first consider three-stack models. We will  consider the
possible realizations of MSSM-like Higgs pairs, and the presence
of baryon and lepton number symmetries. We also indicate the total
number of configurations of a given type. In our search, we can
also include the right-handed neutrino $\n^c$ which may appear as
an open string with both ends on the weak or other branes.

Requiring that the particles have the proper hypercharge there are
two possible ways to embed the Standard Model in this D-brane
system of three stacks, \cite{AD}:
\bea Y=-{1\over 3}Q_{\brn a}-{1\over 2}Q_{\brn
b}~~,~~~~~~~Y={1\over 6}Q_{\brn a}+{1\over 2}Q_{\brn c}\eqp \eea
%

For the first embedding, $Y=-{1\over 3}Q_{\brn a}-{1\over
2}Q_{\brn b}$, we obtain the following allowed spectra, (by
$\tilde R$ we indicate that both the representation $R$ or the
conjugate representation $R^*$ can be a valid choice)
\begin{eqnarray}\nonumber
&Q:&~~~(V,V,0)\\\nonumber
&u^c:&~~~(A,0,0)\\ \nonumber
&d^c:&~~~(V^*,0,\tilde{V})\\\nonumber
&L:&~~~(0,V^*,\tilde{V}^*)\\ \nonumber
&l^c:&~~~(0,A,0)\\ \nonumber
&H:&~~~(0,V,\tilde{V})\\ \nonumber
&H':&~~~(0,V^*,\tilde{V})
\end{eqnarray}
{}From the above charge assignments we can construct families and
search for triplets of these families which form anomaly-free
models. For that embedding ($Y=-{1\over 3}Q_{\brn a}-{1\over
2}Q_{\brn b}$) there are 10 different anomaly-free spectra that
describe the SM.
If the anti-neutrino $\n^c$ also arises from strings stretching
inside this stack, it will be of the form $(0,0,\tilde{S})$.

For the second embedding $Y={1\over 6}Q_{\brn a}+{1\over 2}Q_{\brn
c}$ we have the following allowed spectra
\begin{eqnarray}\nonumber
&Q:&~~~(V,\tilde{V},0)\\\nonumber
&u^c:&~~~(V^*,0,V^*)\\ \nonumber
&d^c:&~~~(A,0,0)~~~{\rm or}~~~(V^*,0,V)\\ \nonumber
&L:&~~~(0,\tilde{V},V^*)\\ \nonumber
&l^c:&~~~(0,0,A)\\\nonumber
&H:&~~~(0,\tilde{V},V)\\ \nonumber
&H':&~~~(0,\tilde{V},V^*)\eqp
\end{eqnarray}
There are 24 different anomaly-free models.
If the anti-neutrino $\n^c$ also arises from strings stretching
inside this stack, it will be of the form  $(0,\tilde{A},0)$.
Notice the ambiguity of the representations (with tilde) when a
brane does not contribute to the hypercharge and also the two
different possibilities for the charges of $d^c$: $(V^*,0,V)$ or
$(A,0,0)$ in the second case.

The baryon number $B=Q_{\brn a}/3$ is a gauge symmetry only in
models where $d^c$ arises from a string with the two ends onto
different branes.
In none of the models above, lepton number is a symmetry.

\subsection{Four stacks: $U(3)\times U(2)\times U(1)\times U(1)'$ Models}

In this section, we study four-stack realizations of the Standard Model.
We continue with the statistics of fours-stack models.

\subsubsection{Hypercharge $Y=(x-{1\over 3})Q_{\brn a}+(x-{1\over 2})Q_{\brn b}
+xQ_{\brn c}+(x-1) Q_{\brn d}$}\label{XisX}

Notice that in order for $x$ to remain arbitrary, the right-handed
neutrino must necessarily arise in the hidden sector. The
corresponding charge assignments are:
\bea
&Q:&~~~(V,V^*,0,0)\nn\\
&u^c:&~~~(V^*,0,0,V)\nn\\
&d^c:&~~~(V^*,0,V,0)\nn\\
&L:&~~~(0,V,V^*,0)~~~{\rm or}~~~(0,V^*,0,V)\nn\\
&l^c:&~~~(0,0,V,V^*)\nn\\
&H:&~~~(0,V,0,V^*)~~~{\rm or}~~~(0,V^*,V,0)\nn\\
&H':&~~~(0,V,V^*,0)~~~{\rm or}~~~(0,V^*,0,V)\nn\eea
Following the same spirit as in the tree-stack models, we can form
families from the above charge assignments and require that
triplets of them are free of irreducible anomalies. For the
present hypercharge embedding there is only one anomaly-free model
which can describe the SM and given by three copies of the
previous assignments; it is (\ref{Oriented}) shown in the previous
section

\subsubsection{Hypercharge $Y=-{1\over 3}Q_{\brn a}-{1\over 2}Q_{\brn b}
+Q_{\brn d}$}\label{Xis0}

The corresponding charge assignments are:
\bea
&Q:&~~~(V,V^*,0,0)\nn\\
&u^c:&~~~(A,0,0,0)~~~~~{\rm or}~~~(V^*,0,0,V)\nn\\
&d^c:&~~~(V^*,0,\tilde{V},0)\nn\\
&L:&~~~(0,V,\tilde{V},0)~~~~{\rm or}~~~(0,V^*,0,V^*)\nn\\
&l^c:&~~~(0,A^*,0,0)~~~~{\rm or}~~~(0,0,\tilde{V},V^*)\nn\\
&H:&~~~(0,V^*,\tilde{V},0)~~~{\rm or}~~~(0,V,0,V^*)\nn\\
&H':&~~~(0,V^*,0,V)~~~{\rm or}~~~(0,V,\tilde{V},0) \eqp\nn\eea
If $\n^c$ is coming from the hidden sector, there are 302
anomaly-free models which can describe the SM particles. Among
them, there are 62, 72, 96 and 72 models with three, two, one and
none chiral Higgs pairs.

On the other hand, if $\n^c$ is attached onto branes of the above
stacks, it can only be charged under the $U(1)_{\brn c}$ which
does not contribute to the hypercharge. Therefore, it will
transform as $(0,0,\tilde{S},0)$.
In that case, there are 1208 different anomaly-free models which
can describe the SM particles (including $\n^c$). Among them,
there are 240, 384, 288 and 248 models with tree, two, one and
none chiral Higgs pairs.

When $u^c$ is not described by an antisymmetric representation,
the baryon number $B=Q_{\brn a}/3$ is conserved.

\subsubsection{Hypercharge
$Y={2\over 3}Q_{\brn a}+{1\over 2}Q_{\brn b}+Q_{\brn c}$}\label{Xis1}

The corresponding charge assignments are:
\bea
&Q:&~~~(V,V^*,0,0)\nn\\
&u^c:&~~~(V^*,0,0,\tilde{V})\nn\\
&d^c:&~~~(V^*,0,V,0)\nn\\
&L:&~~~(0,V^*,0,\tilde{V})~~~{\rm or}~~~(0,V,V^*,0)\nn\\
&l^c:&~~~(0,A,0,0)~~~~~{\rm or}~~~(0,0,V,\tilde{V})\nn\\
%
&H:&~~~(0,V^*,V,0)~~~{\rm or}~~~(0,V,0,\tilde{V})\nn\\
&H':&~~~(0,V^*,0,\tilde{V})~~~{\rm or}~~~(0,V,V^*,0) \nn\eea
In total, there are 6 different anomaly-free models which can
describe the SM particles with chiral Higgs-pairs.

A $\n^c$ which is a string attached onto these stacks of branes
would be of the form $(0,0,0,\tilde{S})$.
In that case, there are 24 different anomaly-free models
with chiral Higgs-pairs (including $\n^c$) and they all
have baryon number $B=Q_{\brn a}/3$.

\subsubsection{Hypercharge
$Y={1\over 6}Q_{\brn a}+{1\over 2}Q_{\brn c}-{1\over 2}Q_{\brn d}$}\label{Xis1over2}

The corresponding charge assignments are:
\bea
&Q:&~~~(V,\tilde{V},0,0)\nn\\
&u^c:&~~~(V^*,0,V^*,0)~~~{\rm or}~~~(V^*,0,0,V)\nn\\
&d^c:&~~~(A,0,0,0)~~~~~~~{\rm or}~~~(V^*,0,V,0)~~~{\rm or}~~~(V^*,0,0,V^*)\nn\\
&L:&~~~(0,\tilde{V},V^*,0)~~~~{\rm or}~~~(0,\tilde{V},0,V)\nn\\
&l^c:&~~~(0,0,S,0)~~~~~~~{\rm or}~~~(0,0,V,V^*)~~~~{\rm or}~~~(0,0,0,S^*)\nn\\
&H:&~~~(0,\tilde{V},0,V^*)~~~~{\rm or}~~~(0,\tilde{V},V,0)\nn\\
&H':&~~~(0,\tilde{V},0,V)~~~~~~{\rm or}~~~(0,\tilde{V},V^*,0) \nn\eea
In that case, there are 8552 different anomaly-free models
with chiral Higgs pairs which can describe the SM particles.

Some models have lepton number. There are four
independent combinations:
\begin{itemize}
%
\item $Q_L={1\over 2}Q_{\brn a}+{1\over 2}Q_{\brn b}-{1\over
2}Q_{\brn c}- {1\over 2}Q_{\brn d}:$
\bea &&3\times (V,V^*,0,0),\nn\\&&
3\times (V^*,0,V^*,0), \nn\\&&
3\times (V^*,0,0,V^*), \nn\\&&
3\times (0,V,V^*,0), \nn\\&&
3\times (0,V,V,0), \nn\\&&
3\times (0,V,0,V), \nn\\&&
3\times (0,0,S,0).\nn\eea
\item $Q_L=Q_{\brn d}:$
\bea &&3\times (V,V^*,0,0),\nn\\&&
3\times (V^*,0,V^*,0),\nn\\&&
\big\{ m\times (V^*,0,V,0),~n\times (A,0,0,0)\big\}, \nn\\&&
3\times (0,V,0,V), \nn\\&&
3\times (0,V,V,0), \nn\\&&
3\times (0,V,V^*,0), \nn\\&&
3\times (0,0,V,V^*)\nn\eea
where $m,n\in [0,1,2,3]$ and $m+n=3$. Therefore, $d^c$ in each
family can be either a string which is attached onto the ${\brn
a}$ and ${\brn c}$ stacks or a string with both ends on the ${\brn
a}$ stack.
\item $Q_L=-Q_{\brn c}:$
\bea &&3\times (V,V^*,0,0),\nn\\&&
3\times (V^*,0,0,V),\nn\\&&
\big\{ m\times
(V^*,0,0,V^*),~n\times (A,0,0,0)\big\}, \nn\\&&
3\times (0,V,V^*,0), \nn\\&&
3\times (0,V,0,V^*), \nn\\&&
3\times (0,V,0,V), \nn\\&&
3\times (0,0,V,V^*)\nn\eea
where again $m,n\in [0,1,2,3]$ and $m+n=3$.

\item $Q_L= {1\over 2}Q_{\brn a}+{1\over 2}Q_{\brn b}+{1\over
2}Q_{\brn c}+ {1\over 2}Q_{\brn d}$
\bea &&3\times (V,V^*,0,0), \nn\\&& 3\times (V^*,0,0,V),  \nn\\&&
3\times (V^*,0,V,0),  \nn\\&& 3\times (0,V,0,V),  \nn\\&& 3\times
(0,V,0,V^*),  \nn\\&& 3\times (0,V,V^*),  \nn\\&& 3\times
(0,0,0,S^*)\nn\eea.
\end{itemize}

If the right-handed neutrino $\n^c$ is attached onto the SM
branes, it can be described by $(0,\tilde{A},0,0)$ or
$(0,0,\tilde{V},\tilde{V})$.
Including $\n^c$, there are 150672 different anomaly-free models.
Among them, there are 29360, 61344, 48800 and 11168 models with
tree, two, one and none chiral Higgs pairs.

If $d^c$ is not described by an antisymmetric representation,
there is baryon number $B=Q_{\brn a}/3$.

\subsubsection{Hypercharge
$Y={1\over 6}Q_{\brn a}+{1\over 2}Q_{\brn c}-{3\over 2}Q_{\brn
d}$}\label{Xis1over2plusF}

The corresponding charge assignments are:
\bea
&Q:&~~~(V,\tilde{V},0,0)\nn\\
&u^c:&~~~(V^*,0,V^*,0)\nn\\
&d^c:&~~~(V^*,0,V,0)~~~{\rm or}~~~(A,0,0,0)\nn\\
&L:&~~~(0,\tilde{V},V,0)\nn\\
&l^c:&~~~(0,0,V^*,V)~~~{\rm or}~~~(0,0,S,0)\nn\\
%
&H:&~~~(0,\tilde{V},V,0)\nn\\
&H':&~~~(0,\tilde{V},V^*,0) \nn\eea
In that case, there are 4 different anomaly-free models with chiral
Higgs pairs which can describe the SM.

A $\n^c$ which is stretched between the four stacks can be of the
form $(0,\tilde{A},0,0)$.
Including $\n^c$, the number of different charge assignments is
24 (8 of them have two chiral Higgs pairs and the other 16 have
non chiral Higgs pairs).
Half of these states have baryon number $Q_B=Q_{\brn a}/3$ and in none
lepton number is a symmetry. All models have one non-anomalous
$U(1)$.

\subsubsection{Hypercharge $Y=-{1\over 3}Q_{\brn a}-{1\over 2}Q_{\brn b}$\label{Xis1over2andCC}}

The corresponding charge assignments are:
\bea
&Q:&~~~(V,V^*,0,0)\nn\\
&u^c:&~~~(A,0,0,0)\nn\\
&d^c:&~~~(V^*,0,\tilde{V},0)~~~{\rm or}~~~(V^*,0,0,\tilde{V})\nn\\
&L:&~~~(0,V^*,\tilde{V},0)~~~{\rm or}~~~(0,V^*,0,\tilde{V})\nn\\
&l^c:&~~~(0,A^*,0,0)\nn\\
%
&H:&~~~(0,V,\tilde{V},0)\nn\\
&H':&~~~(0,V^*,\tilde{V},0) \nn\eea
with 936 anomaly-free models. Among them, there are 256, 120,
120 and 440 models with tree, two, one and none chiral Higgs pairs.

A $\n^c$ which will be stretched between the four branes will be
of the form $(0,0,0,\tilde{S})$ or $(0,0,\tilde{S},0)$ or
$(0,0,\tilde{V},\tilde{V})$.
Including $\n^c$, there are 106792 different anomaly-free models.
Among them, there are 15072, 32332, 36228 and 23160 models with
tree, two, one and none chiral Higgs pairs.

\subsubsection{Hypercharge
$Y=-{5\over 6}Q_{\brn a}-{Q_{\brn b}}-{1\over 2}Q_{\brn c}+{3\over
2}Q_{\brn d}$ \label{Xisminus1over2}}

The above hypercharge embedding is allowed only in cases where the
right-handed neutrino is coming from the hidden sector. The
corresponding charge assignments are:
\bea
&Q:&~~~(V,V^*,0,0)\nn\\
&u^c:&~~~(V^*,0,0,V)\nn\\
&d^c:&~~~(V^*,0,V,0)\nn\\
&L:&~~~(0,V^*,0,V)~~~{\rm or}~~~(0,V,V^*,0)\nn\\
&l^c:&~~~(0,0,S^*,0)~~~~{\rm or}~~~(0,0,V,V^*)\nn\\
&H:&~~~(0,V^*,V,0)~~~~{\rm or}~~~(0,V,0,V^*)\nn\\
&H':&~~~(0,V^*,0,V)~~~{\rm or}~~~(0,V,V^*,0) \nn\eea
In that case, there are 2 different anomaly-free models
which can describe the SM:
\bea &&3\times (V,V^*,0,0), \nn\\&&
3\times (V^*,0,0,V),  \nn\\&&
3\times (V^*,0,V,0),  \nn\\&&
6\times (0,V,V^*,0),  \nn\\&&
3\times (0,V,0,V^*),  \nn\\&&
\big\{3\times (0,0,S^*,0)~~~{\rm or}~~~3\times (0,0,V,V^*)\big\}\nn\eea
and they have baryon number $Q_B=Q_{\brn a}/3$. Lepton number
is not a symmetry.

\subsubsection{Hypercharge
$Y={7\over 6}Q_{\brn a}+{Q_{\brn b}}+{3\over 2}Q_{\brn c}+{1\over
2}Q_{\brn d}$\label{Xis3over2}}

The above hypercharge embedding is allowed only in cases where the
right-handed neutrino is coming from the hidden sector. The
corresponding charge assignments are:
\bea
&Q:&~~~(V,V^*,0,0)\nn\\
&u^c:&~~~(V^*,0,0,V)\nn\\
&d^c:&~~~(V^*,0,V,0)\nn\\
&L:&~~~(0,V,V^*,0)~~~{\rm or}~~~(0,V^*,0,V)\nn\\
&l^c:&~~~(0,0,0,S)~~~~~{\rm or}~~~(0,0,V,V^*)\nn\\
&H:&~~~(0,V^*,V,0)~~~{\rm or}~~~(0,V,0,V^*)\nn\\
&H':&~~~(0,V,V^*,0)~~~{\rm or}~~~(0,V^*,0,V) \nn\eea
In that case, there are 2 different anomaly-free models
which can describe the SM particles:
\bea &&3\times (V,V^*,0,0), \nn\\&&
3\times (V^*,0,0,V),  \nn\\&&
3\times (V^*,0,V,0),  \nn\\&&
6\times (0,V,V^*,0),  \nn\\&&
3\times (0,V,0,V^*),  \nn\\&&
\big\{3\times (0,0,0,S)~~~{\rm or}~~~3\times (0,0,V,V^*)\big\}\nn\eea
and they have baryon number $Q_B=Q_{\brn a}/3$. Lepton number
is not a symmetry.

\section{Top-down configurations and SM spectra\label{TDspectra}}

\subsection{Scope of the top-down search}

The set of models we are able to search in principle consists of
all three and four-stack combinations of all boundaries of all
simple current orientifolds \cite{Fuchs:2000cm} of all simple
current MIPFs \cite{Gato-Rivera:1991ru}\cite{Kreuzer:1993tf} of
the 168 $c=9$ tensor products of $N=2$ minimal models. We denote
these as $(k_1,\ldots,k_m)$, where $k_i$ is the $SU(2)$ level,
which ranges from 1 to $\infty$. The total number of MIPFs is
5403, and the total number of orientifolds 49304. Some of these
have zero-tension O-planes, which means that there is no
possibility of cancelling tadpoles between D-branes and O-planes.
This leaves 33012 orientifold models. Of the 168 Gepner models, 5
are non-chiral $K_3 \times T_2$ compactifications, which need not
be considered because they can never yield a chiral
spectrum\rlap.\footnote{Note that all boundaries we consider
respect the full chiral algebra of the tensor product, and all
partition functions are expressed in terms of the characters of
that algebra, which are space-time non-chiral. One may also
consider orbifold projections of these theories, which reduce the
chiral algebra, and may introduce chiral characters, but our
methods do not apply to that case. We do allow the inverse of
this: a chiral theory with a non-chiral extension. Indeed, we
found some standard model configurations for such theories.} These
non-chiral theories contribute in total 88 MIPFs and 228
orientifolds.

The number of boundary states in a complete set can range from a
few hundred to 108612 for tensor product $(1,5,82,82)$. In that
case the number of unitary brane pairs is 53046 and 52920 for the
two orientifold choices. The number of combinations one needs to
consider for a four-stack configuration grows with the fourth
power of the number of pairs. In \cite{Dijkstra:2004cc} almost all
these cases were searched. This was possible because the standard
model configuration searched for was more limited. For example, no
chiral rank-2 tensors were allowed, reducing the number of choices
for the {\brn a},{\brn b},{\brn c} and {\brn d} branes
dramatically. Furthermore the configuration of
\cite{Ibanez:2001nd} is such that branes {\brn a } and {\brn d}
have a different multiplicity (3 and 1) but identical intersection
numbers with the other branes. This can be used to reduce the
power behavior of the search algorithm essentially from four to
three.

Neither of these shortcuts help us here, and therefore a full
search is practically impossible at present. Here we limit
ourselves to MIPFs with at most 1750 boundaries. This limits us to
4557 of the 5403 MIPFS and 29257 of the 33012 non-zero tension
orientifolds. We can now work out how many brane configurations
exist in total. To do this really correctly, unitary, orthogonal
and symplectic branes must be distinguished.

\begin{table}[h]\caption{Total number of three and four stack configurations of various types.}\label{tbl:Summary}
\begin{center}
~~~~~~~~~~~~~~~~\begin{tabular}{|l|r|r|} \hline
Type & Total & This paper   \\ \hline \hline
UUU&       1252013821335020     &    1443610298034 \\
UUO, UOU&         99914026743414     &     230651325566 \\
UUS, USU&         14370872887312     &     184105326662 \\
USO&          2646726101668     &      74616753980 \\
USS&          1583374270144     &      73745220170 \\
UUUU&   21386252936452225944    &   366388370537778 \\
UUUO&    2579862977891650682    &   105712361839642 \\
UUUS&     187691285670685684     &   82606457831286 \\
UUOO&     148371795794926076     &   19344849644848 \\
UUOS&      17800050631824928     &   26798355134612 \\
UUSS&       4487059769514536    &    13117152729806 \\
USUU&      93838457398899186    &    41211176252312 \\
USUO&      17800050631824928    &    26798355134612 \\
USUS&       8988490411916384    &    26418410786274 \\ \hline
\end{tabular}
\end{center}
\end{table}
Table (\ref{tbl:Summary}) lists the total number of configurations
for all combinations of unitary, orthogonal and symplectic branes,
without taking into account the additional freedom of assigning
Chan-Paton multiplicities. The second column gives the grand total
for all 163  chiral Gepner models and non-zero tension
orientifolds. It is the maximal number of three and four-stack
configurations of given type that we have  at our disposal for
Standard Model searches. The third column gives the size of the
subset actually searched in this paper.

The precise counting is as follows. Denote the number of unitary
brane pairs as $N_U$. Then the total number of UUUU configurations
with distinct {\brn c} and {\brn d} branes is $( 2 N_U)( N_U)
\times \frac12 N_U (N_U -1 )$, etc. The choices for {\brn a},
{\brn b} and {\brn c} are  independent, since we allow all these
stacks to coincide, but if {\brn c} and {\brn d} coincide we
regard it as a three-stack configuration. Furthermore both
conjugates of the {\brn a} brane are counted, because they give
rise to conjugate $SU(3)$ representations, and hence yield
distinct spectra.  Conjugations of the {\brn b}, {\brn c} and
{\brn d} branes can always be compensated by changing the sign of
the coefficients of $Y$, and hence do not yield new possibilities.

Obviously, although we cover a substantial fraction of MIPFs and
orientifolds, only a small fraction of possible brane
configurations has been searched, because the missing MIPFs are
the ones with the largest number of branes. Nevertheless, in our
previous search \cite{Dijkstra:2004cc}, which was more extensive,
the MIPFs we are not considering in the present paper produced
relatively few SM-configurations and tadpole solutions. Part of
the reason for the latter is that probably there are many more
candidate branes in the hidden sector, making the tadpole
equations harder to solve.

\subsection{Standard model brane configurations found}

\begin{table}[h]\caption{Number of standard model configurations sorted by the value of $x$.}
\label{tbl:TableX}
\begin{center}
~~~~~~~~~~~~~~~~
\begin{tabular}{|c|r|r|} \hline
$x$ & Total occurrences & Without  $SU(3)$ tensors\\
\hline \hline
$-1/2$ & 0 & 0 \\
$0$   &  21303612 & 202108 \\
$1/2$ &  124006839 & 115350426\\
$1$ &    12912 & 12912 \\
$3/2$ &         0 & 0 \\
$*$  & 1250080 & 1250080\\ \hline
\end{tabular}
\end{center}
\end{table}

Of the 4557 MIPFs, 1639 contained at least one standard model
spectrum, without taking into account tadpole cancellation. In
table (\ref{tbl:TableX}) we list the total number of brane
configurations with a chiral standard model spectrum sorted
according to $x$. In \cite{Dijkstra:2004cc} only a subset of the
possible $x=\frac12$ models was considered, but for a much larger
set of MIPFs. This produced a total of about 45 million such
configurations, whereas now we find about 124 million, in both
cases before attempting to solve the tadpole conditions. In column
1, a $*$ indicates that the value of $x$ is not fixed by the quark
and lepton charges, as is the case in orientable models. In these
models, the value of $x$ may or may not be fixed by the zero-mass
condition for $Y$. If it is fixed, it can in principle have any
real value. In table (\ref{tbl:TableX}) this distinction is not
taken into account, but we do treat these models as distinct in
the complete list, table (\ref{tbl:Freq}), to be discussed below.

Apart from the $x=*$ cases, all other models are categorized with
the value of $x$ that follows from the quark and lepton charges as
well as the zero mass condition for $Y$. In some cases, the quark
and lepton charges alone might allow more than one value of $x$
even for unorientable models. For example, in $SU(5)$ GUT models
one can get the correct spectrum for $x=0$ (standard $SU(5)$) and
$x=\frac12$ (flipped $SU(5)$). The zero-mass condition for $Y$
always allows the former option (since $Y$ is a generator of the
non-abelian group $SU(5)$) and may or may not allow the latter. If
both are allowed, both are taken into account in table
(\ref{tbl:TableX}). Finally, if a model with $x=*$  gets $x$
fixed to a half-integer value by the $Y$-mass condition, it is
counted once as an $x=*$ model, and once for the actual value of
$x$.

In the third column we list how many of the configurations have no
anti-quarks realized as anti-symmetric $SU(3)$ tensors. As we will
discuss later, it is nearly impossible to get mass terms or Yukawa
couplings for such tensors, and therefore they should be regarded
as implausible. Note that anti-symmetric $SU(3)$ tensors are only
allowed for $x=0$ and $x=1/2$. In the former case, it turns out
that about $99\%$ of the configurations have such tensors, whereas
for $x=1/2$ only a few per cent have them.


\LTcapwidth=14truecm
\begin{center}
\begin{longtable}{|l|l|l|l|r|r|r|r|}\caption{\em Number of standard model configurations and tadpole solutions according to type.}\label{tbl:TableOne}\\
 \hline \multicolumn{1}{|l|}{$x$}
& \multicolumn{1}{l|}{Config.}
& \multicolumn{1}{l|}{stack {\brn c}}
& \multicolumn{1}{l|}{stack {\brn d}}
& \multicolumn{1}{r|}{cases}
& \multicolumn{1}{r|}{Total occ.}
& \multicolumn{1}{r|}{Top MIPFs}
& \multicolumn{1}{r|}{Solved} \\ \hline
\endfirsthead
\multicolumn{8}{c}%
{{\bfseries \tablename\ \thetable{} {\rm-- continued from previous page}}} \\
\hline \multicolumn{1}{|c|}{$x$}
& \multicolumn{1}{l|}{Config.}
& \multicolumn{1}{l|}{stack {\brn c}}
& \multicolumn{1}{l|}{stack {\brn d}}
& \multicolumn{1}{r|}{cases}
& \multicolumn{1}{r|}{Total occ.}
& \multicolumn{1}{r|}{Top MIPFs}
& \multicolumn{1}{c|}{Solved} \\ \hline
\endhead
\hline \multicolumn{8}{|r|}{{Continued on next page}} \\ \hline
\endfoot
\hline \hline
\endlastfoot
1/2 & UUUU &  C,D & C,D & 1732  &  1661111   &    8011  &   110(1,0)$^*$ \\
1/2 & UUUU &   C  & C,D & 2153 &   2087667   &   10394   &  145(43,5)$^*$ \\
1/2 & UUUU &   C &  C &  358  &  586940    &   1957  &   64(42,5)$^*$ \\
1/2 & UUU &  C,D & - &    2  &       28      &    2   &  0 \\
1/2 & UUU &  C  &  -  &  7   &   13310     &    74  &   3(3,2)$^*$ \\
1/2 & UUUN &  C,D &  - &    2  &       60     &     2  &   0 \\
1/2 & UUUN &  C &   - &   11   &    845     &    28   &  0 \\
1/2 & UUUR & C,D & C,D & 1361  &  3242251    &  12107   &  128(1,0)$^*$ \\
1/2 & UUUR &  C &   C,D &  914 &   3697145   &   12294  &   105(72,6)$^*$ \\
1/2 & USUU & C,D & C,D & 1760  &  4138505    &  14829   &  70(2,0)$^*$ \\
1/2 & USUU &  C &   C,D & 1763 &  8232083   &   17928  &   163(47,5)$^*$ \\
1/2 & USUU &  C &    C &  201  &  4491695    &   3155  &   48(39,7)$^*$ \\
1/2 & USU &   C,D &  - &    5   &   13515    &    384   &  5(2,0) \\
1/2 & USU &   C &   - &    2    &    222      &    4   &  0 \\
1/2 & USUN &   C,D &  - &   29  &    46011   &     338   &  2(2,0) \\
1/2 & USUN &   C &  - &   1     &    32     &     1   &  0 \\
1/2 & USUR &  C,D & C,D & 944  & 45877435   &   34233  &   130(4,0)$^*$ \\
1/2 & USUR &   C  &  C,D &  207 &  49917984  &    11722   &  70(54,10)$^*$ \\
0 & UUUU &   C,D  &   C,D  &   20   &    7950   &     110   &  2(2,0)  \\
0 & UUUU &   C  &   C,D &   164  &    50043   &     557   &  8(0,0) \\
0 & UUUU &   D   &  C,D   &   5   &    4512     &    40  &   0 \\
0 & UUUU &   C   &  C  & 1459   & 999122    &   5621   &  119(40,3)$^*$ \\
0 & UUUU &   C  &   D  &   26    &   6830     &    54   &  0 \\
0 & UUU &   C  &  -  &  11  &   17795   &    225   &  3(3,3)$^*$ \\
0 & UUUN &   C  &  -  &   31    &   5989    &    133   &  0 \\
0 & UUUR &   C,D  &   C   &  90   &  195638   &     702   &  4(4,0) \\
0 & UUUR &   C  &   C  & 4411  &  7394459   &   24715   &  392(112,2)$^*$ \\
0 & UUUR &   D   &  C  &   24   &   50752   &     148   &  0 \\
0 & UUR &   C  &  -   &   8   &  233071    &   1222   &  6(6,0) \\
0 & UURN &   C  &  -   &  37  &   260450   &     654   &  4(4,0) \\
0 & UURR &  C   &  C  &  1440 &  12077001   &   15029   &  218(44,0) \\
1 & UUUU &  C,D   &  C,D   &    5   &     212  &        8  &   0 \\
1 & UUUU &   C   &  C,D    &   6   &   7708    &     21  &   0 \\
1 & UUUU &   D   &  C,D    &   4     &  7708   &      11   &  0 \\
1 & UUUR &  C,D  &   D   &    1    &   1024   &       2  &   0 \\
1 & UUUR &  C    &  D    &   1    &    640    &      4   &  0 \\
$*$ & UUUU &  C,D  &  C,D  &  109  &   571472   &    1842   &  19(1,0)$^*$ \\
$*$ & UUUU &  C  &  C,D  &   32  &   521372  &     1199  &   7(7,0) \\
$*$ & UUUU &  D  &  C,D  &    8   &  157232     &   464  &   0 \\
$*$ & UUUU &  C  &  D  &    1     &     4     &     1   &  0
\end{longtable}
\end{center}

Table \ref{tbl:TableOne} summarizes all 19345 top-down distinct spectra we have observed after considering
all three and four stacks counted in the last column of table (\ref{tbl:Summary}).
The spectra are distinguished
on the basis of the chiral numbers of rank-2 tensors and bi-fundamentals, the
decomposition of $Y$, the presence and embedding of additional massless
({\it i.e.} not acquiring mass from axion couplings)
$U(1)$-gauge bosons from the
{\brn a}, {\brn b}, {\brn c}, {\brn d} stacks and brane unification among the {\brn a}, {\brn b}, {\brn c}, {\brn d} branes.
The columns contain the following data:
\begin{itemize}
\item{1. The value of $x$. An asterisk indicates that any value is allowed.
In all other cases the value of $x$
is the one determined from the ``zero $Y$-mass" condition.}
\item{2. Number of participating branes and their property:
\begin{itemize}
\item{U: Unitary (complex)}
\item{S: Symplectic}
\item{R: Real (Symplectic or Orthogonal)}
\item{N: Neutral (see below for a definition)}
\end{itemize}
}
\item{3. Composition of stack {\brn c} in terms of branes of types C and D.}
\item{4. Composition of stack {\brn d} in terms of branes of types C and D.}
\item{5. Total number of distinct (in the sense defined above) spectra of the type
specified in the first four columns. }
\item{6. Total number of spectra of given type. This is the grand total of
all such spectra found after scanning all the three and four brane
configurations in the
last column of table (\ref{tbl:Summary}), and assigning Chan-Paton multiplicities in order to get the Standard Model gauge group and spectrum. }
\item{7. Total number of MIPFs for which spectra of given type were found.}
\item{8. Number of distinct spectra for which tadpole solutions were found. Between parenthesis
we specify how may of these solutions have at most three mirror pairs, three MSSM Higgs pairs and
six singlet neutrinos, and how many have no mirror pairs, at most one Higgs pairs, and precisely three
singlet neutrinos. An asterisk indicates that at least one solution was found without additional hidden branes.}
\end{itemize}

In column 2, ``Neutral" means that this brane does not participate
to Y, and that there are no chiral bi-fundamentals ending on it.
The latter fact implies that there must be chiral rank-2 tensors
in this brane (which in particular implies that it must be
unitary), or otherwise it would violate condition 5b of the search
algorithm. Such a brane can only give singlet neutrinos. We found
a total of 111 such cases. They are anomaly free by having (a
multiple of) $-(N-4)$ symmetric tensors and $(N+4)$ antisymmetric
ones (for $N=4$ the anti-symmetric tensors are actually real, and
should strictly speaking have been omitted.) An N-brane can always
be removed to get a valid three-stack model, which of course
satisfies all our search criteria by itself. Note that branes of
this kind are in any case allowed to exist in the hidden sector,
and therefore from the point of view of classification it is most
natural to view these models as three-stack models with one
additional hidden sector brane. The reason we explicitly allowed
them is that singlet neutrinos from separate branes might be of
interest for understanding the neutrino mass problem (see also
section \ref{nm}). In the following analysis we will omit these
111 cases.

\subsection{Bottom-up versus Top-down}

In table( \ref{tbl:TableTwo}) and (\ref{tbl:TableThree}) we
compare the bottom-up and top-down results. This can only be done
by imposing some restrictions on the spectra. In addition to three
families of quarks and leptons and fully non-chiral matter (which
we ignore) there can be $G_{CP}$-chiral matter that is $G_{SM}$
non-chiral. The possibilities are mirror pairs of fermions,
singlet neutrino's and MSSM Higgs pairs. Denote these three
quantities as $M$, $N$ and $H$. If we leave them unrestricted,
there is an infinite number of bottom up solutions. Given the
current experimental knowledge, the optimal values for getting the
Standard Model would appear to be $M=0$, $N=3$ and $H=1$. However,
if there is a surplus of these particles, one can assume that they
get a standard-model-allowed mass above the weak scale. On the
other hand, if there is a shortage ($H=0$ or $N < 3$), there still
remains a possibility that the missing particles can come from
$G_{CP}$ non-chiral matter, or (in the case of neutrinos) from
additional branes (other than {\brn a}, {\brn b}, {\brn c} or
{\brn d}). Note for example that most of the models of
$\cite{Dijkstra:2004cc}$ have no $G_{CP}$-chiral Higgses, but
usually a large number of fully non-chiral Higgs candidates. Since
we have to impose cuts on $M,N$ and $H$ to make the comparison, we
present the comparison for two cases: a loose cut (with $M \leq 3,
N\leq 6, H\leq 3$) and a tight cut ($M=0, H\leq 1$ and $N=3$). The
former comparison is in table (\ref{tbl:TableTwo}) and the latter
in table (\ref{tbl:TableThree}). In both tables, the number of
bottom-up configurations satisfying the criteria is listed in
column 5. In column 6, we list the number of those bottom-up
configurations that was encountered in our search, and in column 7
the total number of occurrences of the given class\footnote{By
``class" we mean here all brane configurations that match the
criteria in the first four columns.}of configurations, summed over
all three or four brane combination considered in the search. This
is the same information as in column 6 of table
(\ref{tbl:TableOne}), but with the limit on the numbers $M,N$ and
$H$ imposed. In column 8 we list the number of distinct
configurations for which the tadpole conditions were solved. In
these tables the top-down spectra are only distinguished on the
basis of criteria that can be directly compared to the bottom-up
approach. Brane unification is ignored and the masses of $U(1)$
vector bosons are not taken into account. This means that some
models that were distinct in the previous table are considered
identical here, because they merely differ by branes that are not
on top of each other, or by different embeddings of an additional
massless $U(1)$ factor. This affects column 6 and column 8, but
not column 7, which is simply the sum of all occurrences within
the class. Note for example the in the class ($x=*$, UUUU, {\brn
c}=C, {\brn d}=(C,D)) there is a total number of occurrences of
521372 in both tables. This implies that all models satisfy the
constraints on the number of Higgs, mirrors and neutrinos. In
table \ref{tbl:Summary} these models correspond to 32 distinct
cases with 7 distinct solutions, whereas in table
\ref{tbl:TableTwo} they form only 7 distinct models with 3
distinct solutions.


\LTcapwidth=14truecm
\begin{center}
\begin{longtable}{|l|l|l|l|r|r|r|r|}\caption{\em Bottom-up versus Top-down results for spectra with at most three mirror pairs,
at most three MSSM Higgs pairs, and at most six singlet neutrinos.}\label{tbl:TableTwo}\\
 \hline \multicolumn{1}{|l|}{$x$}
& \multicolumn{1}{l|}{Config.}
& \multicolumn{1}{l|}{stack {\brn c}}
& \multicolumn{1}{l|}{stack {\brn d}}
& \multicolumn{1}{r|}{Bottom-up}
& \multicolumn{1}{r|}{Top-down}
& \multicolumn{1}{c|}{Occurrences}
& \multicolumn{1}{c|}{Solved} \\ \hline
\endfirsthead
\multicolumn{8}{c}%
{{\bfseries \tablename\ \thetable{} {\rm-- continued from previous page}}} \\
\hline \multicolumn{1}{|c|}{$x$}
& \multicolumn{1}{c|}{Config.}
& \multicolumn{1}{c|}{stack {\brn c}}
& \multicolumn{1}{r|}{stack {\brn d}}
& \multicolumn{1}{c|}{Bottom-up}
& \multicolumn{1}{c|}{Top-down}
& \multicolumn{1}{c|}{Occurrences}
& \multicolumn{1}{c|}{Solved} \\ \hline
\endhead
\hline \multicolumn{8}{|r|}{{Continued on next page}} \\ \hline
\endfoot
\hline \hline
\endlastfoot
1/2 & UUUU &   C,D & C,D &        27     &   9   &  5194  &    1 \\
1/2 & UUUU &   C & C,D &    103441   &   434 & 1056708 &      31 \\
1/2 & UUUU &   C  & C &  10717308   &   156 &   428799     &       24 \\
1/2 & UUUU &   C  & F &    351      &    0  &   0          &     0 \\
1/2 & UUU &   C,D  &  - &         4     &   1    &   24       &        0 \\
1/2 & UUU &   C &  - &       215    &    5  &   13310   &     2 \\
1/2 & UUUR &  C,D & C,D &        34   &     5  &   3888   &      1 \\
1/2 & UUUR &  C &  C,D &    185520   &   221  & 2560681   &   31 \\
1/2 & USUU & C,D & C,D &       72   &     7   &  6473       &        2\\
1/2 & USUU & C & C,D &    153436    &  283 & 3420508     &       33 \\
1/2 & USUU & C &  C &  10441784   &   125 & 4464095     &       27 \\
1/2 & USUU & C &  F &    184         &     0 &     0       &       0 \\
1/2 & USU &  C  & - &       104     &   2  &    222   &           0 \\
1/2 & USU &  C,D & - &        8    &   1   &  4881   &            1\\
1/2 & USUR &   C & C,D &    54274   &    31 & 49859327    &       19 \\
1/2 & USUR &   C,D & C,D &        36  &      2 &  858330       &        2 \\
0 & UUUU & C,D & C,D &        5     &   5 &    4530      &        2 \\
0 & UUUU & C & C,D &     8355   &    44  &  54102       &        2 \\
0 & UUUU & D & C,D &        14     &   2  &   4368      &         0 \\
0 & UUUU & C  & C &   2890537    &  127 &  666631    &        9 \\
0 & UUUU & C  &  D &     36304    &   16   &  6687     &         0 \\
0 & UUU & C &  - &       222    &    2  &  15440     &     1 \\
0 & UUUR & C,D & C &      3702   &    39 &  171485     &        4 \\
0 & UUUR & C  & C &  5161452  &    289 &  4467147     &       32 \\
0 & UUUR & D &  C &      8564    &   22   & 50748      &      0 \\
0 & UUR & C  &  - &        58     &   2  & 233071     &        2 \\
0 & UURR &   C  &  C &     24091 &      17  &  8452983   &          17 \\
1 & UUUU &  C,D   &  C,D   &    4   &     1  &       1144  &   1 \\
1 & UUUU &   C   &  C,D    &  16   &  5    &    10714  &   0 \\
1 & UUUU &   D   &  C,D    &   42     &  3   &      3328   &  0 \\
1 & UUUU &   C   &  D    &   870     &  0   &      0   &  0 \\
1 & UUUR &  C,D  &   D   &    34    &   1   &       1024  &   0 \\
1 & UUUR &  C    &  D    &   609    &    1    &      640   &  0 \\
3/2 & UUUU & C &  D  &         9    &  0        & 0      &     0   \\
3/2 & UUUU & C,D & D &         1     &   0       &   0       &    0  \\
3/2 & UUUU & C, D  & C &        10     &   0       &   0       &    0    \\
3/2 & UUUU & C,D & C,D  &         2     &   0  &      0   &     0 \\
$*$ & UUUU &  C,D  &  C,D  &  2  &   2   &    5146   &  1 \\
$*$ & UUUU &  C  &  C,D  &   10  &   7  &     521372  &   3\\
$*$ & UUUU &  D  &  C,D  &    1   &  1     &   116  &  $0$ \\
$*$ & UUUU &  C  &  D  &    3     &     1     &     4   &  $0$
\end{longtable}
\end{center}
Some bottom-up solutions can exist for more than one value of $Y$.
The most obvious example is the class $x=*$, which can exist for
all values of $Y$. In making the comparison we have used the
actual massless linear combination of $Y$ allowed by the
axion-gauge boson couplings in the top-down Gepner model. Only for
the $x=*$ case we have ignored  the precise form of $Y$, because
this would split this class into an indefinite number of
subclasses. However, in those cases where $Y$ was of the form
corresponding to $x=0,\frac12$ or 1, we have compared those
top-down models twice: once in the $x=*$ class, and once in the
class given by $Y$. This explains the tadpole solution indicated
in the last column of table (\ref{tbl:TableTwo}) for an $x=1$
model. Actually, this model has $x=*$, but $x$ is fixed to 1 by
the $Y$-mass condition.

The bottom-up numbers in these tables cannot be directly compared
with those in section \ref{StatBU} because here we allow several
branes of types C and D on the same stack, whereas in section
\ref{StatBU} we assumed that stack {\brn c} consists only of a
single type-C brane, and stack {\brn d} of a single type-D brane.
Furthermore in section \ref{StatBU} both $G_{\rm CP}$ chiral and
$G_{\rm CP}$ non-chiral Higgses are counted. We do not do that
here because the top-down search $G_{\rm CP}$ non-chiral Higgses
were ignored.


\LTcapwidth=14truecm
\begin{center}
\begin{longtable}{|l|l|l|l|r|r|r|r|}\caption{\em Bottom-up versus Top-down results for spectra without mirror pairs,
at most one MSSM Higgs pair, and precisely three singlet neutrinos. Only cases that
have been found in the top-dow search are shown.}\label{tbl:TableThree}\\
 \hline \multicolumn{1}{|l|}{$x$}
& \multicolumn{1}{l|}{Config.}
& \multicolumn{1}{l|}{stack {\brn c}}
& \multicolumn{1}{l|}{stack {\brn d}}
& \multicolumn{1}{r|}{Bottom-up}
& \multicolumn{1}{r|}{Top-down}
& \multicolumn{1}{c|}{Occurrences}
& \multicolumn{1}{c|}{Solved} \\ \hline
\endfirsthead
\multicolumn{8}{c}%
{{\bfseries \tablename\ \thetable{} {\rm-- continued from previous page}}} \\
\hline \multicolumn{1}{|c|}{$x$}
& \multicolumn{1}{c|}{Config.}
& \multicolumn{1}{c|}{stack {\brn c}}
& \multicolumn{1}{r|}{stack {\brn d}}
& \multicolumn{1}{c|}{Bottom-up}
& \multicolumn{1}{c|}{Top-down}
& \multicolumn{1}{c|}{Occurrences}
& \multicolumn{1}{c|}{Solved} \\ \hline
\endhead
\hline \multicolumn{8}{|r|}{{Continued on next page}} \\ \hline
\endfoot
\hline \hline
\endlastfoot
1/2 & UUU &  C  &  - &         8   &     2  &  13242       &        1\\
1/2 & UUUU &   C  & C &     10670   &    16 & 81985      &        4\\
1/2 & UUUU &   C & C,D &       148   &     8  & 378418      &        3\\
1/2 & UUUR &  C & C,D &       495    &   13  & 641485      &        3 \\
1/2 & USUU & C & C,D &       314   &     6 & 2757164   &       3 \\
1/2 & USUU & C &  C &     10816     &   6 & 4037872     &       4 \\
1/2 & USUR & C &  C,D &       434    &    3 & 47689675     &        3 \\
0 & UUUU &  C & C,D &        23     &   1    &    6        &        0\\
0 & UUUU &  C  & C &      1996    &    5   & 17301      &        2\\
0 & UUUU &  C &  D &        91     &   4  &   4227        &        0\\
0 & UUU & C  & - &         9    &    1  &  15282      &        1 \\
0 & UUUR & C  & C &      5136     &  15 &  63051      &     1
\end{longtable}
\end{center}


Table (\ref{tbl:Freq}) contains all 19345 distinct models we
found. Unfortunately the full table would be more than 500 pages,
and is too long to include, so we have only displayed the top and
some entries of interest.\footnote{However, the full list is
available on request.} The table is ordered according to the total
number of occurrences (listed in column 2) of a given spectrum.
Column 3 gives the number of MIPFs for which it occurs. This gives
some more indication how rare a certain spectrum is. In column 4
we give the Chan-Paton group, with factors combined if some of the
branes are on the same position. In column 5 we give a rough
indication of the spectrum. Here ``V" means that a CP-factor only
contributes bi-fundamentals, ``S"(``A") that there is at least one
(anti)-symmetric tensor and ``T" that both occur. Column 6 gives
the value of $x$, and the last column indicates if a solution to
the tadpole conditions was found (``Y"), and if a solution was
found without additional branes (``Y!").


\LTcapwidth=20truecm
\begin{center}
\begin{longtable}{|l|l|l|l|r|l|l|}
\caption{\em The list of 19345 models sorted according to frequency}\label{tbl:Freq} \\
\hline \multicolumn{1}{|c|}{nr}
& \multicolumn{1}{c|}{Total occ.}
& \multicolumn{1}{c|}{MIPFs}
& \multicolumn{1}{c|}{Chan-Paton Group}
& \multicolumn{1}{c|}{spectrum}
& \multicolumn{1}{c|}{x}
& \multicolumn{1}{c|}{Solved} \\ \hline
\endfirsthead
\multicolumn{7}{c}%
{{\bfseries \tablename\ \thetable{} {\rm-- continued from previous page}}} \\
\hline \multicolumn{1}{|c|}{nr}
& \multicolumn{1}{c|}{Total occ.}
& \multicolumn{1}{c|}{MIPFs}
& \multicolumn{1}{c|}{Chan-Paton Group}
& \multicolumn{1}{c|}{Spectrum}
& \multicolumn{1}{c|}{x}
& \multicolumn{1}{c|}{Solved} \\ \hline
\endhead
\hline \multicolumn{7}{|r|}{{Continued on next page}} \\ \hline
\endfoot
\hline \hline
\endlastfoot
1 &   9801844 &        648          & $U(3)\times Sp(2)\times Sp(6)\times U(1)$ &     VVVV & 1/2 & Y! \\
2 &      8479808(16227372) &  675 & $U(3)\times Sp(2)\times Sp(2)\times U(1)$ & VVVV & 1/2 & Y! \\
3 &       5775296 &  821 & $U(4)\times Sp(2)\times Sp(6)$ & VVV & 1/2 & Y! \\
4 &    4810698  & 868 & $U(4)\times Sp(2)\times Sp(2)$ & VVV & 1/2 & Y! \\
5 &     4751603 &  554 & $U(3)\times Sp(2)\times O(6)\times U(1)$ & VVVV & 1/2 & Y! \\
6 &     4584392  & 751 & $U(4)\times Sp(2)\times O(6)$ & VVV &  1/2 & Y \\
7 &     4509752(9474494) &  513 & $U(3)\times Sp(2)\times O(2)\times U(1)$ & VVVV &  1/2 & Y! \\
8 &      3744864 &  690 & $U(4)\times Sp(2)\times O(2)$ & VVV &  1/2 & Y! \\
9 &    3606292 &  467 & $U(3)\times Sp(2)\times Sp(6)\times U(3)$ & VVVV &  1/2 & Y \\
10 &      3093933 &  623 & $U(6)\times Sp(2)\times Sp(6)$ & VVV &  1/2 & Y \\
11 &     2717632  & 461 & $U(3)\times Sp(2)\times Sp(2)\times U(3)$ & VVVV &  1/2 & Y! \\
12 &      2384626  & 560 & $U(6)\times Sp(2)\times O(6)$ & VVV &  1/2 & Y  \\
13 &      2253928 &  669 & $U(6)\times Sp(2)\times Sp(2)$ & VVV &  1/2 & Y! \\
14 &     1803909 &  519 & $U(6)\times Sp(2)\times O(2)$ & VVV &  1/2 & Y! \\
15 &      1676493 &  517 & $U(8)\times Sp(2)\times Sp(6)$ & VVV &  1/2 & Y  \\
16 &     1674416 &  384 & $U(3)\times Sp(2)\times O(6)\times U(3)$ & VVVV &  1/2 & Y  \\
17 &     1654086 &  340 & $U(3)\times Sp(2)\times U(3)\times U(1)$ & VVVV &  1/2 & Y \\
18 &    1654086  & 340 & $U(3)\times Sp(2)\times U(3)\times U(1)$ & VVVV &  1/2 & Y \\
19 &    1642669 &  360 & $U(3)\times Sp(2)\times Sp(6)\times U(5)$ & VVVV &  1/2 & Y  \\
20 &      1486664 &  346 & $U(3)\times Sp(2)\times O(2)\times U(3)$ & VVVV &  1/2 & Y! \\
21 &     1323363  & 476 & $U(8)\times Sp(2)\times O(6)$ & VVV &  1/2 & Y  \\
22 &     1135702 &  350 & $U(3)\times Sp(2)\times Sp(2)\times U(5)$ & VVVV &  1/2 & Y! \\
23 &      1050764 &  532 & $U(8)\times Sp(2)\times Sp(2)$ & VVV &  1/2 & Y  \\
24 &      956980 &  421 & $U(8)\times Sp(2)\times O(2)$ & VVV &  1/2 & Y  \\
25 &      950003 &  449 & $U(10)\times Sp(2)\times Sp(6)$ & VVV &  1/2 & Y  \\
26 &      910132  &  51 & $U(3)\times U(2)\times Sp(2)\times O(1)$ & AAVV &  0 & Y  \\
$\ldots$ &  &   &  & &   &     \\
34 & 869428(1096682) & 246 & $U(3)\times Sp(2)\times U(1)\times U(1)$ & VVVV &  1/2 & Y!  \\  
153 & 115466 & 335 & $U(4)\times U(2)\times U(2)$ & VVV &  1/2 & Y   \\ 
225 & 71328 & 167 & $U(3)\times U(3)\times U(3)$ & VVV &  1/3 &     \\
303 & 47664  & 18 & $U(3) \times U(2) \times U(1) \times U(1)$ & AAVA &  1/2 & Y   \\ 
304 & 47664  & 18 & $U(3) \times U(2) \times U(1) \times U(1)$ & AAVA &  0  & Y   \\ 
343 & 40922(49794) & 63 & $U(3)\times Sp(2)\times U(1)\times U(1)$ & VVVV &  1/2 & Y!  \\ 
411 & 31000 & 17 & $U(3) \times U(2) \times U(1) \times U(1)$ & AAVA &  0  & Y   \\ 
417 & 30396 & 26 & $U(3) \times U(2) \times U(1) \times U(1)$ & AAVS &  0  & Y   \\ 
495 & 23544 & 14 & $U(3) \times U(2) \times U(1) \times U(1)$ & AAVS &  0  &     \\ 
509 & 22156 & 17 & $U(3) \times U(2) \times U(1) \times U(1)$ & AAVS &  0  & Y   \\ 
519 & 21468 & 13 & $U(3) \times U(2) \times U(1) \times U(1)$ & AAVA &  0  & Y   \\ 
543 & 20176(*) & 38 & $U(3) \times U(2) \times U(1) \times U(1)$ & VVVV &  1/2 & Y   \\ 
617 & 16845 & 296 & $U(5)\times O(1)$ & AV &  0 & Y   \\
671 & 14744(*) & 29 & $U(3) \times U(2) \times U(1) \times U(1)$ & VVVV &  1/2 &     \\  
761 & 12067 & 26 & $U(3)\times U(2)\times U(1)$ & AAS &  1/2 & Y!  \\
762 & 12067 & 26 & $U(3)\times U(2)\times U(1)$ & AAS &  0 & Y!  \\
1024 & 7466 & 7 & $U(3) \times U(2) \times U(2) \times U(1)$ & VAAV &1 &     \\ 
1125 & 6432 & 87 & $U(3)\times U(3)\times U(3)$ & VVV &  * & Y   \\
1201 & 5764(*) & 20 & $U(3) \times U(2) \times U(1) \times U(1)$ & VVVV &  1/2 &     \\ 
1356 & 5856(*) & 10 & $U(3) \times U(2) \times U(1) \times U(1)$ & VVVV &  1/2 & Y   \\ 
1725 & 2864 & 14 & $U(3) \times U(2) \times U(1) \times U(1)$ & VVVV &  1/2 & Y   \\ 
1886 & 2381 & 115 & $U(6) \times Sp(2)$ & AV &  1/2 & Y!  \\
1887 & 2381 & 115 & $U(6) \times Sp(2)$ & AV &  0 & Y!  \\
1888 & 2381 & 115 & $U(6) \times Sp(2)$ & AV &  1/2 & Y!  \\
2624 & 1248 & 3 & $U(3) \times U(2) \times U(2) \times U(3)$ & VAAV &  1 &     \\ 
2880 & 1049 & 34 & $U(5)\times U(1)$ & AS &  1/2 & Y!  \\
2881 & 1049 & 34 & $U(5)\times U(1)$ & AS &  0 & Y!  \\
2807 & 1096(*)  & 8 & $U(3) \times U(2) \times U(1) \times U(1)$ & VVVV &  1/2 &     \\ 
2919 & 1024 & 2 & $U(3) \times U(2) \times U(2) \times O(3)$ & VAAV &  1 &     \\ 
4485 & 400(*) & 2 & $U(3) \times U(2) \times U(1) \times U(1)$ & VVVV &  1/2 &     \\ 
4727 & 352 & 3 & $U(3) \times U(2) \times U(1) \times U(1)$ & VVVV &  1/2 &     \\ 
4825 & 332  &  20  & $U(4)\times U(2)\times U(2)$ & VAS & 1/2 & Y! \\
4902 & 320(*) & 1 & $U(3) \times U(2) \times U(1) \times U(1)$ & VVVV &  1/2 & Y   \\ 
4996 & 304 & 30 & $U(3)\times Sp(2)\times U(1)\times U(1)$ & VVVV &  1/2 & Y   \\ 
6993 & 128(**) & 1 & $U(3)\times U(2) \times U(2) \times U(1)$ & VVVV &  1/2 &     \\ 
7053 &  124 &    4 & $U(3)\times U(2) \times U(2) \times U(1)$  & VASV & 1/2 &    Y! \\
7241 & 116(**) & 4 & $U(3)\times U(2) \times U(2) \times U(1)$ & VVVV &  1/2 &     \\ 
7280 & 114 & 3 & $U(3)\times Sp(2)\times U(1)$ & AVS &  1/2 &     \\
7464 & 108 & 1 & $U(3)\times Sp(2)\times U(1)$ & VVT &  1/2 &     \\
7905 & 96(*) & 1 & $U(3) \times U(2) \times U(1) \times U(1)$ & VVVV &  1/2 &     \\ 
8747 & 68(**) & 3 & $U(3) \times U(2) \times U(1) \times U(1)$ & VVVV &  1/2 &     \\ 
8773 & 68 & 4 & $U(3) \times U(2) \times U(1) \times U(1)$ & VVVV &  1/2 &     \\ 
11347 & 32(**) & 1 & $U(3) \times U(2) \times U(1) \times U(1)$ & VVVV &  1/2 &     \\ 
11462 & 32(*) & 1 & $U(3) \times U(2) \times U(1) \times U(1)$ & VVVV &  1/2 &     \\  
12327 & 24 & 1 & $U(3)\times U(3)\times U(3) $ & VVV &  1/2 &     \\
15824 & 8 & 1 & $U(3) \times U(2) \times U(1) \times U(1)$ & VVVV &  0 &     \\  
15846 & 8 & 1 & $U(3) \times U(2) \times U(1) \times U(1)$ & VVVV &  1/2 &     \\ 
16674 & 6 & 1 & $U(3)\times U(2)\times U(1)$ & AVT  & 1/2 & Y!  \\
17055 & 4 & 1 & $U(3) \times U(2) \times U(1) \times U(1)$ & VVVV &  * &     \\  
19345 & 1 & 1 & $U(5) \times U(2)\times O(3)$ & ATV &  0 &
\end{longtable}
\end{center}

The first 25 models are all relatives of the $U(3)\times Sp(2)
\times U(1) \times U(1)$ models that dominated the search results
of \cite{Dijkstra:2004cc}. The variations include replacing the
third factor by $O(2)$ or $Sp(2)$, absorbing the family
multiplicity of some of the quarks or leptons in the Chan-Paton
multiplicities of the {\brn c} and {\brn d} branes, unifying the
baryon and lepton brane to get a Pati-Salam-like structure, and
other brane unifications. Models 17 and 18 occur with the same
frequency because they are closely related. They only differ by a
traceless generator $\diag(\frac13,\frac13,-\frac23)$ from the
$U(3)$ factor contributing to $Y$, changing the distribution of
some quarks and leptons. There are several other cases of closely
related models with identical frequencies, and one such set, nrs.
$1886\ldots 1888$ will be discussed in more  detail in section
\ref{Curios}. In the bottom part of the table we display several
lines of special interest, which will be discussed in more detail
below.

Entry nr. 26 in the table is the first one that cannot be viewed
as a relative of the ``Madrid model". It has $x=0$ and three
anti-symmetric tensors on the QCD and the weak brane. It can be
viewed as a broken $SU(5)$ model.

There exist several infinite series of models. In the top of the
list one can observe the beginning of the series $U(2n)\times
Sp(2) \times  G, n > 2$, where $G$ can be $O(2)$, $O(6)$, $Sp(2)$
or $Sp(6)$, with a chiral spectrum consisting of $\frac{6}{N_c}
(V,0,V)+3 (V,V,0)$.

In column 2 we indicate between parentheses if a certain type of
model was searched for in \cite{Dijkstra:2004cc}, and how often it
was found. It is interesting to compare this with table
(\ref{tbl:Summary}). Observe that the number of four-stack
configurations we consider in the present paper is considerably
smaller than in  \cite{Dijkstra:2004cc}, but nevertheless we
recover a large fraction of the standard model configurations of
that paper. For example, in \cite{Dijkstra:2004cc}, $2.8 \times
10^{15}$ configurations of type USUS were examined, in the present
paper only $26 \times 10^{14}$, ten times less. Nevertheless, we
have already found about half of the standard model
configurations. This is because the number of brane configurations
is dominated by cases with a large number of branes, but very few
standard model spectra. This in particular true for the charge
conjugation invariant (the simplest case, for which the boundary
coefficients were derived by Cardy \cite{Cardy:1989ir}) which in
essentially all cases has by far the largest number of boundaries.
The explanation may be that a non-trivial MIPF tends to fold over
a Calabi-Yau manifold several times, thus increasing the typical
intersection numbers, and causing the number three to occur more
frequently.

There are in total three cases with an $SU(3) \times Sp(2) \times
U(1) \times U(1)$ Chan-Paton group and only bi-fundamentals,
namely nr. 30, nr. 343 and nr. 4996. The first two were also
searched for in \cite{Dijkstra:2004cc}, and we find most of them
back. They are distinguished by having a massless (nr. 30) or
massive (nr. 343) $B-L$ gauge boson. The third one differs in the
way quarks and leptons end on branes {\brn c} and {\brn d}. It
does not have a lepton number symmetry, and was not considered in
\cite{Dijkstra:2004cc}. We show this case in more detail in the
next section, as a curiosity.

The remaining models considered in \cite{Dijkstra:2004cc} have a
$U(2)_{\brn b}$ group instead of  $Sp(2)_{\brn b}$. Here a direct
comparison is harder, because this splits into many subclasses,
which differ in the way the doublets are divided into $(2)$ and
$(2^*)$ representations of $U(2)$. The cases indicated by a single
$(*)$ are models considered in \cite{Dijkstra:2004cc} that have a
massless $B-L$ boson. In total 131704 such configurations were
found in that paper. For three of them we found tadpole solutions;
they correspond to the three ``type-1" models in table 4 of
\cite{Dijkstra:2004cc}. The ones indicated by $(**)$ have a
massive $B-L$ boson. Only 1306 of these were found in
\cite{Dijkstra:2004cc}, and in no case the tadpole conditions
could be solved.

Perhaps the most standard Chan-Paton group for standard model
realizations is $U(3) \times U(2) \times U(1) \times U(1)$. The
total number of spectra with that CP-group on the complete list is
281. Of these, 19 have a purely bi-fundamental spectrum, and among
these 19 there are 17 with $x=\frac12$, one with $x=0$ and one
with $x=*$. Of the 17 $x=\frac12$ models, 13 are variations on the
``Madrid" model, discussed above. The fourth $x=\frac12$ model
with a tadpole solution is discussed below in section
\ref{Curios}. All these 19 purely bi-fundamental models are shown
in table (\ref{tbl:Freq}). In addition we show all $U(3) \times
U(2) \times U(1) \times U(1)$ configurations that occur more
frequently than the first purely bi-fundamental model, nr. 543.
These are models with anti-symmetric $U(3)$ tensors. Note that
they occur more frequently despite the fact that models with
rank-2 tensors are suppressed, as will be discussed below. All of
them are broken $SU(5)$ models, except nr. 303, which is a broken
flipped $SU(5)$ variation of nr. 304.


\subsection{Standard model brane configurations not found\label{NotPresent}}

\begin{figure}[ht]\caption{Chiral tensor distribution for all standard model configurations.}\label{fig:TensorDistribution}
\begin{center}
\epsfig{file=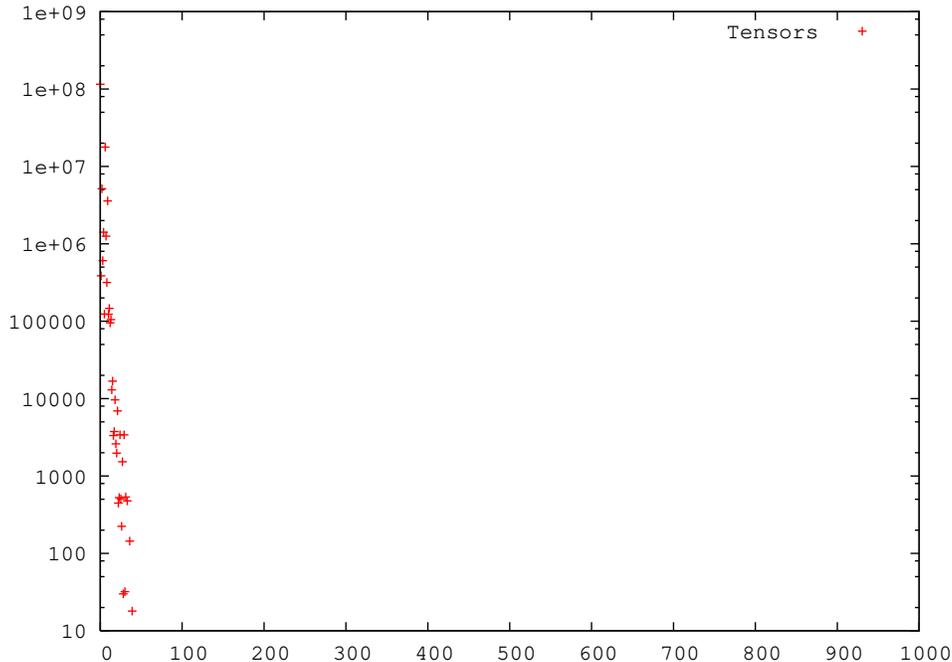,width=131mm}
\end{center}
\end{figure}
%

\begin{figure}[ht]\caption{Number of chiral tensors and bi-fundamentals for a selection  of branes.}\label{fig:Tensors}
\begin{center}
\epsfig{file=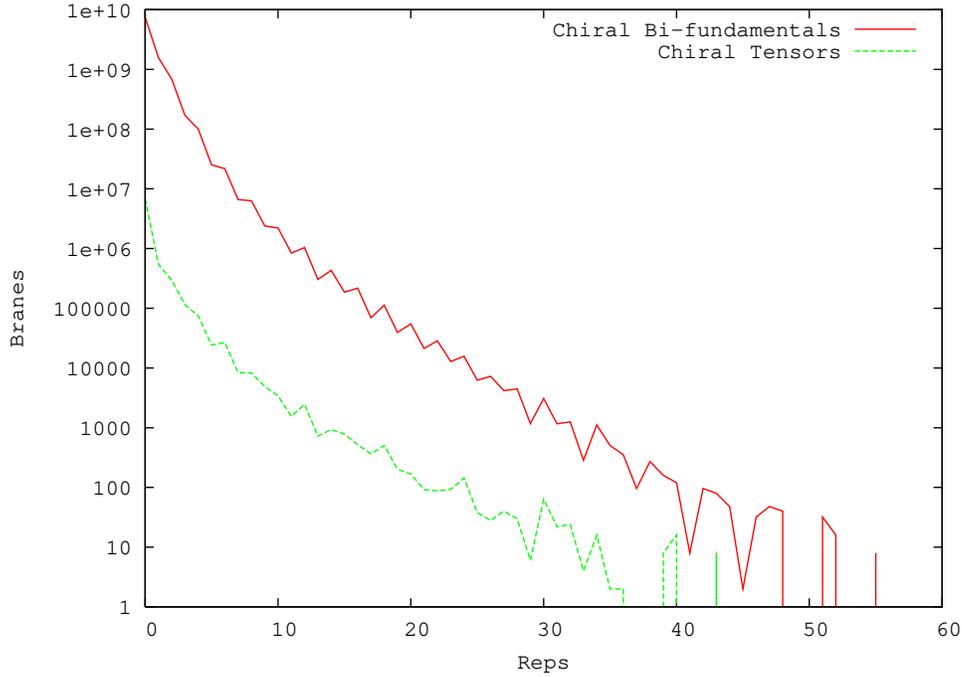,width=131mm}
\end{center}
\end{figure}
%

\begin{figure}[ht]\caption{Number of chiral  and non-chiral tensors
 for all single branes.}\label{fig:TensorsTwo}
\begin{center}
\epsfig{file=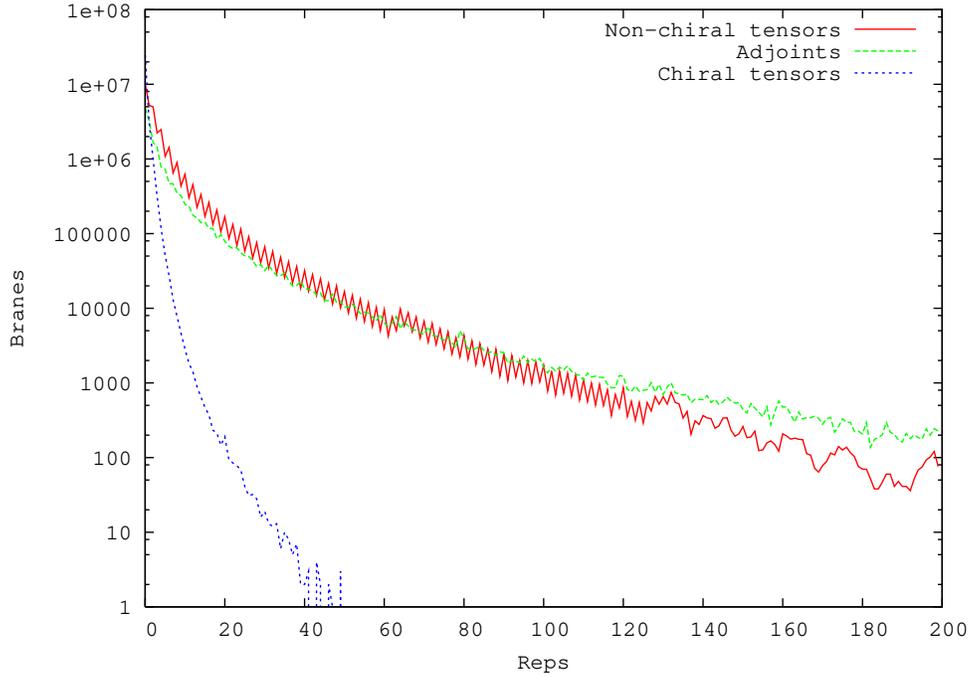,width=131mm}
\end{center}
\end{figure}
%

Note that only a very small fraction of the allowed bottom-up
models is actually realized as top-down
configurations\rlap.\footnote{All results in this section concern
brane configurations prior to tadpole cancellation.} This can be
explained in part by the fact that the bottom up models can have
several chiral tensors instead of chiral bi-fundamentals. In
figure (\ref{fig:TensorDistribution}) we plot the distribution of
the number of standard model top-down configurations we have found
versus the total number of chiral tensors in the spectrum. This
distribution is sharply peaked at zero. This implies that models
in which some quarks and leptons are realized as rank-2 tensors
are considerably harder to find in the part of the landscape we
are exploring here. In itself, this does not mean much for the
actual realization of the standard model in our universe. After
all, the suppression of models with tensors is by several factors
of ten only, and this does not seem very significant in comparison
to the total number of models in the landscape.

A partial understanding of this strong chiral tensor suppression
can be gained as follows. In fig. (\ref{fig:Tensors})  we plot for
all branes of a sample of 18001 orientifolds the distribution of
chiral bi-fundamentals and chiral tensors. On the horizontal axis
is the absolute value of the chirality, and on the vertical axis
the total number of occurrences. Clearly -- and not unexpectedly
-- the number bi-fundamentals is much greater than the number of
chiral tensors. This can be intuitively understood by realizing
that a brane has a much bigger chance intersecting with {\it any}
brane yielding a bi-fundamental than intersecting with one
specific brane (namely itself), yielding a chiral tensor.

One can also make an interesting observation regarding the
occurrence of chiral tensors in comparison to non-chiral ones. In
fig. (\ref{fig:TensorsTwo}) we list for all branes in all 33012
non-zero tension orientifolds the distribution of chiral and
non-chiral tensors (separately for adjoints and the other rank-2
tensors). Note that this includes {\it all} branes in all Gepner
orientifolds with non-zero-tension O-planes, not just those
considered in the present paper. Clearly the chiral distribution
falls off much faster than the non-chiral ones.

Although some other qualitative observations can be made, we do
really not have a good understanding of the absence of certain
models. Hypercharge embeddings with $x=-1/2,~3/2$ were not found
at all. The full list of 19345 configurations does contain some
genuine $x=1$ models, with $x$ fixed to that value by the quark
and lepton charges. There is a total of 17 distinct ones (for none
of these we found a solution to the tadpole conditions).  Only one
of these, nr. 2919, has an orthogonal group on the $\brn d$-stack,
but it is not identical to one of the simple models written down
in section \ref{XisOne}. It has a Chan-Paton group $U(3)\times
U(2)\times U(2)\times O(3)$, with both a C and a D brane on stack
{\brn c}. This model was found a total of 1024 times for just two
MIPFs. The purely unitary $x=1$ models 1024 and 2624 occur more
frequently. Another noteworthy absence in this class is the type
B,B' model introduced in \cite{dsm1}. These models have a
Chan-Paton group $U(3)\times U(2)\times U(1) \times U(1)$, and the
type-B model only has bifundamentals, whereas type-B' has
anti-symmetric tensor on $U(2)_{\brn b}$. However, all $x=1$
models we found have a $U(2)$ group on brane {\brn c}, and all
have anti-symmetric tensors both on branes {\brn b} and {\brn c}.
Some of these are similar to the models of \cite{dsm1}, but not
identical. Note that the type B,B' models of \cite{dsm1}, in to
order to be free of  cubic anomalies in the two $U(1)$ factors and
the $U(2)$, need $U(2)_{\brn b}$-chiral Higgs pairs and
anti-symmetric $U(1)$ tensors, as discussed in section
\ref{XisOne}. This suppresses their statistical likelihood.

Another model proposed in the literature that did not emerge in
our search is model C of \cite{AD}. This is a $U(3)\times U(2)
\times U(1)$ model with three $G_{\rm CP}$-chiral neutrinos
appearing as anti-symmetric tensors of $U(2)$. However, model nr
7464 in table (\ref{tbl:Freq}) is similar to it. It has exactly
the same structure as model C of \cite{AD}, after replacing $U(2)$
by $Sp(2)$. Then such neutrinos necessarily become non-chiral, and
the anomaly cancellation condition for the $U(2)$ factor becomes
irrelevant, increasing the chances of finding an example. Model
nr. 7464 occurred only 108 times (and without tadpole solutions).
Its presence suggests that there is no fundamental obstacle to
finding model C, but that it is simply statistically disfavored.
In other situations, replacing $U(2)$ by $Sp(2)$ increases the
number of occurrences by factors of about 40 to 80, and hence we
would expect at most a few examples of model C. This is consistent
with finding none.

On the full list of 19345 models there are 150 of the class $x=*$.
All of them are truly orientable, {\it i.e} the possibility of
having anti-symmetric $U(1)$ tensors that do not contribute
massless states does not occur. Only one has Chan-Paton group
$U(3)\times U(2) \times U(1) \times U(1)$. It is indeed precisely
the model (\ref{Oriented}) shown in section \ref{ClassBU}.
Amazingly this simple model occurs only four times (nr. 17055),
and just for one MIPF (and without any tadpole solution to tadpole
cancellation). This is especially surprising since there are many
other $U(3)\times U(2) \times U(1) \times U(1)$ configurations
with only bi-fundamentals that do occur much more frequently, as
discussed above. For example nr. 543 in table (\ref{tbl:Freq})
occurs 20176 times. This is a standard ``Madrid"-type
configuration.


\subsection{Higgs, neutrino and mirror distributions}

\begin{figure}[ht]\caption{Higgs pair distribution for all standard model configurations.}\label{fig:HiggsDistribution}
\begin{center}
\epsfig{file=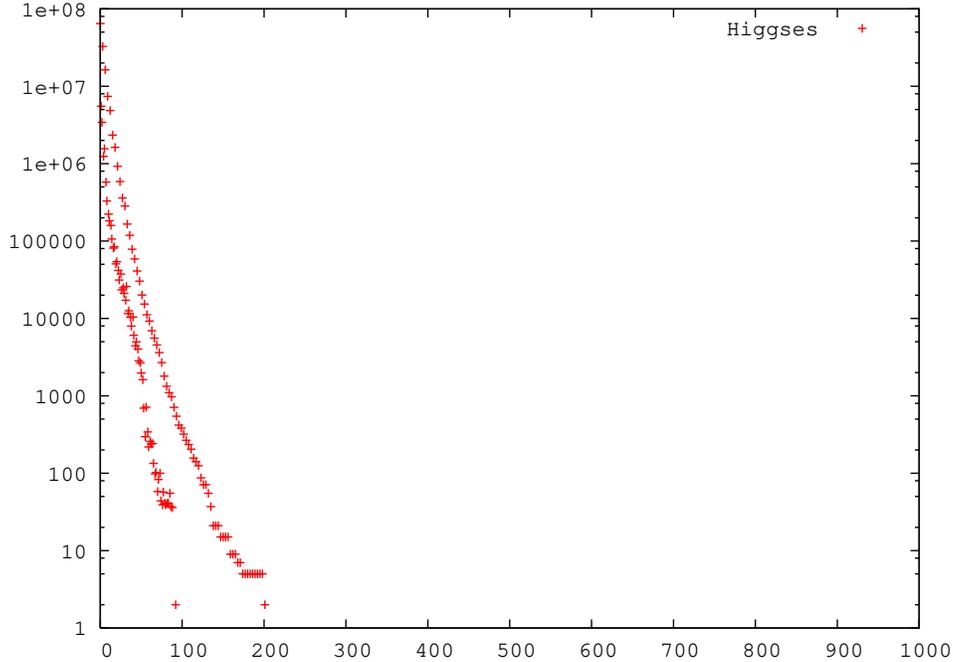,width=131mm}
\end{center}
\end{figure}
%

\begin{figure}[ht]\caption{Right-handed neutrino distribution for all standard model configurations.}\label{fig:NeutrinoDistribution}
\begin{center}
\epsfig{file=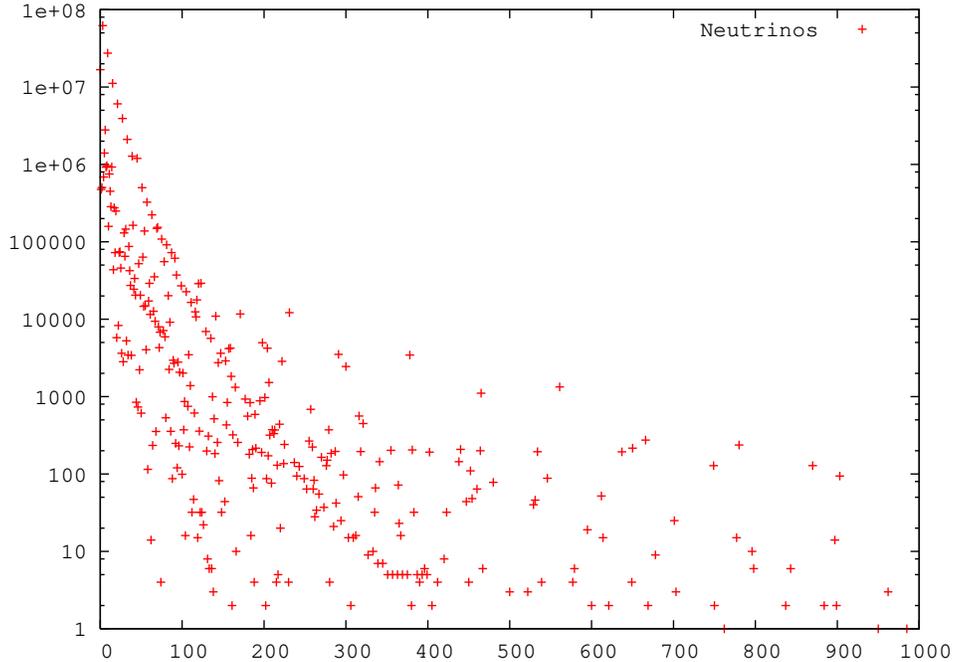,width=131mm}
\end{center}
\end{figure}
%

\begin{figure}[ht]\caption{Mirror distribution for all standard model configurations.}\label{fig:MirrorDistribution}
\begin{center}
\epsfig{file=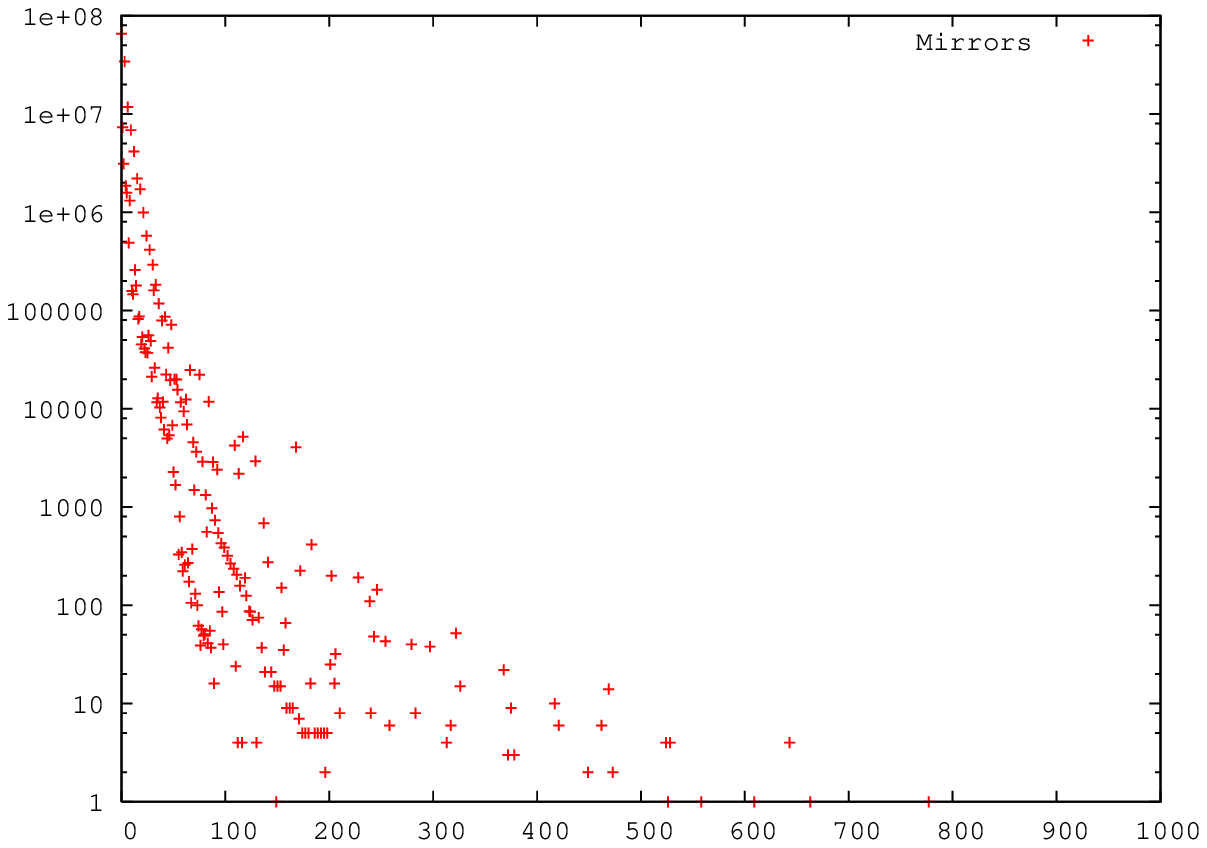,width=131mm}
\end{center}
\end{figure}
%

Figures (\ref{fig:HiggsDistribution}),
(\ref{fig:NeutrinoDistribution}) and
(\ref{fig:MirrorDistribution}) and show the distribution in terms
of the number of Higgs, right-handed neutrinos and mirror pairs.
On the vertical axis we show the total number of three and
four-brane configurations that have a chiral standard model
spectrum, plus the number of Higgses/neutrinos/mirrors indicated
on the horizontal axis. Just as all data in this section, these
numbers refer to brane configurations prior to tadpole
cancellation. The Higges/neutrinos/mirrors are $G_{\rm CP}$ chiral
but of course $G_{\rm SM}$ non-chiral. In addition to these
particles, the massless spectrum may contain  $G_{\rm
CP}$-non-chiral particles with the same standard model
transformation properties. Since we classify models modulo full
non-chiral matter, we have no general information about such
particles. The mirror count is the total of all mirror pairs of
quark and charged lepton weak singlets, as well as quark doublets
(in this case mirrors can occur only for $x=\frac12$). The Higgs
count refers to $(1,2,\frac12)+(1,2,-\frac12)$ pairs; for example
the MSSM has one such pair. Note that these pairs could also be
viewed as lepton doublet mirror pairs. The distinction can be made
in models with a well-defined lepton number, but since we are not
insisting on that we simply count all such pairs as candidate
Higgs. Once one (or more) of these candidates acquires a v.e.v,
one may discuss if lepton number violation is absent or acceptably
small.

Finally fig. (\ref{fig:NeutrinoDistribution}) shows the
distribution of the total number of standard model singlets  in
the $G_{\rm CP}$-chiral spectrum.

In all three plots two lines are visible. The top line corresponds
to multiplicities that are $0 {\rm~mod~} 3$, and the lower to
multiplicities that are not $0 {\rm~mod~} 3$. The former occur
more frequently due to anomaly cancellation and the fact that we
require the presence of three chiral families. In some classes of
models this imposes a mod 3 constraint on the multiplicities of
Higgses, mirror or neutrinos. This feature is clearest in the
Higgs plot, because the Higgs is in a definite,  and non-trivial
standard model representation with few $G_{CP}$ realizations. It
is less clear in the neutrino plot, because there are often many
ways of making neutrinos. The models with huge numbers of
(right-handed) neutrino candidates usually contain a large factor
$G_{\brn c}$ or $G_{\brn d}$, with neutrinos coming from rank-2
tensors.

\section{Solutions to the tadpole conditions\label{TadSol}}

In this section we present some examples of solutions to complete
set of tadpole solutions that we have found. All solutions that we
present also satisfy the probe brane constraints for the absence
of global anomalies \cite{Uranga:2000xp}, as discussed in
\cite{Gato-Rivera:2005qd} for this class of models. We emphasize
that we have collected at most two tadpole solutions for each
chiral model, one with additional branes, and one without
additional branes. This means, for example, that as soon as one
solution was found for one of the 9785532 $SU(3) \times SU(2)
\times Sp(6) \times U(1)$ models that appears as nr. 1 in table
(\ref{tbl:Freq}), no further attempt was made for any of the
others with the same chiral spectrum. This is a very different
strategy than the one of \cite{Dijkstra:2004cc}, where all tadpole
solutions were collected for models with distinct {\it non}-chiral
spectra. In the examples below we present the full massless
spectrum of the actual tadpole solution, including non-chiral
states. The non-chiral states are however specific to the example
we present, and solutions with different non-chiral multiplicities
for a given chiral multiplicity certainly exist. Indeed, for
spectrum nr 2 in table (\ref{tbl:Freq}), which was included in the
search presented in \cite{Dijkstra:2004cc}, more than 100000
non-chirally distinct samples with tadpole solutions were found.

We only present a small selection of the 1900 tadpole solution we
have collected. They should be viewed merely as existence proofs
of a certain type of model, and not as a statement that one of
these is likely to survive further phenomenological constraints.
Whenever possible, we present examples without hidden branes, not
because we believe these are more viable (indeed, hidden sector
branes may be required for a variety of phenomenological reasons),
but simply because they can be written down more easily.

\subsection{Hypercharge embeddings of the tadpole solutions }


Let us first make a few more comments on the models that do or do
not occur in  the list of 1900 tadpole solutions. We have seen in
the previous section that most  bottom-up  models of section
\ref{ClassBU} and \ref{StatBU} do not occur on the list of brane
configuratios, and it is therefore clear that most are also absent
from the list of tadpole solutions (see section \ref{NotPresent}).
Furthermore, in many top-down tadpole solutions, the hypercharge
appears to be a combination of more than one of the hypercharge
embeddings of the bottom-up  models in section \ref{StatBU}. First
consider the ``pure" models
\begin{itemize}
\item 762 top-down configurations have hypercharge of the form
$Y=-{1\over 3}Q_{\brn a}-{1\over 2}Q_{\brn b}$. This is related to
a small subclass of bottom-up models in section
\ref{Xis1over2andCC}. In table \ref{tbl:TableTwo} these
models have $x=0$ and both ${\brn c}$, ${\brn d}$ branes are of
the C type (or are real, or absent).

\item 1095 top-down configurations have hypercharge of the form
$Y=-{1\over 6}Q_{\brn a}+ {1\over 2}Q_{\brn c} -{1\over 2}Q_{\brn
d}$ which is related to a subclass of the bottom-up models in
section \ref{Xis1over2}.
\end{itemize}
The rest of the configurations appear with the hypercharge to be
described by two different embeddings. This is due to the
contribution of traceless generators in the hypercharge. These
``mixed" models are distributed as follows:
\begin{itemize}
\item 17 top-down configurations
have a combined
hypercharge of the type: $Y=-{1\over 3}Q_{\brn a}-{1\over
2}Q_{\brn b}+Q_{\brn d}$ (section \ref{Xis0} and corresponding to models A,A' in \cite{dsm1,dsm3,d-effective}) and $Y=-{1\over
3}Q_{\brn a}-{1\over 2}Q_{\brn b}$ (section \ref{Xis1over2andCC}).
These are hypercharges with $x=0$ but ${\brn c}$ and ${\brn d}$
branes are of the type C and D, C respectively.
\item 2 top-down configurations appear with a combined hypercharge with $x=1$
(section \ref{Xis1} and corresponding to models B,B' in \cite{dsm1,dsm3,d-effective}), but ${\brn c}$ and ${\brn d}$
 branes are of the
type C, D and D respectively.
\end{itemize}

Here we used the hypercharge values as determined from the quark
and lepton charges as well as the $Y$-mass condition. The two
mixed $x=1$ models mentioned above actually have $x=*$, with $x$
fixed to 1 by the $Y$-mass condition. One of those appears in
(\ref{tbl:TableTwo}); the other has too many neutrinos and lies
outside the limits used for that table. A total of 20 out of the
1900 tadpole solutions have $x=*$, but $x$ fixed to a
non-canonical value by the $Y$-mass condition. Finally there are 4
with $x$ completely unfixed by any condition.

\subsection{Notation}
The notation of the examples is as follows. Minimal model tensor
products are denoted as $(k_1,\ldots,k_m)$, where $k_i$ is the
$SU(2)$ level. Their modular invariant partition functions are
labelled by an integer, which is assigned sequentially as they are
computed. This labelling can be resolved in terms of more precise
data: the simple current subgroup and the rational matrix $X$
defining the MIPF (as defined in
\cite{Gato-Rivera:1991ru}\cite{Kreuzer:1993tf}). We omit these
data here, but they are available on request. To help identify the
MIPF we will provide the Hodge numbers of the corresponding
Calabi-Yau manifold, and the number of singlets that occur in the
spectrum of heterotic strings compactified on such a manifold.
Orientifolds are also labelled by a sequential integer assigned by
the computer program.

Representations are denoted as $(r_a,...r_d,...)$, where each entry refers to
one of the branes ({\brn a}, {\brn b}, {\brn c}, {\brn d} and hidden), and $r$ can be $V$ for vector, $A$
for anti-symmetric tensor, $S$ for symmetric tensor and $Adj$ for Adjoint.
An asterisk indicates complex conjugation.
All representations refer to left-handed fermions.
Multiplicities of complex
representations are denoted as
$$ N \times (r_a,....)_M $$
where $N$ is the total number of times a representation plus its
conjugate appears, and $M$ is the chirality, the difference of the
multiplicity of the representation that is listed, and its
conjugate. The subscript is omitted for non-chiral
representations.

\subsection{$U(3) \times U(2) \times U(1)$ models}

Here we list all tadpole solutions we found with a Chan-Paton
group which is exactly $U(3) \times U(2) \times U(1)$ (or less, if
some combinations of the unitary phase factors -- other than $Y$
-- get a mass from axion couplings).

The first two examples are nr. 761 and 762 from the list. They
are respectively broken versions of $SU(5)$ and flipped $SU(5) \times U(1)$
unifications, with $SU(5)$ broken by splitting the stack of five branes into
three plus two. These models occurred for MIPF 31 of $(1,1,1,1,7,16)$ (there
is just one orientifold choice). The $U(3) \times U(2) \times U(1)$ spectrum is
\begin{eqnarray*}
      3  &\times& ( A ,0 ,0 )_3\\
      3  &\times& ( 0 ,A ,0 )_3\\
      5  &\times& ( V ,V ,0 )_3\\
     25  &\times& ( 0 ,0 ,S )_3\\
      9  &\times& ( V ,0 ,V )_{-3}\\
      3  &\times& ( 0 ,V ,V )_{-3}\\
      4  &\times& ( Ad,0 ,0 ) \\
      1  &\times& ( 0 ,Ad,0 ) \\
     16  &\times& ( 0 ,0 ,Ad) \\
      6  &\times& ( 0 ,0 ,A ) \\
      8  &\times& ( S ,0 ,0 ) \\
     14  &\times& ( V ,0 ,V^*) \\
      4  &\times& ( 0 ,V ,V^*)
\end{eqnarray*}
The possible choices for $Y$ are the $SU(5)$ embedding $Y=-\frac13
Q_{\brn a} + \frac12 Q_{\brn b}$ and the flipped embedding
$Y=\frac16 Q_{\brn a} + \frac12 Q_{\brn c}$ for nr. 562 and 561
respectively. In both cases an additional $U(1)$, the independent
linear combination of these two, also remains massless.

There is a second, far less standard example of a $U(3) \times
U(2) \times U(1)$ model, which occurred for invariant 28 of
441010, orientifold 0. This is nr 16674 on the list, which
occurred only six times in total (and only for this MIPF), but
against all odds a tadpole solution was found for at least one of
the six occurrences. The embedding of $Y$ is as for the flipped
$SU(5)$ model above, but only two of the three down quarks are due
to anti-symmetric tensors, and there are no anti-symmetric tensors
in $U(2)$. Furthermore there are three candidates for Higgs
bosons, but unfortunately no symmetry like lepton number to
distinguish them from the lepton doublets. This implies that there
are no singlet neutrino candidates from the standard model branes,
and that with a suitable Higgs boson chosen from the three
candidates mentioned above, all up quarks and one of the down
quarks can acquire a mass. The exact spectrum is as follows
\begin{eqnarray*}
      9 &\times& ( 0 ,V ,V^*)_{-3}\\
      3 &\times& ( 0 ,0 ,S )_3\\
      6 &\times& ( A ,0 ,0 )_2\\
      3 &\times& ( 0 ,0 ,A )_1\\
      6 &\times& ( 0 ,V ,V )_{-6}\\
      7 &\times& ( V ,V ,0 )_{3}\\
      7 &\times& ( V ,0 ,V^*)_{-1}\\
      3 &\times& ( V ,0 ,V )_{-3}\\
      3 &\times& ( Ad,0 ,0 )\\
      6 &\times& ( 0 ,A ,0 )\\
      7 &\times& ( 0 ,Ad,0 )\\
      8 &\times& ( 0 ,S ,0 )\\
      8 &\times& ( V ,V^*,0 )\\
      4 &\times& ( 0 ,0 ,Ad)
\end{eqnarray*}
The gauge group is exactly $SU(3)\times SU(2)\times U(1)$, because
all abelian gauge bosons other than $Y$ acquire a mass.

Somewhat surprisingly, there were no tadpole solutions for
$U(3)\times Sp(2) \times U(1)$ models, even though usually
replacing $U(2)$ by $Sp(2)$ greatly increases the frequency of a
model.

\subsection{Unification}

In general we can speak of (partial) unification if some of the
stacks {\brn a}, {\brn b}, {\brn c} and {\brn d} coincide. One can
distinguish the following possibilities

\begin{enumerate}
\item{{\brn a} = {\brn b}.
In this case the bi-fundamentals that yield quark doublets must
necessarily come from anti-symmetric tensors on the combined stack. There must
therefore be three anti-symmetric tensors, and the combined gauge group
is $U(5)$. Hence this leads to $SU(5)$ GUT models. The $SU(3)$ anti-symmetric tensors
can be $u^c$ or $d^c$ quarks. The first case corresponds to standard $SU(5)$, the
second to flipped $SU(5)$. There must be at least one more brane stack to accommodate the
anti-quarks of the other charge. Hence these models can be realized with just two stacks.}
\item{{\brn a} = {\brn c}. In this case the weak brane remains
separate, but the QCD brane is extended. The best-known example is
the Pati-Salam model, where $U(3)_{\brn a}$ is extended with a
lepton-number $U(1)$. The Pati-Salam model requires three stacks,
but it is possible to realize unifications of this type with just
two stacks. An example is (one of the variations of) the $U(6)
\times Sp(2)$ discussed below.}
\item{{\brn b} = {\brn d}. In this case the weak brane is part of
a larger group. An example is trinification: here $U(2)_{\brn b}$
is embedded in a $U(3)$. Without loss of generality, we may choose
stack {\brn d} as the one that merges with the weak brane. The
trinification model then needs one additional brane stack,
$U(3)_{\brn c}$. All models in this class must in fact have a
third brane stack, in order to get anti-quarks as bi-fundamentals;
at least one of the two anti-quarks charges must be realized as a
bi-fundamental.}
\item{{\brn a} = {\brn b} = {\brn d}. An example will be given
below.}
\end{enumerate}
Here it is assumed that no more branes coincide than those
indicated. If {\brn c} and {\brn d} coincide this would be
regarded as a single stack denoted {\brn c}. If {\brn c} coincides
with {\brn a} or {\brn b} we switch the r\^oles of {\brn c} and
{\brn d}. This limits the possibilities to those listed here.

\subsubsection{SU(5) models}

The following is an example of an $SU(5)$ model. It is item 617 in
table (\ref{tbl:Freq}) and despite having a hidden sector, this
model has as its gauge group precisely $SU(5)$ and nothing more!
The standard model part consists of an $U(5)$ complex stack and a
single real $O(1)$ brane. This is needed for the endpoints of the
strings yielding the representation $(5^*)$. In addition this
example has one extra $O(1)$ brane that serves as a hidden sector.
The example occurs for tensor product (1,4,4,4,4) and MIPF nr. 63
in our classification, which is characterized by Hodge numbers $
(h_{21},h_{11})=(7,31)$, and yields 237 singlets if one uses this
MIPF to construct a heterotic string. The total number of
boundaries is 246. The orientifold is the one with maximal O-plane
tension. The precise spectrum is as follows
\begin{eqnarray*}
      3 &\times& (A ,0 ,0 )_3\\
     11 &\times& (V ,V ,0 )_{-3}\\
      8 &\times& ( S ,0 ,0 ) \\
      3 &\times& ( Ad,0 ,0 ) \\
      1 &\times& ( 0 ,A ,0 ) \\
      3 &\times& ( 0 ,V ,V ) \\
      8 &\times& ( V ,0 ,V ) \\
      2 &\times& ( 0 ,S ,0 ) \\
      4 &\times& ( 0 ,0 ,S ) \\
      4 &\times& ( 0 ,0 ,A )
\end{eqnarray*}

We emphasize that this just one sample of many such models.  There
are 16845 configurations of this kind ({\it i.e.} with the same
first two CP-factors $U(5) \times O(1)$ and the same chiral
spectrum). The other 16844 configurations may differ from the one
shown here by having, for example, different numbers of $U(5)$
adjoints or $(V,V)$ mirror pairs. Some of these 16845
configurations are identical to the one shown here, because of
surviving discrete symmetries of the $(1,4,4,4,4)$ tensor product.
But the fact that this chiral spectrum was found for 296 different
MIPFs essentially guarantees that many different versions exist.

This model has one hidden sector brane. According to our strategy,
outlined in the beginning of this section, none of the remaining
models of this type was checked for tadpole cancellation {\it
with} hidden branes after this tadpole solution was found.  All
16845 configurations were checked for tadpole cancellation {\it
without} hidden branes, and no solutions were found. It is
straightforward to re-examine all these 16845 model and check for
further possibilities of tadpole cancellation, in order to obtain
different non-chiral spectra or different hidden sectors. But
there are many other models of potential interest, including many
more $SU(5)$ models.

\subsubsection{Flipped SU(5) models}

The simplest flipped $SU(5)$ we found occurs for
for invariant 52 of (1,4,4,4,4), orientifold 0, with
characteristics (3,51,253). It solves all tadpole equations
with just two brane stacks, the minimal number needed to
realize flipped $SU(5)$. The full Chan-Paton group is
$U(5) \times U(1)$, and the spectrum is
\begin{eqnarray*}
     11 &\times& ( 0 ,S )_3\\
      3 &\times& ( A ,0 )_3\\
      5 &\times& ( V ,V )_{-3}\\
      8 &\times& ( S ,0 ) \\
      9 &\times& ( Ad,0 ) \\
      5 &\times& ( 0 ,Ad) \\
      4 &\times& ( 0 ,A ) \\
     12 &\times& ( V ,V^*)
\end{eqnarray*}
In terms of ({\brn a}, {\brn b}, {\brn c}, {\brn d}) branes this
model is of the form $U(3)_a \times U(2)_b  \times U(1)_c$ with
{\brn a} $=$ {\brn b} and no {\brn d} brane, and
$Y=\frac16(1,0,3)$. The way the $U(1)$ anomalies cancel is
noteworthy. Per family, there are five $U(1)$ anti-vector
representations, contribution -5 to the cubic anomaly. This
anomaly is cancelled by a symmetric tensor, which contributes $+5$
in a $U(1)$ theory. The chiral part of the spectrum yields exactly
the standard model spectrum, with 3 right-handed neutrinos from
the three chiral symmetric tensors. There are no $G_{\rm
CP}$-chiral Higgs candidates.

This is model nr. 2880 in table (\ref{tbl:Freq}). As explained
earlier, such a flipped $SU(5)$ model always has a standard
$SU(5)$ counterpart, because the masslessness of the extra $U(1)$
of flipped $SU(5)$ is an additional constraint not needed for
standard $SU(5)$. This is model nr 2881 in table (\ref{tbl:Freq}).

To the best of our knowledge, these are the first exact chiral,
supersymmetric $SU(5)$ and flipped $SU(5)$ models in the
literature. Their chiral spectrum, directly obtained in string
theory, without postulating further Higgs effects or
non-perturbative physics, is exactly $3 \times (10) + 3 \times
(5^*)$. By contrast, the models found in \cite{cvetic} contain
additional $(15)$'s of $SU(5)$. The models found recently in
\cite{Chen:2006ip} have $G_{\rm CP}$ mirror pairs of $(5)$ and
$(5^*)$, which must be made massive by postulating an additional
Higgs mechanism breaking part of the additional gauge symmetry. We
emphasize that the mirror pairs shown above in the explicit
spectrum are non-chiral with respect to the full Chan-Paton group,
and hence require no gauge symmetry breaking to acquire a mass.

In addition, the model shown above is obviously the simplest one
possible, apart from the $U(5) \times O(1)$ of the previous
subsection, if one could find a realization without hidden sector.

However, both the standard and the flipped $SU(5)$ model have a
serious problem with either the $(u,c,t)$ or $(d,s,b)$ Yukawa
coupling. We will discuss this in detail in section
\ref{UFiveMasses}.

For other work discussing aspects of (flipped) $SU(5)$ model
building along similar lines, see
\cite{Blumenhagen:2001te}\cite{Ellis:2002ci}
\cite{floratos},\cite{cvetic}\cite{nano}\cite{Chen:2006ip}. For
other issues in $SU(5)$ model building with branes and the
associated problems see \cite{Antebi:2005hr}, \cite{bere}.

\subsubsection{Pati-Salam models}

The simplest Pati-Salam model is nr. 4 on the list, and
is therefore one of the most frequent ones. A tadpole solution was
found for invariant 57 of (2,10,10,10), orientifold 3.
The gauge group is $U(4) \times Sp(2) \times Sp(2)$, and
the spectrum is as follows
\begin{eqnarray*}
      5  &\times& ( V ,0 ,V )_{-3}\\
      3  &\times& ( V ,V ,0 )_3\\
      2  &\times& ( Ad,0 ,0 )  \\
      2  &\times& ( 0 ,A ,0 )  \\
      7  &\times& ( 0 ,0 ,A ) \\
      4  &\times& ( A ,0 ,0 )  \\
      2  &\times& ( 0 ,S ,0 )  \\
      5  &\times& ( 0 ,0 ,S )  \\
      7  &\times& ( 0 ,V ,V )
\end{eqnarray*}
The embedding of $Y$ is as $Y=\frac16 Q_{\brn a}-\frac12 Q_{\brn
d} + W_{\brn c}$, where  $W_{\brn c}=\frac12 \sigma_3$. Brane
{\brn a} and {\brn d} are unified to $U(4)$.

The following model is of interest because it is a $U(4) \times
U(2) \times U(2)$ Pati-Salam model that satisfies all tadpole
conditions without hidden branes, because it has some chiral
rank-2 tensors in its spectrum, and because it occurs for a MIPF
related to the ``quintic" Calabi-Yau, namely MIPF 6 of
(3,3,3,3,3), the trivial orientifold (the only one possible). It
is nr. 4825 on the list (\ref{tbl:Freq}). It has precisely one
$G_{\rm CP}$ chiral MSSM Higgs pair, plus a $G_{\rm CP}$-chiral
charged lepton mirror pair, and four right-handed neutrinos. There
is one massless $U(1)$ in addition to $Y$, namely the diagonal
combination of the phase factors of the $U(2)$'s. The Chan-Paton
group is $U(4)\times U(2) \times U(2)$, and the representations
are
\begin{eqnarray*}
      3 &\times& ( V ,0 ,V^*)_{-1}\\
      2 &\times& ( V ,0 ,V )_{-2}\\
      1 &\times& ( 0 ,0 ,S )_{1} \\
      5 &\times& ( 0 ,A ,0 )_{1}\\
      5 &\times& ( V ,V^*,0 )_{1}\\
      6 &\times& ( V ,V ,0 )_{2}\\
      3 &\times& ( 0 ,V ,V )_{-1}\\
      4 &\times& ( 0 ,S ,0 )\\
      4 &\times& ( S ,0 ,0 )\\
      3 &\times& ( Ad,0 ,0 )\\
      5 &\times& ( 0 ,Ad,0 )\\
      1 &\times& ( 0 ,0 ,Ad)\\
      2 &\times& ( 0 ,V ,V^*)
\end{eqnarray*}
There also exist a broken version of this model, with $U(4)$ split
into $U(3) \times U(1)$ already in the exact string theory. This
is nr. 7053 in (\ref{tbl:Freq}).

There is also a $U(4)\times U(2) \times U(2)$ Pati-Salam model
(nr. 153) which has a standard, purely bi-fundamental spectrum.
For this model we only found a tadpole solution with hidden
branes, which is a bit too complicate to display here. It has a
hidden sector group $U(6) \times U(2)^3 \times O(2)^2 \times
Sp(2)$.

Orientifolds exhibiting a Pati-Salam realization of the SM have
been considered before, \cite{ps1,ps2,ps3}. Bottom-up
configurations, investigating also gauge couplings and the issue
of masses, have been also considered, \cite{lr,lr1}.

\subsubsection{Trinification models}

Trinification models are built out of three factors $SU(3)$ with
purely bi-fundamental matter. At first sight this would seem to be
an ideal configuration for intersecting brane models, but in fact
it is surprisingly rare.

In a genuine trinification model the generator $Y$ is embedded in
$SU(3)_{\brn a} \times SU(3)_{\brn b} \times SU(3)_{\brn d}$ as
$Y=\frac16 W_{\brn b}-\frac13 W_{\brn d}$, where $W_{\brn
b}=W_{\brn d}=\rm{diag}(1,1,-2)$. However, a trinification model
is in our classification a model with $x=*$, which allows
arbitrary shifts in the choices of $Y$. For any other choice of
$Y$ this implies that a combination of the unitary phases
contributes to $Y$. The canonical choice of $Y$ has no
contribution from $U(3)_{\brn a}$ and hence would correspond to
$x=\frac13$, a non-standard choice. Although the quark and lepton
charges do not fix $x$, this may be done by the zero $Y$-mass
condition.

In table (\ref{tbl:Freq}) three distinct models with this
characteristic appear. The most frequent one, nr. 225, has a fixed
value of $Y$ of the canonical trinification type, with
$x=\frac13$. However, we did not find solutions to the tadpole
conditions for any of these 71328 models. The second one, nr.
1125, has a completely free $Y$; even the zero mass condition for
$Y$ does not fix it.  This type of model occurred 6432 times and
for at least one of these we found a solution to all tadpole
conditions. The third one, nr. 12327, occurred only 24 times, and
for none of them the tadpoles were solved. It has $Y$ fixed to a
value which does not correspond to standard trinification
($x=\frac12$).

The aforementioned tadpole solution occurred for invariant 11 of
tensor product $(1,16,16,16)$, orientifold 0 (with
$(h_{21},h_{11},S)=(9,111,481)$). It has a rather large hidden
sector gauge group $U(3)\times U(3) \times U(3) \times O(4) \times
O(2) \times U(6) \times U(12) \times O(12) \times U(12) \times
O(4)$, with respect to which the spectrum is as follows:
\begin{eqnarray*}
      3 &\times& ( V ,V ,0 ,0 ,0 ,0 ,0 ,0 ,0 ,0 )_3\\
      3 &\times& ( V ,0 ,V ,0 ,0 ,0 ,0 ,0 ,0 ,0 )_{-3}\\
      3 &\times& ( 0 ,V ,V^*,0 ,0 ,0 ,0 ,0 ,0 ,0 )_{-3}\\
      1 &\times& ( 0 ,0 ,0 ,V ,0 ,V ,0 ,0 ,0 ,0 )_{-1}\\
      1 &\times& ( 0 ,0 ,0 ,0 ,0 ,S ,0 ,0 ,0 ,0 )_1\\
      5 &\times& ( 0 ,0 ,0 ,0 ,0 ,0 ,0 ,V ,V ,0 )_1\\
      3 &\times& ( 0 ,0 ,0 ,0 ,0 ,0 ,0 ,0 ,S ,0 )_1\\
      1 &\times& ( 0 ,0 ,0 ,0 ,0 ,A ,0 ,0 ,0 ,0 )_{-1}\\
      2 &\times& ( 0 ,0 ,0 ,0 ,0 ,0 ,0 ,0 ,A ,0 )_{-2}\\
      1 &\times& ( 0 ,0 ,0 ,V ,0 ,0 ,0 ,0 ,V ,0 )_1\\
      1 &\times& ( 0 ,0 ,0 ,0 ,V ,0 ,0 ,0 ,V ,0 )_1\\
      1 &\times& ( 0 ,0 ,0 ,0 ,0 ,V ,0 ,V ,0 ,0 )_1\\
      1 &\times& ( 0 ,0 ,0 ,0 ,0 ,V ,0 ,0 ,V ,0 )_{-1}\\
      1 &\times& ( 0 ,0 ,0 ,0 ,0 ,0 ,V ,V ,0 ,0 )_1\\
      1 &\times& ( 0 ,0 ,0 ,0 ,0 ,0 ,V ,0 ,V ,0 )_{-1}\\
      1 &\times& ( 0 ,0 ,0 ,0 ,0 ,V ,0 ,0 ,0 ,V )_{-1}\\
      1 &\times& ( 0 ,0 ,0 ,V ,V ,0 ,0 ,0 ,0 ,0 )\\
      1 &\times& ( 0 ,0 ,0 ,0 ,S ,0 ,0 ,0 ,0 ,0 )\\
      1 &\times& ( 0 ,0 ,0 ,0 ,0 ,Ad,0 ,0 ,0 ,0 )\\
      1 &\times& ( 0 ,0 ,0 ,0 ,0 ,0 ,Ad,0 ,0 ,0 )\\
      3 &\times& ( 0 ,0 ,0 ,0 ,0 ,0 ,0 ,S ,0 ,0 )\\
      3 &\times& ( 0 ,0 ,0 ,0 ,0 ,0 ,0 ,0 ,Ad,0 )\\
      1 &\times& ( 0 ,0 ,0 ,0 ,0 ,0 ,0 ,0 ,0 ,S )\\
      2 &\times& ( 0 ,0 ,0 ,0 ,V ,V ,0 ,0 ,0 ,0 )\\
      1 &\times& ( 0 ,0 ,0 ,0 ,V ,0 ,0 ,V ,0 ,0 )\\
      2 &\times& ( 0 ,0 ,0 ,0 ,0 ,V ,0 ,0 ,V^*,0 )\\
      2 &\times& ( 0 ,0 ,0 ,0 ,0 ,0 ,V ,0 ,V^*,0 )\\
      1 &\times& ( 0 ,0 ,0 ,0 ,V ,0 ,0 ,0 ,0 ,V )\\
      1 &\times& ( 0 ,0 ,0 ,0 ,0 ,0 ,0 ,V ,0 ,V )
\end{eqnarray*}

Bottom-up trinification models and their phenomenology has been
discussed in \cite{d-trinif}.

\subsection{Curiosities\label{Curios}}

\subsubsection{A non-standard $U(3) \times Sp(2) \times U(1) \times U(1)$ model}

The following spectrum was found for 17, orientifold 2 of the
tensor product $(2,2,2,6,6)$. It has a hidden sector group $U(2)$
which is completely decoupled from all massless matter: both OH as
HH matter is absent. The main reason for listing it here is
however that it is an alternative to the standard lepton-number
conserving configurations. This is nr. 4996 in (\ref{tbl:Freq}).

The full Chan-Paton group is $U(3) \times Sp(2) \times U(1)\times U(1) \times U(2)$,
with the following spectrum
\begin{eqnarray*}
      3  &\times & ( V ,V ,0 ,0 ,0 )_{3} \\
      3  &\times & ( 0 ,0 ,V ,V ,0 )_{-3} \\
      1  &\times & ( V ,0 ,0 ,V^*,0 )_{-1} \\
      2  &\times & ( V ,0 ,V ,0 ,0 )_{-2} \\
      2  &\times & ( 0 ,V ,0 ,V ,0 )_{2} \\
      3  &\times & ( V ,0 ,0 ,V ,0 )_{-1} \\
      3  &\times & ( 0 ,V ,V ,0 ,0 )_{1} \\
      2  &\times & ( V ,0 ,V^*,0 ,0 )_{-2} \\
      1  &\times & ( 0 ,0 ,V ,V^*,0 )_{1} \\
      4  &\times & ( A ,0 ,0 ,0 ,0 )  \\
      2  &\times & ( 0 ,0 ,0 ,S ,0 )
\end{eqnarray*}
The $Y$-embedding is $Y=\frac16 Q_{\brn a} -\frac12 Q_{\brn c}
-\frac12 Q_{\brn d}$. There is no additional massless $U(1)$
factor from the standard model branes (we did not compute the mass
of the abelian factor of $U(2)$). Note that the endpoints of the
quarks and lepton doublet bi-fundamentals are distributed over the
{\brn c} and {\brn d} branes, making it impossible to assign a
lepton number. Indeed, there are perturbatively allowed
lepton-number violating couplings of the type $(Q,L,d^c)$ or
$(L,L,l^c)$, but further CFT computations would be needed to
verify if these couplings do indeed occur. The $G_{\rm CP}$-chiral
spectrum has no Higgs candidates and just one right-handed
neutrino candidate.

We have also found a similar model with $U(2)_{\brn b}$ instead of
$Sp(2)_{\brn b}$, and a slightly more complicated hidden sector.
It combines two features not encountered together in
\cite{Dijkstra:2004cc}: a group $U(3)\times U(2)\times U(1)\times
U(1)$ of which only $U(1)_Y$ survives as an abelian vector boson.
Unfortunately this is achieved at a price that is presumably to
high: the reason is that lepton number cannot be written in terms
of the brane charges. As a result, no linear combination of $B$
and $L$ is anomaly free. Model nr. 1725 is of the same kind, but
with $Sp(2)$ replaced by $U(2)$. A tadpole solution exists for
that model with an $O(2) \times O(2)$ hidden sector.

\subsubsection{A $U(6)$ model}

The following examples were found for invariant 79 of (1,4,4,4,4),
orientifold 0, corresponding to an orientifold with Calabi-Yau
characteristics (6,60,288). These are exact standard model
realizations with just two branes stacks, a complex and a real
one. In fact, this single model can accommodate the standard model
spectrum in three distinct ways. The unified gauge group is
$U(6)\times Sp(2)$. The spectrum is as follows
\begin{eqnarray*}
      9 &\times& ( A ,0 )_3\\
      9 &\times& ( V ,V )_{-3}\\
      8 &\times& ( Ad,0 ) \\
      1 &\times& ( 0 ,A ) \\
      7 &\times& ( 0 ,S )
\end{eqnarray*}

The first standard model realization is obtained by splitting
$U(6)$ so that the full ({\brn a},{\brn b},{\brn c},{\brn d})
configuration becomes $U(3)_{\brn a}\times U(2)_{\brn b} \times
Sp(2)_{\brn c} \times U(1)_{\brn d}$, with {\brn a}, {\brn b} and
{\brn d} belonging to the same stack. The choice of $Y$ is
$\frac16(1,0,0,-3)+W_{\brn c}$, where $W_{\brn c}$ is the diagonal
Pauli matrix $\frac12 \sigma_3$ in $Sp(2)_{\brn c}$. The first
term of $Y$ is part of the non-abelian group $SU(4)$ formed by the
{\brn a} and {\brn d} branes, and hence automatically massless. If
the breaking pattern is interpreted as $U(6) \rightarrow U(5)
\times U(1) \rightarrow U(3)_{\brn a}\times U(2)_{\brn b} \times
U(1)_{\brn c}$ the second step is a flipped $SU(5)$ model; if the
breaking is interpreted as $U(6) \rightarrow U(4) \times U(2)
\rightarrow U(3)_{\brn a}\times U(2)_{\brn b} \times U(1)_{\brn
c}$ the intermediate stage is Pati-Salam-like.

The second realization appears if we split $U(6)$ in the same way
as $U(3)_{\brn a}\times U(2)_{\brn b} \times Sp(2)_{\brn c} \times
U(1)_{\brn d}$, but now with $Y$ is $(-\frac13,\frac12,0,0)$. This
amounts to a standard $SU(5)$ embedding of the standard model. The
$Sp(2)$ group does not contribute to $Y$ in this case.

Finally there is the possibility of using $Sp(2)$ as a {\brn
b}-type stack for the weak interactions. To achieve this we split
$U(6)$ as $U(3)_{\brn a} \times U(3)_{\brn c}$, and write $Y$ as
$\frac16(1,0,-1)+W_{\brn c}$, where $W_{\brn c}$ is the
$SU(3)$-generator $(\frac23,-\frac13,-\frac13)$. There is no {\brn
d}-stack.

All three models have three candidate Higgs pairs an three down
quarks mirror pairs, as well as six right-handed neutrino
candidates, which are chiral with respect to $U(6)$. The first two
are nrs. 1886 and 1887 in table (\ref{tbl:Freq}), and the third
one is nr. 1888.

\section{Phenomenological implications and the problem of masses\label{Pheno}}

In this section we will address, in a rather general fashion, some
phenomenological aspects of SM brane configurations. In
particular, we are going to discuss the problem of masses in
theories with anti-quarks in the antisymmetric representation of
SU(3), as well as the nature of potential family symmetries and
neutrino masses.

\subsection{Antisymmetric anti-quarks  and the problem of quark masses\label{mass}}

There is a generic potential phenomenological problem, when one of
the anti-quarks originate from anti-symmetrized strings starting
and ending on the color branes. Although for SU(3),
$\Yboxdim8pt\yng(1,1)={\overline{\Yboxdim8pt\yng(1)}}$, the
antisymmetric representation has charge 2, under the U(1)$_{\bf
a}$ instead of the -1 for ${\overline{\Yboxdim8pt\yng(1)}}$.

We are using the language of left-handed fermions where
\be \bar \psi^c_R\equiv \psi_L^T~C\sp
C^{-1}\gamma^{\m}C=-(\gamma^{\m})^{T} \ee
where $C$ is the charge conjugation matrix. $\bar \psi^c_R$ is a
right-handed Weyl fermion transforming in the same representation
of the gauge group as $\psi_L$. The mass terms can be therefore be
written in terms of fermion bilinears
\be \bar \psi^c_R~\chi_L+h.c. \ee

Consider  the (color singlet) quark mass operator\footnote{We work
with left-handed spinors only. $\bar Q^c$ is the proper conjugate
of a left-handed spinor.}  $(\bar Q^c)^I_aq^J$, where $Q$ denotes
the quarks (3,2) and $q$ stands for the anti-quarks in the
$\Yboxdim8pt\yng(1,1)$ of SU(3). $I,J$ indices from now on will
collectively indicate any other index except color and weak
indices. $a$ is a weak doublet index. $({\bar Q^c})^Iq^J$
transforms as a weak doublet, and has charge 3 under U(1)$_{\bf
a}$. Therefore it must be coupled to a weak Higgs doublet that
should also carry charge -3. However a single field in
orientifolds cannot carry charge -3. Therefore, a product of
scalar fields must be involved. The minimal case involves scalars
$H^I_a$ transforming as ($\bar 3$,2,-1) under $SU(3)\times
SU(2)\times U(1)_{\bf a}$ and $A^K$ transforming as
($\overline{\Yboxdim8pt\yng(1,1)}$,1,-2). The putative mass term
would then be
\be \delta {\cal L}_1=h_{I,J,K,L}~((\bar Q^c)^I_aQ^J)(H^{Ka}A^L)
\label{hh1}\ee
where the parentheses indicate the color contractions. Non-minimal
couplings would include
\be \delta {\cal L}_2=\tilde h_{I,J,K,L}~((\bar
Q^c)^I_aQ^J)G^a(F^KA^L)\sp \delta {\cal L}_2=\hat
h_{I,J,K,L,M}~((\bar Q^c)^I_aQ^J)G^a(F^KF^LF^M) \label{hh2}\ee
where $G^a$ is a standard Higgs (weak doublet), $F^I$ transforms
as ($\bar 3$,1,-1) and in the last case an antisymmetric color
coupling of three triplets is implied. There might also be
additional constraints, due to the fact that the $F_I$ scalars
come from strings that have one end point in the $c,d$ branes.

The crucial point is that in order to generate the quark mass
terms, the scalar combinations in (\ref{hh1}) and (\ref{hh2}) must
acquire expectation values. This necessarily implies that the
scalars $H^{Ia}$ or $F^I$ or $A^I$ must have vevs, and this
necessarily breaks the color symmetry to SU(2)$_{\rm color}$
(along with $U(1)_{\bf a}$ of course). This seems incompatible
with current data. Moreover, this conclusion is robust, and is
valid independent of the presence or not of
supersymmetry.\footnote{A related fact is that a U(N) D-brane on a
CY manifold is generically expected to have it gauge symmetry
broken to U(N-1) because of the D-terms. The gauge symmetry  may
be enhanced back to U(N) at orbifold points.}

There are two a priori possibilities in order to avoid  the
previous impasse. The first is that non-perturbative effects break
the associated global symmetry. It is well known that anomalous
U(1)'s have always mixed anomalies with non-abelian groups.
Therefore, there are always gauge instantons and their string
theory generalization, that violate the global symmetry
non-perturbatively (see \cite{d-review}). There are two distinct
possibilities, but only one is relevant here: the case when the
non-abelian gauge group is unbroken at low energy \footnote{The
other case concerns a spontaneously broken group. This is
qualitatively distinct since more terms in the effective action
can be generated.}. This is indeed the case with the color group.
In this case only terms involving a minimum number of fermions can
be generated. This minimum number is required in order to soak up
the zero modes of instantons. It is always larger than two in
realistic situations. Therefore, it is not relevant for generating
mass terms for the fermions.

The other option is to start from a higher gauge-group, that is
eventually broken to the color SU(3), giving masses to the
standard quarks. Let us entertain first the case of SU(4). We
should use the following facts, \cite{li}: A scalar in the adjoint
of SU(4) obtaining a vev may break the gauge symmetry $SU(4)\to
SU(2)\times U(2)$ or $SU(3)$ depending on the type of vev. A
scalar in the $\Yboxdim8pt\yng(2)$, breaks the gauge symmetry as
$SU(4)\to O(4)$ or $SU(3)$ depending on the type of vev. A scalar
in the $\Yboxdim8pt\yng(1,1)$ breaks the gauge symmetry as
$U(4)\to Sp(4)$ or $SU(4)\to SU(2)$ depending on the type of vev.
Finally a scalar transforming in the (4,2) of $SU(4)\times SU(2)$
breaks the symmetry as $SU(4)\times SU(2)\to SU(3)$, or
$SU(2)\times SU(2)$ depending on the type of vev. Although this
may be acceptable from the color point of view, the breaking of
the weak SU(2) group is acceptable only if the bi-fundamental
scalar carries the correct SM hypercharge.

Therefore the scalar vevs that preserve an SU(3) color
subgroup SU(4) transformations are\footnote{We use greek indices from the beginning of the alphabet
for color.}
\be
{\rm adjoint} \sim {\Phi_{\a}}^{\beta}\sim \left(\matrix{1&0&0&0\cr
0&0&0&0\cr
0&0&0&0\cr
0&0&0&0\cr}\right)\sim \phi_{\a\b}\sim\Yboxdim8pt\yng(2)
\label{hh3}\ee
\be
\Yboxdim8pt\yng(1)\sim F_{\a}\sim  \left(\matrix{1\cr
0\cr
0\cr
0\cr}\right)\sp (4,2)\sim H^a_{\a}\sim \left(\matrix{1&0&0&0\cr
0&1&0&0\cr}\right)
\label{hh4}\ee
The last operator breaks also to the SU(2). and they must be
aligned. This poses strong constraints on the appropriate scalar
potentials. In particular, no antisymmetric vev is allowed.

We may now go through the potential mass terms and show that none is acceptable.
We suppress all other indices but SU(4) color and write
\be
{\cal O}_1=(\bar Q^c)_{\a}q_{\b\g}F^{\a}F_I^{\b}F_J^{\g}\sp
{\cal O}_2=(\bar Q^c)_{\a}q_{\b\g}\phi^{\a\b}F^{\g}
\label{hh5}\ee
\be {\cal O}_3=\e^{\a\b\g\d}(\bar Q^c)_{\a}q_{\b\g}F_{\d}~
\epsilon_{\a'\b'\g'\d'}F_I^{\a'}F_J^{\b'}F_K^{\g'}F_L^{\d'}
\label{hh6}\ee
where a lower SU(4) index transforms as $\Yboxdim8pt\yng(1)$ and
an upper one as $\overline{\Yboxdim8pt\yng(1)}$. The operators
${\cal O}_i$ moreover transform as weak doublets and have
$U(1)_{\bf a}$ charge zero. There are also operators which involve
adjoint scalars but they have no new features. It is
straightforward to check that operators ${\cal O}_{1,2}$ fail to
provide mass operators for any of the fermions after the breaking
$SU(4)\to SU(3)$ by the vevs in (\ref{hh3}) and (\ref{hh4}).
Operator ${\cal O}_{3}$ gives masses to the standard SU(3) quarks,
but leaves the rest massless. One of the fundamentals in ${\cal
O}_i$ can be substituted with the $H^a_{\a}$ scalar. This will
provide a weak singlet. Moreover as we have seen this vev breaks
$SU(4)\times SU(2)\to SU(3)$, and if the hypercharge of the scalar
is 1/2, then it will provide at the same time the proper,
electroweak symmetry breaking. However, the same considerations as
above indicate than no reasonable mass terms are generated.

The final case to be considered is the possibility to include a
scalar vev in the antisymmetric representation, $R^{\a\b}$. In
this case we must start from SU(5), which the vev will break to
SU(3). Upon choosing a convenient basis this vev is
\be
\Yboxdim8pt\yng(1,1)\sim R^{\a\b}\sim \left(\matrix{\phantom{-}0&1&0&0&0\cr
-1&0&0&0&0\cr
\phantom{-}0&0&0&0&0\cr
\phantom{-}0&0&0&0&0\cr
\phantom{-}0&0&0&0&0\cr}\right)
\ee
We also assume that there are fundamentals $F^{\a}$ with a vev in the 4 and 5 directions,
so that it does not break SU(3) further.
Then we may write the following operators
\be
{\cal O}_4=(\bar Q^c)_{\a}q_{\b\g}F^{\a}R^{\b\g}\sp
{\cal O}_5=\e^{\a\b\g\d\e}(\bar Q^c)_{\a}q_{\b\g}\rho_{\d\e}~
\epsilon_{\a'\b'\g'\d'\e'}F^{\a'}R^{\b'\g'}R^{\d'\e'}
\label{hh7}\ee
The operator ${\cal O}_4$ provides masses for the various singlets
after the breaking. Operator ${\cal O}_5$ provides masses for the
standard quarks. However the two extra triplets emerging from the
$\Yboxdim8pt\yng(1,1)$ of SU(5) will remain massless.

It therefore seems that orientifold models with anti-quarks in
antisymmetric representations are phenomenologically untenable.

\subsection{Masses in SU(5) and flipped SU(5) vacua\label{UFiveMasses}}

The case of standard U(5) group deserves special
attention\footnote{Several of the remarks below were independently
put forward recently in \cite{bere}.}. The SM particles are in the
antisymmetric representation $\psi^{\a\b}$ as well as the
anti-fundamental, $\psi_{\a}$. The minimal set of scalar needed for
symmetry breaking is an adjoint ${\Phi^{\a}}_{\b}$ whose expectation
value $\diag (2V,2V,2V,-3V,-3V,-3V)$ breaks $SU(5)\to SU(3)\times
SU(2)\times U(1)_{Y}$ and a fundamental, $H^{\a}$ whose expectation
value $(0,0,0,0,v)$ breaks $SU(2)\times U(1)_Y\to U(1)_{em}$ The
standard mass terms
\be {\cal O}_1\sim (\bar \psi^c)_{\a}\psi^{\a\b}H_{\b}\sp {\cal
O}_{2}\sim \e_{\a\b\g\d\e}(\bar\psi^c)^{\a\b} \psi^{\g\d}H^{\e}
\ee
give masses to all SM fermions. However here, ${\cal O}_2$ which
gives masses to up-type quarks is not allowed, since it carries
charge +5 under the overall U(1) of the U(5). This charge can be
cancelled by multiplication by
$\epsilon^{\a\b\g\d\e}H^I_{\a}H^J_{\b} H^K_{\g}H^K_{\d}H^L_{\e}$,
which however requires the presence of 5 fundamental Higgs scalars
with vevs that are aligned, and of the order of the electroweak
scale. However, such a mass is suppressed by a factor
$\prod_{I=1}^5{v_I\over M_s}$. Since all $v_I\lesssim M_Z$, we
obtain an unacceptable suppression factor of $10^{-50}$. The other
possibility is the presence of symmetric or antisymmetric scalars
that acquire vevs. An antisymmetric vev cannot preserve the
$SU(3)\times U(1)_{em}$ group of low energy physics.
 A symmetric one, $R^{\a\b}$ is fine provided
it is aligned as in (\ref{hh3}). Its vev $\tilde V$ must be smaller
than the EW vev as it contributes to the W,Z masses. Again, although
${\cal O}_2$ can be neutralized, it gives too small a contribution
to up quark masses. There are new operators we may write now like
\be {\cal O}_3\sim (\bar\psi^c)_{\a\b}\psi_{\g\d}R^{\a\g}R^{\b\d}
\ee
However, such an operator does not contribute to fermion masses.

We can imagine of two non-perturbative loopholes to the previous
arguments. A first non-perturbative possibility is based on
breaking the offending U(1) symmetry by a vacuum condensate. An
example will be a Chan-Paton group that contains U(5)$\times$
SO(5)$\times$ SO(5), and we have extra scalars (denoted
Q$_{\a}^A$) in the representation (V,V,0)+(V$^*$,0,V), so that the
U(5) anomalies cancel. If the dynamics is favorable,  we may
imagine that one of the SO(5)'s creates a composite out of five
scalars, of the form
\be
\epsilon^{\a\b\g\d\e} \epsilon_{ABCDE} Q^A_{\a}Q^B_{\b}\cdots Q^E_{\e},
\ee
where $\a,\b,\cdots$ are SU(5) indices and $A,B,\cdots $ SO(5)
indices. If  the condensate gets a vev at the GUT scale, but not
the individual fields $Q^A_{\a}$, it breaks the U(1) of U(5), and
upon coupling to the U(5) quarks and leptons can generate the
appropriate masses. It should be however mentioned that such a
dynamical setup seems unlikely.

The final possibility, is a non-perturbative breaking of the
overall (anomalous)  U(1) symmetry because of instantons. In the
case of unbroken SU(5) symmetry, the numerous zero modes of the
instantons do not produce mass terms. They could though upon
symmetry breaking. Whether mass terms can be generated in this
case requires a detailed analysis of the instanton induced terms
and will not be attempted here.

\subsection{Family symmetries}

We have allowed extra non-abelian groups to participate in the
local SM collection of branes. In particular standard model
particles are charged under such groups. This setup is very
reminiscent of the idea of family symmetries. The purpose of the
introduction of family symmetry in the past was to
explain/organize the existence of three generations and the
hierarchy of masses of the SM particles. There are two relevant
questions in this context:

(a) Can such symmetries play the role of family symmetries? Can
they help achieve realistic mass matrices for the SM particles?

(b) Are there cases where the presence of such symmetries forbids
realistic mass matrices?

In following we will make some comments on these two questions.
Although our setup is reminiscent of family symmetries, it
incorporates a radical departure from that idea as well. The
reason is that the quark (3,2) states, cannot be charged under any
other gauge symmetry. This is unlike any other family symmetry
introduced in the literature. Since the quarks are necessarily
not-charged under such symmetries, there are non-trivial
considerations concerning the potential mass matrices and the
existence of realistic patterns.

At this stage, we are not fully prepared to calculate three and
higher point couplings in the superpotential. We can however
derive some selection rules on couplings, especially
renormalizable ones (three-point couplings) that are allowed by
the gauge symmetries. Such selection rules can have non-trivial
consequences because

(i) Extra non-abelian symmetries, although broken, may be more or
less constraining, due to the possible symmetry breaking vevs.

(ii) The presence of several (anomalous) U(1)s provides further
constraints, especially if the corresponding global symmetries
remain intact in perturbation theory.

{}From now on we will call for concreteness the non-abelian group $G$
distinct from SU(2) and S(3), the family symmetry group. Let us
consider the case where the anti-quarks $q_i$ transform in a
non-trivial representation $R$ of the group $G$. Then the potential
mass term $(\bar Q^c)^{I,a}q_i$ transform as a doublet of SU(2) and
as R of $G$ (I is a extra index labeling the three quark
generations, while $a$ is a weak doublet index). At the cubic level
the existence of a scalar $\Phi_a^i$ transforming in the (2,R) of
$SU(2)\times G$, gives rise to the Yukawa coupling
\be
(\bar
Q^c)^{I,a}q_i~\Phi_a^i
\ee
Up to base change there are two types of vevs for $\Phi$,
\cite{li}. The first type breaks the symmetry $SU(2)\times G\to
G'$, with $G'=O(N-1)$ if $G=O(N)$ and $G'=SU(N-1)$ if $G=SU(N)$.
Therefore, the electroweak symmetry is broken while the family
symmetry is not fully broken. For this to be realistic, further
vev's should break both the $U(1)_Y$ symmetry and the leftover
family symmetry. The pattern then becomes complicated and deserves
a detailed study. The second type breaks $SU(2)\times G\to
SU(2)\times G'$ with $G'=O(N-2)$ if $G=O(N)$ and $G'=SU(N-2)$ if
$G=SU(N)$. Here the family symmetry is completely broken if $N=2$.
This is the case for example of a Pati-Salam group. If there is a
leftover family symmetry, further symmetry breaking is necessary.
The $\Phi$ vev identifies the weak and the G index and provides a
mass matrix for quarks that is degenerate. The existence of
several copies of $\Phi$ does not improve the situation.

We can contemplate higher dimension terms involving a weak double $H^a$ and a scalar $\Phi_i$ in the fundamental of $G$
\be (\bar Q^c)^{I,a}q_i~H^a\Phi^i \ee
In such a case a vev of $\Phi$ of the order of $M_{s}$ will give a
mass matrix of order of the electroweak scale but it will be
degenerate. Moreover the $G$ symmetry is partly broken. Several
scalars $H^I_i$ with couplings
\be {g_{IJ}\over M_s}~(\bar Q^c)^{I,a}q_i~H^a\Phi^i_J \ee
could fare better. First non-aligned expectation values can break
a larger portion or all of the G group. Second, for generic
couplings $g_{IJ}$ the mass matrix after electroweak breaking will
be non-degenerate. Therefore, in this case, a reasonable non-zero
mass matrix is viable.

There are more complicated possibilities of the occurrence of
 quasi-family symmetries and the charge assignments of SM particles under them.
We have studied in some indicative examples, the relevant issues
present. A full study of all possibilities is a major task and it
will not be undertaken here.

\subsection{Neutrino masses\label{nm}}

In our search we have not explicitly constrained the presence of
anti-neutrinos. A priori, any SM singlet fermion can play that
role. Of course, for a realistic pattern of masses to emerge,
important constraints on the interactions are appropriate.

There are two mechanisms that so far have been successful in
producing neutrino masses of acceptable magnitude. The first,
relies on the see-saw mechanism and is appropriate for vacua with
high values of the string scale. An important ingredient for its
operations is that lepton number is not conserved. Moreover at
least two (and typically three) antineutrinos are necessary for
accommodating present data. As we have discussed earlier, the
presence of lepton number cannot be directly tracked until a
formal separation of doublets into leptons and Higgses is
possible. Therefore in this context, the question of neutrino
masses remains a question to be addressed in concrete string
ground states.

The second mechanism involves a brane wrapping one (or several)
large dimensions and is necessarily operative in string vacua with
a low string scale. In this context the neutrinos mix with
antineutrinos emerging from the ``bulk" brane , and the masses are
suppressed by the volume of large dimensions. For this mechanism
to succeed large Majorana masses should be forbidden. Therefore it
is important that lepton number is a good symmetry. Moreover, the
minimal implementation involves a single antineutrino and its KK
tower of states and leads to predictions marginally compatible
with current data \cite{dsm3}. More comfortable constructions
involve at least two antineutrinos.

\section{Dependence of the results on the Calabi-Yau topology\label{CY}}

Table \ref{tableCY} lists the MIPFs for which the standard model
spectrum was found, and how often it occurred. The table is
ordered according to standard model frequency, that is the total
number of standard model configurations divided by the total
number of three and four brane configurations. Note that this does
not take into account tadpole cancellation, since we have not
systematically solved the tadpole conditions for all standard
model configurations. Column 2 gives the MIPF id-number using the
same sequential labelling used in \cite{Dijkstra:2004cc}. We can
provide further details on these MIPFs on request. To help
identifying them, we list in columns 3,4 and 5 the resulting
heterotic Calabi-Yau spectrum (Hodge numbers and the number of
$E_6$ singlets). In columns 6,7 and 8 we list the total number of
configurations for each value of $x$. The last column gives the
frequency.
%
%
\LTcapwidth=14truecm
\begin{center}
\begin{longtable}{|l|l|l|l|r|r|r|r|r|}\caption{\em Standard model success rate for various MIPFs.}
\label{tableCY}\\
 \hline \multicolumn{1}{|l|}{Tensor product}
& \multicolumn{1}{l|}{MIPF}
& \multicolumn{1}{l|}{$h_{11}$}
& \multicolumn{1}{l|}{$h_{12}$}
& \multicolumn{1}{r|}{Scalars}
& \multicolumn{1}{r|}{$x=0$}
& \multicolumn{1}{c|}{$x=\frac12$}
& \multicolumn{1}{c|}{$x=*$}
& \multicolumn{1}{c|}{Success rate} \\ \hline
\endfirsthead
\multicolumn{9}{c}%
{{\bfseries \tablename\ \thetable{} {\rm-- continued from previous page}}} \\
 \hline \multicolumn{1}{|l|}{Tensor product}
& \multicolumn{1}{l|}{MIPF}
& \multicolumn{1}{l|}{$h_{11}$}
& \multicolumn{1}{l|}{$h_{12}$}
& \multicolumn{1}{r|}{Scalars}
& \multicolumn{1}{r|}{$x=0$}
& \multicolumn{1}{c|}{$x=\frac12$}
& \multicolumn{1}{c|}{$x=*$}
& \multicolumn{1}{c|}{Success rate} \\ \hline
\endhead
\hline \multicolumn{9}{|r|}{{Continued on next page}} \\ \hline
\endfoot
\hline \hline
\endlastfoot
  (1,1,1,1,7,16) &   30 &   11 &   35 &  207 &    1698 &     388 &       0 & $2.1 \times 10^{-3}$ \\
  (1,1,1,1,7,16) &   31 &    5 &   29 &  207 &     890 &     451 &       0 & $1.35 \times 10^{-3}$ \\
    (1,4,4,4,4) &   53 &   20 &   20 &  150 & 2386746 &  250776 &       0 & $4.27\times 10^{-4}$ \\
    (1,4,4,4,4) &   54 &    3 &   51 &  213 &    5400 &    5328 &    4248 & $3.92\times 10^{-4}$ \\
      (6,6,6,6) &   37 &    3 &   59 &  223 &       0 &  946432 &       0 & $2.79\times 10^{-4}$ \\
  (1,1,1,1,10,10) &   50 &   12 &   24 &  183 &    1504 &     508 &      36 & $2.63\times 10^{-4}$ \\
  (1,1,1,1,10,10) &   56 &    4 &   40 &  219 &     244 &      82 &       0 & $2.01\times 10^{-4}$ \\
   (1,1,1,1,8,13) &    5 &   20 &   20 &  140 &     328 &      27 &       0 & $1.93\times 10^{-4}$ \\
  (1,1,1,1,7,16) &   26 &   20 &   20 &  140 &     157 &      14 &       0 & $1.72\times 10^{-4}$ \\
     (1,1,7,7,7) &    9 &    7 &   55 &  276 &    7163 &     860 &       0 & $1.59\times 10^{-4}$ \\
  (1,1,1,1,7,16) &   32 &   23 &   23 &  217 &     135 &      20 &       0 & $1.56\times 10^{-4}$ \\
    (1,4,4,4,4) &   52 &    3 &   51 &  253 &  110493 &    8303 &       0 & $1.02\times 10^{-4}$ \\
    (1,4,4,4,4) &   13 &    3 &   51 &  250 &  238464 &  168156 &       0 & $1.01\times 10^{-4}$ \\
   (1,1,1,2,4,10) &   44 &   12 &   24 &  225 &     704 &     248 &       0 & $1.01\times 10^{-4}$ \\
  (1,1,1,1,1,2,10) &   21 &   20 &   20 &  142 &       2 &       1 &       0 & $1.00\times 10^{-4}$ \\
   (1,1,1,1,1,4,4) &  124 &    0 &    0 &   78 &     729 &       0 &       0 & $9.8 \times 10^{-5} $ \\
    (4,4,10,10) &   79 &    7 &   43 &  215 &       0 &   57924 &       0 & $9.39 \times 10^{-5} $ \\
    (4,4,10,10) &   77 &    5 &   53 &  232 &       0 & 1068926 &       0 & $8.29 \times 10^{-5} $ \\
    (1,4,4,4,4) &   77 &    3 &   63 &  248 &       0 &    1024 &       0 & $8.12 \times 10^{-5} $ \\
    (4,4,10,10) &   74 &    9 &   57 &  249 &       0 & 1480812 &       0 & $8.06 \times 10^{-5} $ \\
  (1,1,1,1,1,2,10) &   24 &   20 &   20 &  142 &       0 &       0 &       6 & $7.87 \times 10^{-5} $ \\
    (1,2,4,4,10) &   67 &   11 &   35 &  213 &       0 &   14088 &    1008 & $7 \times 10^{-5} $ \\
   (1,1,1,1,5,40) &    5 &   20 &   20 &  140 &     303 &      36 &       0 & $6.73 \times 10^{-5} $ \\
     (2,8,8,18) &    8 &   13 &   49 &  249 &       0 & 1506776 &       0 & $6.03 \times 10^{-5} $ \\
     (1,1,7,7,7) &    7 &   22 &   34 &  256 &    2700 &      68 &       0 & $5.5 \times 10^{-5} $ \\
    (1,4,4,4,4) &   78 &   15 &   15 &  186 &   20270 &    6792 &       0 & $5.39 \times 10^{-5} $ \\
     (2,8,8,18) &   28 &   13 &   49 &  249 &       0 &  670276 &       0 & $5.25 \times 10^{-5} $ \\
    (1,2,4,4,10) &   75 &    5 &   41 &  212 &     304 &     580 &     244 & $4.87 \times 10^{-5} $ \\
     (1,1,7,7,7) &   17 &   10 &   46 &  220 &    1662 &     624 &     108 & $4.76 \times 10^{-5} $ \\
     (2,2,2,6,6) &  106 &    3 &   51 &  235 &       0 &  201728 &       0 & $4.74 \times 10^{-5} $ \\
   (1,1,1,16,22) &    7 &   20 &   20 &  140 &     244 &      19 &       0 & $4.67 \times 10^{-5} $ \\
    (1,2,4,4,10) &   65 &    6 &   30 &  196 &       0 &    1386 &       0 & $4.41 \times 10^{-5} $ \\
    (4,4,10,10) &   66 &    6 &   48 &  223 &       0 &   61568 &       0 & $4.33 \times 10^{-5} $ \\
    (1,4,4,4,4) &   57 &    4 &   40 &  252 &       0 &  266328 &   58320 & $4.19 \times 10^{-5} $ \\
    (1,4,4,4,4) &   80 &    7 &   37 &  200 &       0 &    1968 &    1408 & $4.15 \times 10^{-5} $ \\
      (6,6,6,6) &   58 &    3 &   43 &  207 &       0 &  190464 &       0 & $3.93 \times 10^{-5} $ \\
  (1,1,1,1,10,10) &   36 &   20 &   20 &  140 &     266 &      26 &       6 & $3.82 \times 10^{-5} $ \\
    (1,1,1,4,4,4) &  125 &   12 &   24 &  214 &     351 &       0 &       0 & $3.62 \times 10^{-5} $ \\
    (4,4,10,10) &   14 &    4 &   46 &  219 &       0 &  114702 &       0 & $3.3 \times 10^{-5} $ \\
  (1,1,1,1,10,10) &   33 &   20 &   20 &  140 &      47 &       5 &       0 & $3.21 \times 10^{-5} $ \\
   $\ldots$     &      &     &   &    &         &       &         & $\ldots$ \\
    (3,3,3,3,3) &    6 &   21 &   17 &  234 &       0 &     192 &       0 & $6.54\times 10^{-6}$\\
   $\ldots$     &      &     &   &    &         &       &         & $\ldots$ \\
     (3,3,3,3,3) &    4 &    5 &   49 &  258 &       0 &      24 &       0 & $8.17\times 10^{-7}$\\
   $\ldots$     &      &     &   &    &         &       &         & $\ldots$ \\
     (3,3,3,3,3) &    2 &   49 &    5 &  258 &       6 &      27 &       6 & $1.65\times 10^{-9}$ \\
   $\ldots$     &      &     &   &    &         &       &         & $\ldots$ \end{longtable}
\end{center}

The complete table has 1639 cases with non-zero frequency.
Therefore we only present the top of the table here, which starts
with a frequency as high as $.2\%$. The last three entries are
modular invariants of the tensor $(3,3,3,3,3)$, corresponding to
the quintic. They occur much further down the list, but are shown
here because the quintic is a well-studied Calabi-Yau manifold.
The lowest non-zero frequency we encountered is $3.5 \times
10^{-12}$ (for a total of 4 configurations found).

In column 2 an asterisk indicates that at least one tadpole
solution was found for that MIPF in \cite{Dijkstra:2004cc}. Note
that we did not perform an exhaustive  search for tadpole
solutions in the present work. Indeed, if all brane configurations
occurring for a given MIPF are of a type for which the tadpoles
have already been solved before (for a different MIPF), no further
attempts are made to solve them. Therefore we cannot make
definitive conclusions about the non-existence of tadpole
solutions for a given MIPF from our present results.

Note the presence of models with Hodge numbers $(20,20)$. The
corresponding Calabi-Yau manifolds are in fact of the form
$K_3\times T_2$. There is also a case with $h_{11}=h_{12}=0$,
which is in fact a torus compactification. The fact that these are
(partly) torus compactifications is not in contradiction with the
fact that the spectrum is chiral. Each MIPF can be thought of as a
an extension of the chiral algebra of the original tensor product,
modified by an automorphism. This extension may lead to a
non-chiral torus compactification. However the boundary states
that are admitted are a complete set with respect to the original
unextended chiral algebra, which always corresponds to a chiral
compactification (except for five non-chiral tensor products that
we do not consider). Hence a non-chiral bulk extension may have
chiral boundary states. It is possible that the $K_3\times T_2$
models are related to models discussed in
(\cite{Aldazabal:2006nz}); this will require further
investigation. In any case we did not find tadpole solutions for
any of these torus or $K_3\times T_2$ models (but again with the
caveat that we did not search for them exhaustively).

We did find tadpole solutions for one of the MIPFs of the quintic,
namely MIPF nr. 6. These solutions are the broken and unbroken
Pati-Salam $U(4)\times U(2)\times U(2)$ models discussed above.

\newpage
\vskip 2cm
\centerline{\bf Acknowledgements}
\addcontentsline{toc}{section}{Acknowledgements}
\vskip 2cm

We would like to thank I. Antoniadis, M. Bianchi,  R. Blumenhagen, M. Douglas,
 F. Gmeiner, G. Honecker, D. L\"ust, and H. Verlinde for discussions.
This work was partially supported by ANR grant NT05-1-41861, INTAS
grant, 03-51-6346, RTN contracts MRTN-CT-2004-005104 and
MRTN-CT-2004-503369, CNRS PICS 2530 and 3059 and by a European
Excellence Grant, MEXT-CT-2003-509661. P. A. was supported by the
research program ``Pythagoras II'' (grant 70-03-7992) of the Greek
Ministry of National Education, partially funded by the European
Union. The work of A.N.S. has been performed as part of the
program FP 52 of the Foundation for Fundamental Research of Matter
(FOM), and the work of T.P.T.D. and A.N.S. has been performed as
part of the program FP 57 of FOM. The work A.N.S. has been
partially supported by funding of the Spanish ``Ministerio de
Ciencia y Tecnolog\'\i a", Project BFM2002-03610.

\newpage

\appendix

\vskip 10mm
 \renewcommand{\theequation}{\thesection.\arabic{equation}}
\centerline{\Large\bf Appendices}
\addcontentsline{toc}{section}{Appendices}

\section{The unbiased search algorithm\label{Algo}}

We introduce the following notation (in the following $a,b,\ldots$
are generic boundary state labels, not to be confused with the
specific labels {\brn a}, {\brn b} for the QCD, weak and other
standard model branes)
$$ N_{ab} =\sum_i A^i_{~ab} \chi_i (m = 0, L) $$
where $A^i_{~ab}$ are the unoriented annulus coefficients and $
\chi_i (m = 0, L)$ is the character of representation $i$,
restricted to massless, left-handed fermions. The boundary
conjugates of $a$ and $b$ are denoted by $a^c$ and $b^c$ . If we
consider two complex boundaries $a$ and $b$, there is $a$ total of
four quantities relevant for the massless spectrum, namely
$N_{ab}$ , $N_{ab^c}$, $N_{a^cb}$ and $N_{a^cb^c}$ . The chiral
information is contained in two quantities, namely
$$\Gamma_{ab} = N_{ab}-N_{a^cb^c} $$
and
$$\Delta_{ab} = N_{ab^c}-N_{a^cb} $$

If either $a$ or $b$ are real, we set $\Delta  = 0$. If both $a$
and $b$ are real $\Gamma  =  \Delta  = 0$. Furthermore we define
the chiral numbers of anti-symmetric and symmetric tensors
$$ A_a=\frac12(N_{aa}-N_{aa^c} -M_a +M_{a^c}) $$
and
$$ S_a=\frac12(N_{aa}-N_{aa^c} +M_a -M_{a^c}) $$
where $M$ is the Moebius contribution
$$ M_a=\sum_i  M^i_{~a} \chi_i (m = 0, L) \ . $$

\noindent Our search procedure is  as follows
\begin{enumerate}
\item{Consider all orientifold choices that have non-zero tension.
We will label them by an integer $\ell$. This sequential label
corresponds to some choice of the discrete parameters of the RCFT,
called ``Klein bottle currents" and ``crosscap signs"
\cite{Fuchs:2000cm}. The sign of the tension of the corresponding
O-plane is a free parameter in RCFT constructions, and we choose
it negative. We denote its value as $T_O ^{\ell}$.}

\item{
For each $\ell$, consider all candidates for brane {\brn a} subject to the conditions
\begin{enumerate}
\item{Brane {\bf a} is complex.}
\item{The brane tension $T_{\brn a}$ satisfies $6T_{\brn a} + T^{\ell}_O < 0$, because the complete
configuration must satisfy the dilaton tadpole condition $\sum_x T_x + T^{\ell}_O=0$, and
all $T_x$ are positive. This is needed in order to accommodate further branes. }
\item{There are no chiral symmetric tensors.}
\end{enumerate}
}
\item{For each $\ell$ and {\brn a}, consider all candidates for brane {\brn b} that satisfy the following conditions
\begin{enumerate}
\item{The CP group associated with {\brn b} is not orthogonal.}
\item{The brane tension $T_{\brn b}$ satisfies $6T_{\brn a} + 2T_{\brn b} + T^{\ell}_O < 0$, if {\brn b} is real,
$6T_{\brn a} + 4T_{\brn b} + T^{\ell}_O < 0$ if {\brn b} is
complex.}
\item{There are three chiral bi-fundamentals $(3, 2)$. These are only counted chirally, i.e additional mirror pairs are allowed.
 If brane {\brn b} is complex, the chiral total of $(3,2)$ and $(3,2^*)$ must be three.}
\item{There are no chiral symmetric tensors.
This is the an application of the condition mentioned in section \ref{WWALF}, that $R^{\rm chir}_{CP}$ should not yield anything
more exotic than mirrors. It is not
absolutely essential here, but it gives a useful early limitation on the number of solutions.}
\end{enumerate}
}
\item{ For each $\ell$, {\brn a} and {\brn b} consider all candidates {\brn c} that satisfy:
\begin{enumerate}
\item{Brane {\brn c} is allowed at least once by the dilaton tension constraint.}
\item{We need weak singlet anti-quarks. They can come from anti-symmetric tensors of brane {\brn a}
or from bi-fundamentals between brane {\brn a}
and either branes {\brn c} or {\brn d}. Since the anti-symmetric tensors can have only one charge, at least
three anti-quarks must come from bi-fundamentals.
There is no {\it a priori} ordering between branes {\brn c} and {\brn d}.
To prevent double-counting,
we will impose the condition that brane {\brn c} must provide more anti-quarks plus mirrors than brane
{\brn d}. More precisely, we will impose the condition
$N_{{\brn c}}(| \Gamma_{{\brn a}{\brn c}} | + | \Delta_{{\brn a}{\brn c}} |)
\geq N_{{\brn d}} (| \Gamma_{{\brn a}{\brn d}} | + | \Delta_{{\brn a}{\brn d}} |)$.
This ordering condition can only be imposed once we have determined branes {\brn c} and {\brn d} as well as
their CP multiplicities $N_{{\brn c}}$ and $N_{{\brn d}}$, but at this
stage it already implies that
 ($| \Gamma_{{\brn a}{\brn c}} | + | \Delta_{{\brn a}{\brn c}} |) >0 \ .$}
\end{enumerate}
}
\item{
Given $\ell$ {\brn a}, {\brn b} and {\brn c} there may be a value for $N_{\brn c}$ (the CP multiplicity of brane {\brn c}) and a hypercharge choice so that the standard model is already obtained for just three stacks. However, in general a fourth stack is needed (even if there is a
valid three-stack solution we will continue looking for a fourth one). Hence we consider
all labels  {\brn d} that satisfy:
\begin{enumerate}
\item{At least one of the stacks {\brn b}, {\brn c} and {\brn d} is complex. Otherwise it would be impossible to
obtain chiral leptons.}
\item{At least one of the quantities $S_{\brn d}$ , $A_{\brn d}$ , $\Gamma_{{\brn a}{\brn d}}$, $\Gamma_{{\brn b}{\brn d}}$ ,
$\Gamma_{{\brn c}{\brn d}}$ , $\Delta_{{\brn a}{\brn d}}$, $\Delta_{{\brn b}{\brn d}}$ , and $\Delta_{{\brn c}{\brn d}}$ is non-zero. Otherwise
brane {\brn d} can be regarded as part of the hidden sector.}
\end{enumerate}
}
\item{
Now we have collected an orientifold and four branes {\brn a}, {\brn b}, {\brn c}, {\brn d} and we have to determine the
CP multiplicities of the last two branes. Because, by assumption, any further branes are in the
hidden sector and cannot contribute chiral states to the four CP groups, all cubic anomalies
must now cancel. This gives at least two and at most four equations for the two quantities
$N_{\brn c}$ and
$N_{\brn d}$ . The following things can happen:
\begin{enumerate}
\item{There are two independent equations that fix $N_{\brn c}$ and $N_{\brn d}$. Both are positive integers, and
are even for symplectic groups. Now we can move on to the next stage, and compute the
Y-charge combination (see below).}
\item{The equations are inconsistent, do not have positive integer solutions, or have a solution
with an odd CP multiplicity for a symplectic group. In all these cases the configuration
$(\ell,{\brn a},{\brn b},{\brn c},{\brn d})$ must be rejected.}
\item{There is only one independent equation. This means that only a linear combination
$f_{\brn c} N_{\brn c}+f_{\brn d} N_{\brn d}$
is fixed. If this happens we consider all values of $N_{\brn d}$ or $N_{\brn c}$ (if $f_{\brn c}=0$)
between 1 and
the maximum allowed by the dilaton tadpoles, and attempt the next stage (computing
$Y$) for all of them.}
\item{There is no equation at all. This means that all anomalies cancel independent of $N_{\brn c}$ and
$N_{\brn d}$. This can only happen if $A_{\brn a} = 6$. If $A_{\brn a} \not= 6$, there must be chiral bi-fundamentals
giving rise to anti-quarks, and their contribution to the $SU(3)$ anomaly depends on $N_{\brn c}$ or $N_{\brn d}$, or both.
In this case we consider all allowed values of $N_{\brn c}$ as well as
$N_{\brn d}$ and attempt to determine $Y$.}
\end{enumerate}
 }
\item{The next step is to compute the standard  model $Y$-charge. In general
it is a linear combination of the form $Y=\sum_{\alpha}  t_{\alpha} Q_{\alpha} + W_{\brn c} + W_{\brn d}$,
where  $Q_{\alpha}$ is the $U(1)$ charge of one of the unitary brane stacks,
with $\alpha=({\brn a},{\brn b},{\brn c},{\brn d})$.
Real stacks have $Q_{\alpha}=0$. The  last two terms
are simple Lie-algebra generators in the CP-factors of branes {\brn c} and {\brn d}, in
other words generators of $SU(N)$, $O(N)$ or $Sp(N)$. They can be brought
to diagonal form and may therefore be parametrized as traceless diagonal
matrices, which in the case of $O(N)$ and $Sp(N)$ must have equal numbers of
 eigenvalues of opposite sign. We first determine the coefficients $t_{\alpha}$.
We do this by solving one of the following sets of equations:
\begin{itemize}
\item{Trace Equations: These are obtained by taking the trace for each of the $SU(3)\times SU(2)$
representations $(3,2)$, $(3^*,1)$, $(2,1)$ and $(1,1)$. On the phenomenological
side, any non-chiral mirror pairs do not contribute to these traces, and
on the string theory side $W_{\brn c}$ and $W_{\brn d}$ do not contribute. Therefore
this gives four equations for at most four variables  $t_{\alpha}$.}
\item{Axion Equations: Require absence of
axion-Y bilinear couplings. This gives a condition for every axion, and yields
in general far more conditions than there are variables.
Note that  $W_{\brn c}$ and  $W_{\brn d}$
do not couple to any axions.
Since  we want
rational  solutions $t_{\alpha}$ and since the axion couplings are
real  numbers, the solutions have to be converted to rational numbers of the
form $p/q$. We perform that conversion assuming $|q| \leq 1024$.}
\item{Exact Charge Equations: Write  down equations
for the actual charges (rather than the traces) for each non-zero coefficient
$A, S, \Gamma$ or $\Delta$. We write these equations for the
maximal eigenvalue in the {\brn c} and {\brn d} sectors, {\it i.e.} for the maximal
eigenvalue of $x_{\brn c}=t_c Q_{\brn c} + W_{\brn c}$ or $x_{\brn d}=t_d Q_{\brn d} + W_{\brn d}$. The right hand side
of such an equation must be a valid (mirror) quark or lepton charge, and
is determined up to at most an integer $0, \pm 1$. These linear equations can be solved,
and limit $x_{\brn c}$ and $x_{\brn d}$ to a definite range of integers and half-integers.
To determine $t_{\brn c}$ (and analogously $t_{\brn d}$) we consider all possible multiplicities
of the eigenvalue $W_{\brn c}^{\rm max}$, between $1$ and $N_{\brn c}$. Given this
multiplicity, and the fact that $W_{\brn c}$ is traceless, we can determine $t_{\brn c}$.
Taking into account all these possibilities (the integer ambiguities and the
number of maximal  eigenvalues) then gives a set of possible variables
for $t_c$ and $t_d$.}
\end{itemize}
These methods are used successively as needed.
The first is the simplest and usually sufficient, and only in rare cases the third method
is needed.
The $Y$-mass constraint is in any case
checked as a condition, if it was not used as equation. Note that the exact charge equations
cannot fix all $t_{\alpha}$ if the brane configuration is orientable. In that case these equations
have a one-dimensional kernel, and only the axion equations might fully determine $t_{\alpha}$.
To summarize, we have
following possibilities:
\begin{enumerate}
\item{The trace equations completely fix all $t_{\alpha}$.
In that  case the axion-$Y$  bilinear
couplings are computed for this  particular $Y$. If they all vanish, we move on
to the next step. }
\item{The equations do not fix all $t_{\alpha}$. In that case we combine the trace equations
with the axion equations. }
\item{The trace and axion equations still do not all fix $t_{\alpha}$.
In that case we use the exact charge equations to determine the missing
coefficient(s) up to a finite set of rational numbers. }
\item{The trace plus axion equations do not determine all $t_{\alpha}$ completely, and neither do the
exact charge equations.
In this case both sets of equations have a non-trivial kernel
and there are two possibilities:
\begin{enumerate}
\item{The kernel vector of the exact charge equations
is in the kernel
of the trace and axion equations. This means that we can add a set
of coefficients $x_{\alpha}$ to $t_{\alpha}$ without affecting the quark and
lepton charges, nor the axion couplings. This is a genuine ambiguity, which
cannot be resolved by any conditions at our disposal. We fix this
ambiguity by setting one of the missing coefficients  to  a chosen
``canonical" value ($\frac16$, 0,$ -\frac12$, $ -\frac12$ for $t_a,\ldots t_c$
respectively).}
\item{The kernel vector of the exact charge equations
is not in the kernel
of the trace and axion equations. In this case the equations can be solved
by combining them. There is a minor complication due to the fact that the
exact charge equations have a range of rational numbers as their right hand side.
To deal with this we consider a set of rational
values $p/q$ for the missing $t_{\alpha}$.  For $q$ we use the  smallest
common multiple of $24, N_{\brn c}$ and $N_{\brn d}$, and we allow all values for $p$ so
that $-1 \leq p/q \leq 1$. Since the kernels of the two set of equations
do not overlap, there  will be at most a few solutions.}
\end{enumerate}
}
\end{enumerate}
}
All possibilities described above do actually occur.
\item{Determining $W_{\brn c}$ and $W_{\brn d}$, given $t_{\alpha}$. This is  now easy, because
the eigenvalues of these generators must lower or raise  the value of $Y$  to
an allowed quark  or lepton charge. Hence at most two distinct
eigenvalues are allowed. Since the generators must be traceless, this
fixes them completely. If the {\brn c} or {\brn d} groups are orthogonal or symplectic,
the two eigenvalues must be equal in number and opposite.
 Note that for $SU(3)\times SU(2)$ singlets we allow
three charges,  $0,\pm 1$, but if there is  an equal  number of charges
$+1$ and $-1$ this just adds non-chiral pairs. This is a degeneracy, that can
be fixed by setting all paired charges to $0$. Hence also in this case at most
two distinct $W_{\brn c}$ or $W_{\brn d}$ eigenvalues are needed. }
\item{Finally we count the  quarks and leptons, to check that the correct
particle multiplicities are obtained.}
\end{enumerate}

There is some potential over-counting in the procedure, due to the following
reasons
\begin{enumerate}
\item{If the {\brn b}-brane is complex, one can interchange {\brn b} and ${\brn b}^c$}
\item{The choice of {\brn c} and {\brn d} is interchangeable.}
\item{The choice of {\brn c} and ${\brn c}^c$  is interchangeable.}
\item{The choice of {\brn d} and ${\brn d}^c$  is interchangeable.}
\end{enumerate}
These degeneracies are fixed as follows. The first one can be dealt with
by requiring that there are more chiral representations  $(3,2)$ than $(3,2^*)$.
Since their total must be three, they cannot be equal. The second one can
be fixed by requiring that  brane {\brn c} produce a larger total number of anti-quarks than brane {\brn d},
{\it i.e.}\
$N_{{\brn c}}(| \Gamma_{{\brn a}{\brn c}} | + | \Delta_{{\brn a}{\brn c}} |)
\geq N_{{\brn d}} (| \Gamma_{{\brn a}{\brn d}} | + | \Delta_{{\brn a}{\brn d}} |)$.
If that still yields equality, we require that brane {\brn c} produce more chiral
anti-quarks than brane {\brn d}. A few further constraints of this type may be used to
fix the ambiguity completely. To fix  the conjugation
ambiguities of the {\brn c} and {\brn d} branes we require that certain chiral
quantities associated with these branes are positive.

\section{Gauge coupling ranges of various hypercharge embeddings\label{gg}}

The range of possible variation of the string scale in orientifold
vacua is a very interesting question. Its extreme values, close to
the Planck scale on th high side and in the TeV range for the low
side, both have phenomenological merits and problems. It is the
purpose of this appendix to give a rough idea on the range of
string scale values allowed in various hypercharge embeddings
described in this paper. Although, branes $\bf c$ and $\bf d$ may
be non-abelian, we will assume here for simplicity that they carry
U(1) factors. Moreover we will allow their associated gauge
couplings to vary in the perturbative regime between zero and one.

Using the string prediction for the hypercharge embedding, we can
evaluate the hypercharge coupling at the string scale. In the
standard normalization of the non-abelian couplings the U(1)
coupling normalization is  $g_i^2/2i$ where $i$ is the number of
``colors":
\bea {1\over g_Y^2}={6x_{\brn a}^2\over g_{\brn a}^2}+{4x_{\brn
b}^2\over g_{\brn b}^2} +{2x_{\brn c}^2\over g_{\brn c}^2}
+{2x_{\brn d}^2\over g_{\brn d}^2} \label{OnStringScale}\eea
%
%
%
where $x_i$ are the coefficients in (\ref{YLinComb}).
The couplings of the $U(1)_{\brn a}$ and $U(1)_{\brn b}$ are
directly related to the strong and weak coupling constants. We
take the  $U(1)_{\brn c}$, $U(1)_{\brn d}$, couplings to be a
priori independent.

The one-loop coupling evolution is given by:
\bea {1\over \a_i(m)}={1\over \a_i(M_Z)} -{b_i\over 2\pi}
\ln{m\over M_Z}\label{Running}\eea
where $\a_i=g^2_i/4\pi$ and:
\bea \a_3(M_Z)=0.1172 (\pm 0.003)~~&,&~~~~~~ \a_{em}(M_Z)=1/127.934 \nn\\
M_Z=0.0911876~ TeV ~~&,&~~~~~~ \sin^2\theta_Z=0.23113
\label{Values}\eea
where $\sin^2\theta_Z=\a_{em}(M_Z)/\a_2(M_Z)$.


Inserting (\ref{Running}) in (\ref{OnStringScale}), we have
\bea
\ln {M_Z \over M_s}={-{1\over \a_Y(M_Z)}+{6x_{\brn a}^2\over
\a_{\brn a}(M_Z)} +{2x_{\brn b}^2\over \a_{\brn
b}(M_Z)}+\sum_{i}{2x_{\brn i}^2\over \a_{\brn i}(M_s)}\over
-{b_Y\over 2\pi}+{6x_{\brn a}^2b_{\brn a}\over 2\pi}
+{2x_{\brn b}^2b_{\brn b}\over 2\pi}}
\label{MaxMstring}\eea
The equation above expresses the string scale as a function of the
three SM couplings $\a_3$, $\a_2$, $\a_Y$ evaluated at the
$Z$-mass, the values of the $U(1)_{c,d}$ couplings at the string
scale, the one-loop $\beta$-function coefficients and the
coefficients of the hypercharge embedding $x_i$. We use the
one-loop $\beta$-functions for non-supersymmetric and
supersymmetric SM to be:
\bea(b_3,b_2,b_Y)_{SM}=(-7,-3,7)~~~,~~~~~~~
(b_3,b_2,b_Y)_{MSSM}=(-3,1,11)~.\eea
Moreover we put a uniform threshold at the supersymmetric case around 1 TeV.

By varying the U(1) couplings $\a_{i={\bf c,d}}$ between zero and
one we obtain a range of allowed values for $M_s$. In particular,
by choosing small values for $\a_{i={\bf c,d}}$ arbitrarily small
values for the string scale are obtained. The maximum value for
the string scale occurs for values of the couplings at the
boundary of the strong coupling region that we take by convention
to be $\a_{{\bf c}}=\a_{{\bf d}}=1$.

The maximum value of the string scale $M_s$ is obtained for
strongly coupled $U(1)$ branes. It is interesting to evaluate this
maximum value for all models, since it provides an upper bound for
the string scale and makes models with the maximum $M_s$ in the
few TeV range particularly attractive.

The indicative maximum value of the string scale is tabulated
below for the various choices of hypercharge embeddings. In column
1 we list the value of $x$; One of the $x=\frac12$ models has a
brane of type F on position {\brn d}, as indicated. $SU(5)$ models
are those with $x=0$ and one or two extra branes of type C.

We observe that only two hypercharge embeddings do not allow a
large string scale.

\begin{center}
\begin{tabular}{rlrrr}
\hline $x$&& $Y~~~~~~~~~~~~~$
&   $M_{max}$ No-SUSY (TeV)    &     $M_{max}$ SUSY (TeV)    \\
\hline \hline
0&&$-\frac13 Q_{\brn a}-\frac12 Q_{\brn b}+Q_{\brn d}$
&$2.133\times 10^{10}$&   $7.168\times 10^{12}$\\
1&&$\frac23 Q_{\brn a}+\frac12 Q_{\brn b}+Q_{\brn c}$
&$1419.91$&  433114. \\
${1\over 2}$&&$\frac16 Q_{\brn a}+\frac12 Q_{\brn c}-\frac12Q_{\brn d}$
&$1.041\times 10^{31}$&   $5.314\times 10^{21}$\\
${1\over 2}$&(F)&$\frac16 Q_{\brn a}+\frac12 Q_{\brn c}-\frac32Q_{\brn d}$
&$4.797\times 10^{29}$&    $5.975\times 10^{20}$\\
0 &(C)&$-\frac13 Q_{\brn a}-\frac12 Q_{\brn b}$
&$5.024\times 10^{10}$&    $2.043\times 10^{13}$\\
$-{1\over 2}$&&$-\frac56 Q_{\brn a}-Q_{\brn b}-\frac12Q_{\brn c} +\frac32Q_{\brn d} $
&$1.041\times 10^{31}$&   $5.314\times 10^{21}$\\
${3\over 2}$&&$\frac76 Q_{\brn a}+Q_{\brn b}+\frac32Q_{\brn c} +\frac12Q_{\brn d} $
&$3.802\times 10^{-5}$&    $5.828\times 10^{-10}$\\
\hline
\end{tabular}
\end{center}

\newpage



\addcontentsline{toc}{section}{References}
\begin{thebibliography}{99}




\bibitem{Susskind:2003kw}
 L.~Susskind,
 {\em ``The anthropic landscape of string theory,''}
 \hre{hep-th}{0302219}.

\bibitem{Schellekens:2006xz}
 A.~N.~Schellekens,
 {\em ``The landscape 'avant la lettre',''}
 \hre{physics}{0604134}.

\bibitem{Denef:2006ad}
 F.~Denef and M.~R.~Douglas,
 {\em ``Computational complexity of the landscape. I,''}
 \hre{hep-th}{0602072}.

 \bibitem{dsm1}
 I.~Antoniadis, E.~Kiritsis and T.~N.~Tomaras,
 {\em ``A D-brane alternative to unification,''}
 Phys.\ Lett.\ B {\bf 486} (2000) 186
 [\hre{hep-ph}{0004214}];\\
 {\em ``D-brane Standard Model,''}
 Fortsch.\ Phys.\  {\bf 49} (2001) 573
 \hre{hep-th}{0111269}.


 \bibitem{aiqu}
 G.~Aldazabal, L.~E.~Ibanez, F.~Quevedo and A.~M.~Uranga,
 {\em ``D-branes at singularities: A bottom-up approach to the string embedding of
 the standard model,''}
 JHEP {\bf 0008} (2000) 002
 \hre{hep-th}{0005067}.

 \bibitem{Ibanez:2001nd}
 L.~E.~Ibanez, F.~Marchesano and R.~Rabadan,
 {\em ``Getting just the standard model at intersecting branes,''}
 JHEP {\bf 0111} (2001) 002
 \hre{hep-th}{0105155}.

\bibitem{AD}
 I.~Antoniadis and S.~Dimopoulos,
 {\em ``Splitting supersymmetry in string theory,''}
 Nucl.\ Phys.\ B {\bf 715} (2005) 120
 \hre{hep-th}{0411032}.

\bibitem{Fuchs:2000cm}
 J.~Fuchs, L.~R.~Huiszoon, A.~N.~Schellekens, C.~Schweigert and J.~Walcher,
 {\em ``Boundaries, crosscaps and simple currents,''}
 Phys.\ Lett.\ B {\bf 495} (2000) 427\newline \hre{hep-th}{0007174}.


\bibitem{Gato-Rivera:1991ru}
  B.~Gato-Rivera and A.~N.~Schellekens,
  Commun.\ Math.\ Phys.\  {\bf 145} (1992) 85.

%
\bibitem{Kreuzer:1993tf}
  M.~Kreuzer and A.~N.~Schellekens,
  Nucl.\ Phys.\ B {\bf 411}, 97 (1994)
  [arXiv:hep-th/9306145].




\bibitem{Dijkstra:2004ym}
 T.~P.~T.~Dijkstra, L.~R.~Huiszoon and A.~N.~Schellekens,
 {\em ``Chiral supersymmetric standard model spectra from orientifolds of Gepner
 models,''}
 Phys.\ Lett.\ B {\bf 609} (2005) 408
 \hre{hep-th}{0403196}.

\bibitem{Dijkstra:2004cc}
 T.~P.~T.~Dijkstra, L.~R.~Huiszoon and A.~N.~Schellekens,
 {\em ``Supersymmetric standard model spectra from RCFT orientifolds,''}
 Nucl.\ Phys.\ B {\bf 710} (2005) 3\newline
 \hre{hep-th}{0411129}.



\bibitem{Angelantonj:1996uy}
 C.~Angelantonj, M.~Bianchi, G.~Pradisi, A.~Sagnotti and Y.~S.~Stanev,
 {\em ``Chiral asymmetry in four-dimensional open- string vacua,''}
 Phys.\ Lett.\ B {\bf 385}, 96 (1996)
 \hre{hep-th}{9606169}.

 \bibitem{Blumenhagen:2005mu}
 R.~Blumenhagen, M.~Cvetic, P.~Langacker and G.~Shiu,
 {\em ``Toward realistic intersecting D-brane models,''}
\hre{hep-th}{0502005}.

 \bibitem{Honecker:2004kb}
 G.~Honecker and T.~Ott,
 {\em ``Getting just the supersymmetric standard model at intersecting branes  on
 the Z$_6$-orientifold,''}
 Phys.\ Rev.\ D {\bf 70}, 126010 (2004)
 [Erratum-ibid.\ D {\bf 71}, 069902 (2005)]
 \hre{hep-th}{0404055}.

%
\bibitem{Angelantonj:1996mw}
 C.~Angelantonj, M.~Bianchi, G.~Pradisi, A.~Sagnotti and Y.~S.~Stanev,
 {\em ``Comments on Gepner models and type I vacua in string theory,''}
 Phys.\ Lett.\ B {\bf 387} (1996) 743
 \hre{hep-th}{9607229}.


\bibitem{Blumenhagen:1998tj}
 R.~Blumenhagen and A.~Wisskirchen,
 {\em ``Spectra of 4D, N = 1 type I string vacua on non-toroidal CY threefolds,''}
 Phys.\ Lett.\ B {\bf 438}, 52 (1998)
 \hre{hep-th}{9806131}.

\bibitem{Aldazabal:2003ub}
 G.~Aldazabal, E.~C.~Andres, M.~Leston and C.~Nunez,
 {\em ``Type IIB orientifolds on Gepner points,''}
 JHEP {\bf 0309}, 067 (2003)
 \hre{hep-th}{0307183}.

%
\bibitem{Brunner:2004zd}
 I.~Brunner, K.~Hori, K.~Hosomichi and J.~Walcher,
 {\em ``Orientifolds of Gepner models,''}
 \hre{hep-th}{0401137}.


%
\bibitem{Blumenhagen:2004cg}
 R.~Blumenhagen and T.~Weigand,
 {\em ``Chiral supersymmetric Gepner model orientifolds,''}
 JHEP {\bf 0402} (2004) 041
 \hre{hep-th}{0401148}.


\bibitem{Aldazabal:2004by}
 G.~Aldazabal, E.~C.~Andres and J.~E.~Juknevich,
 {\em ``Particle models from orientifolds at Gepner-orbifold points,''}
 JHEP {\bf 0405}, 054 (2004)
 \hre{hep-th}{0403262}.

\bibitem{Gmeiner:2005vz}
 F.~Gmeiner, R.~Blumenhagen, G.~Honecker, D.~Lust and T.~Weigand,
 {\em ``One in a billion: MSSM-like D-brane statistics,''}
 JHEP {\bf 0601} (2006) 004
 \hre{hep-th}{0510170}.

 \bibitem{dsm3}
 I.~Antoniadis, E.~Kiritsis, J.~Rizos and T.~N.~Tomaras,
 {\em ``D-branes and the standard model,''}
 Nucl.\ Phys.\ B {\bf 660} (2003) 81
 \hre{hep-th}{0210263}.

 \bibitem{au1}
 I.~Antoniadis, E.~Kiritsis and J.~Rizos,
 {\em ``Anomalous U(1)s in type I superstring vacua,''}
 Nucl.\ Phys.\ B {\bf 637} (2002) 92
 \hre{hep-th}{0204153}.



 \bibitem{KN}
B. Kors and P. Nath,
"A St\"{u}ckelberg extension of the standard model",
{\it Phys. Lett. B} {\bf 586} (2004) p. 366,
\hre{hep-ph}{0402047}\\
{ ``A supersymmetric Stueckelberg U(1) extension of the MSSM,''}
 JHEP {\bf 0412} (2004) 005
  \hre{hep-ph}{0406167}\\
``Aspects of the St\"{u}ckelberg extension",
\hre{hep-ph}{0503208}



\bibitem{leigh}
 D.~Berenstein, V.~Jejjala and R.~G.~Leigh,
 {\em ``The standard model on a D-brane,''}
 Phys.\ Rev.\ Lett.\  {\bf 88} (2002) 071602
 \hre{hep-ph}{0105042}.

\bibitem{trinification}
A. de R\'{u}jula, H. Georgi, and S. L. Glashow,
in \em{Fifth Workshop on Grand Unification}, edited by K.
Kang, H. Fried, and P. Frampton (World Scientific, Singapore,
1984); Y. Achiman and B.
Stech, in \em{New Phenomena in Lepton-Hadron Physics}, edited
by D. E. C. Fries and J. Wess (Plenum, New York, 1979).


\bibitem{d-trinif}
 G.~K.~Leontaris and J.~Rizos,
 {\em ``A D-brane inspired U(3)$_C \times  $U(3)$_L \times $U(3)$_R$ model,''}
 Phys.\ Lett.\ B {\bf 632} (2006) 710
 \hre{hep-ph}{0510230};\\
 {\em ``A D-brane inspired trinification model,''}
\hre{hep-ph}{0603203}.


 \bibitem{Cardy:1989ir}
 J.~L.~Cardy,
 {\em ``Boundary Conditions, Fusion Rules And The Verlinde Formula,''}
 Nucl.\ Phys.\ B {\bf 324}, 581 (1989).

 \bibitem{Uranga:2000xp}
 A.~M.~Uranga,
 {\em ``D-brane probes, RR tadpole cancellation and K-theory charge,''}
 Nucl.\ Phys.\ B {\bf 598} (2001) 225
 \hre{hep-th}{0011048}.

\bibitem{Gato-Rivera:2005qd}
 B.~Gato-Rivera and A.~N.~Schellekens,
 {\em ``Remarks on global anomalies in RCFT orientifolds,''}
 Phys.\ Lett.\ B {\bf 632} (2006) 728
 \hre{hep-th}{0510074}.


 \bibitem{d-effective}
 C.~Corian\'o, N.~Irges and E.~Kiritsis,
 {\em ``On the effective theory of low scale orientifold string vacua,''}
 \hre{hep-ph}{0510332}.


 \bibitem{cvetic}
 M.~Cvetic, I.~Papadimitriou and G.~Shiu,
 {\em ``Supersymmetric three family SU(5) grand unified models from type IIA
 orientifolds with intersecting D6-branes,''}
 Nucl.\ Phys.\ B {\bf 659} (2003) 193
 [Erratum-ibid.\ B {\bf 696} (2004) 298]
 \hre{hep-th}{0212177}.

\bibitem{Chen:2006ip}
 C.~M.~Chen, T.~Li and D.~V.~Nanopoulos,
 {\em ``Flipped and unflipped SU(5) as type IIA flux vacua,''}
 \hre{hep-th}{0604107}.

 %
\bibitem{Blumenhagen:2001te}
 R.~Blumenhagen, B.~Kors, D.~Lust and T.~Ott,
 {\em ``The standard model from stable intersecting brane world orbifolds,''}
 Nucl.\ Phys.\ B {\bf 616} (2001) 3
 \hre{hep-th}{0107138}.

\bibitem{Ellis:2002ci}
 J.~R.~Ellis, P.~Kanti and D.~V.~Nanopoulos,
 {\em ``Intersecting branes flip SU(5),''}
 Nucl.\ Phys.\ B {\bf 647} (2002) 235
 \hre{hep-th}{0206087}.

\bibitem{floratos}
 M.~Axenides, E.~Floratos and C.~Kokorelis,
 {\em ``SU(5) unified theories from intersecting branes,''}
 JHEP {\bf 0310} (2003) 006
 \hre{hep-th}{0307255}.


\bibitem{nano}
 C.~M.~Chen, G.~V.~Kraniotis, V.~E.~Mayes, D.~V.~Nanopoulos and J.~W.~Walker,
 {\em ``A K-theory anomaly free supersymmetric flipped SU(5) model from
 intersecting branes,''}
 Phys.\ Lett.\ B {\bf 625} (2005) 96
 \hre{hep-th}{0507232}.

\bibitem{Antebi:2005hr}
 Y.~E.~Antebi, Y.~Nir and T.~Volansky,
 {\em ``Solving flavor puzzles with quiver gauge theories,''}
 Phys.\ Rev.\ D {\bf 73} (2006) 075009
 \hre{hep-ph}{0512211}.

\bibitem{bere}
 D.~Berenstein,
 {\em ``Branes vs. GUTS: Challenges for string inspired
phenomenology,''}
 \hre{hep-th}{0603103}.

 \bibitem{ps1}
 R.~Blumenhagen, L.~Gorlich and T.~Ott,
 {\em ``Supersymmetric intersecting branes on the type IIA T$^6$/$Z_4$ orientifold,''}
 JHEP {\bf 0301} (2003) 021
 \hre{hep-th}{0211059}.

 \bibitem{ps2}
 G.~Honecker,
 {\em ``Chiral supersymmetric models on an orientifold of $Z_4 \times Z_2$ with
 intersecting D6-branes,''}
 Nucl.\ Phys.\ B {\bf 666} (2003) 175
 \hre{hep-th}{0303015}.

 \bibitem{ps3}
 M.~Cvetic, T.~Li and T.~Liu,
 {``Supersymmetric Pati-Salam models from intersecting D6-branes: A road to
 the standard model,''}
 Nucl.\ Phys.\ B {\bf 698} (2004) 163\newline
 \hre{hep-th}{0403061}.

\bibitem{lr}
 G.~K.~Leontaris and J.~Rizos,
 {\em ``A Pati-Salam model from branes,''}
 Phys.\ Lett.\ B {\bf 510} (2001) 295
 \hre{hep-ph}{0012255}.

 \bibitem{lr1}
 T.~Dent, G.~Leontaris and J.~Rizos,
 {\em ``Fermion masses and proton decay in string-inspired SU(4)$\times$SU(2)$^2\times$
 U(1)$_X$,''}
 Phys.\ Lett.\ B {\bf 605} (2005) 399\newline
 \hre{hep-ph}{0407151}.

\bibitem{d-review}
 E.~Kiritsis,
 {\em ``D-branes in standard model building, gravity and cosmology,''}
 Fortsch.\ Phys.\  {\bf 52} (2004) 200
 [Phys.\ Rept.\  {\bf 421} (2005) 105]
 \hre{hep-th}{0310001}.

\bibitem{li}
 L.~F.~Li,
 {\em ``Group Theory Of The Spontaneously Broken Gauge Symmetries,''}
 Phys.\ Rev.\ D {\bf 9} (1974) 1723.



\bibitem{Aldazabal:2006nz}
 G.~Aldazabal, E.~Andres and J.~E.~Juknevich,
 {\em ``On SUSY standard-like models from orbifolds of D = 6 Gepner orientifolds,''}
\hre{hep-th}{0603217}.




















%













\end{thebibliography}
\end{document}